\documentclass{article}

\usepackage[english]{babel}

\usepackage[letterpaper,top=2cm,bottom=2cm,left=3cm,right=3cm,marginparwidth=1.75cm]{geometry}

\usepackage{amsmath}
\usepackage{graphicx}
\usepackage[colorlinks=true, allcolors=blue]{hyperref}
\DeclareMathOperator*{\argmax}{arg\,max}
\DeclareMathOperator*{\argmin}{arg\,min}
\usepackage{multirow}
\usepackage{float}
\usepackage{arydshln}

\providecommand{\keywords}[1]{\textbf{\textit{Keywords: }} #1}

\title{To MCMC or not to MCMC: Evaluating non-MCMC methods for Bayesian penalized regression}

\author{
  $^*$Florian D. van Leeuwen \\
  Department of Methods and Statistics \\
  Utrecht University \\
  \texttt{f.d.vanleeuwen@uu.nl} \\
  Sara van Erp \\
  Department of Methods and Statistics \\
  Utrecht University \\
}

\begin{document}

\maketitle
\def\thefootnote{*}\footnotetext{Corresponding Author}\def\thefootnote{\arabic{footnote}}

\begin{abstract}

Markov Chain Monte Carlo (MCMC) sampling is computationally expensive, especially for complex models. Alternative methods make simplifying assumptions about the posterior to reduce computational burden, but their impact on predictive performance remains unclear. This paper compares MCMC and non-MCMC methods for high-dimensional penalized regression, examining when computational shortcuts are justified for prediction tasks.

We conduct a comprehensive simulation study using high-dimensional tabular data, then validate findings with empirical datasets featuring both continuous and binary outcomes. An in-depth analysis of one dataset provides a step-by-step tutorial implementing various algorithms in R.

Our results show that mean-field variational inference consistently performs comparably to MCMC methods. In simulations, mean-field VI exhibited 3-90\% higher MSE across scenarios while reducing runtime by 7-30× compared to Hamiltonian Monte Carlo. Empirical datasets revealed dramatic speed-ups (100-400×) in some cases with similar or superior predictive performance. However, performance varied: some cases showed over 100× MSE increases with only 30× speed-ups, highlighting the context-dependent nature of these trade-offs.

\end{abstract}


\section{Introduction}

In the realm of prediction modeling, many potential sources of uncertainty exist, including measurement bias, sampling variability, model choices, and missing or incorrect data \cite{gruber_sources_2023}. Most predictive models disregard uncertainty by only providing a point estimate of the predicted value. When making decisions, using point estimates might lead to different choices than when the uncertainty is included in the decision-making process \cite{holt_risk_2002}. In addition, modeling uncertainty can result in important information not otherwise obtained. For example, in forecasting extreme climate events such as floods, quantifying different sources of uncertainty is important to decide where improvements in the measuring systems can be made \cite{biondi_bayesian_2012}. Climate models are often very complex, in the sense that there are many parameters in the prediction model, sometimes even exceeding the number of samples (high-dimensional data). A common problem is then overfitting, where the model focuses on trends in the given data too much, so it does not generalize well, as it becomes hard to distinguish (non) influential parameters of a model \cite{johnstone_statistical_2009}.

Bayesian statistics is a framework often used for uncertainty quantification that enables us to obtain a probability distribution for our parameters of interest. In many models, the posterior distribution of the parameters can be combined with new data to obtain a distribution of predicted values (posterior predictive distribution). This makes it possible to obtain uncertainty estimates for predictions, even for very complicated non-parametric models \cite{papamarkou_position_2024}. Another benefit of the Bayesian framework is that a specific class of priors, called shrinkage priors, can be used to avoid overfitting. Shrinkage priors assist in reducing small coefficients towards zero, and keeping large coefficients (mainly) untouched. Commonly used shrinkage priors are the spike-and-slab \cite{mitchell_bayesian_1988, george_variable_1993} and the (regularized) horseshoe \cite{carvalho_handling_2009, piironen_sparsity_2017}.

The advantage of the Bayesian framework in terms of uncertainty quantification does come at a cost: to obtain the full posterior distribution, Markov Chain Monte Carlo (MCMC) sampling is often needed. MCMC methods for posterior estimation rely on an iterative process of drawing samples, converging to the true posterior distribution as the number of iterations increases. Popular MCMC algorithms are Metropolis-Hastings \cite{metropolis_equation_1953,hastings_monte_1970}, Gibbs sampling \cite{geman_stochastic_1984} and Hamiltonian Monte Carlo \cite{brooks_mcmc_2011}. Unfortunately, it can take many iterations for the MCMC algorithms to converge to the posterior, resulting in extensive running time, especially for complex models (e.g., \cite{izmailov_what_2021, rainforth_modern_2024, gunapati_variational_2022, ballnus_comprehensive_2017, dang_stochastic_2019}). To circumvent this problem, alternatives to MCMC methods have been proposed, such as Laplace approximation \cite{rue_bayesian_2017} or variational inference \cite{blei_variational_2017}. In these methods, some assumptions about the posterior are made to simplify the estimation procedure. These crude approximations lead to significant reductions in computational costs, but might result in less accurate estimates \cite{izmailov_what_2021, blei_variational_2017}.

As researchers try to make predictions based on more high-dimensional data, it becomes increasingly important to investigate the accuracy of fast Bayesian algorithms that enable uncertainty quantification. Therefore, this paper aims to evaluate the performance of three popular non-MCMC methods (variational inference, Laplace approximation, and pathfinder) and compare them against Hamiltonian Monte Carlo sampling. The paper is organized as follows: Bayesian penalization and specific shrinkage priors are first discussed (Section 2). Furthermore, an overview of methods that can be used to obtain posterior distributions is provided. For methods based on MCMC, Metropolis-Hastings, Gibbs sampling and Hamiltonian Monte Carlo are discussed (Section 3). For non-MCMC methods, the focus is on variational inference and Laplace approximation (Section 4). We also discuss a hybrid method that uses the approximate Pathfinder algorithm to obtain initial values for an MCMC method. Convergence metrics for both MCMC and non-MCMC methods are presented and discussed. Different algorithms are then compared in a simulation study for high-dimensional tabular data (Section 5). The metrics to assess the performance include convergence, predictive performance, and estimation speed. Empirical datasets are used to substantiate the findings from the simulation study (Section 6). One empirical dataset is used to create a tutorial on how to use and assess the performance of MCMC and non-MCMC methods in R. The results are discussed in Section 7.


\section{Bayesian penalization}

Bayesian statistics is based on Bayes' theorem:

\begin{equation}
p(\boldsymbol{\theta} | D) = \frac{p(D| \boldsymbol{\theta}) \cdot p(\boldsymbol{\theta})}{p(D)},
\label{eq:bayes}
\end{equation}

\noindent
where $\boldsymbol{\theta}$ is a vector of parameters, $D$ is the data, $p(D | \boldsymbol{\theta})$ is the likelihood of the data, $p(\boldsymbol{\theta})$ is the prior, $p(D)$ is a normalizing constant, and $p(\boldsymbol{\theta} | D) $ is the posterior. In this paper, $\boldsymbol{\theta}$ denotes quantities referred to as either parameters or latent variables; the two terms are used interchangeably. Using the law of total probability, $p(D)$ can be expanded to:

\noindent
\begin{equation}
    p(D) = \int p(D|\boldsymbol{\theta}) p(\boldsymbol{\theta}) d\boldsymbol{\theta}.
    \label{eq:bayes2}
\end{equation}

$p(D)$ is often intractable to compute analytically, and when $\boldsymbol{\theta}$ has many dimensions, numeric integration becomes imprecise and computationally intractable \cite[p.~115]{lambert_students_2018}. 

\subsection{Likelihood}
The likelihood represents the probability of observing the data given the model parameters. In the case of predicting a continuous outcome, a common model choice is linear regression. The model for linear regression is given by:

\begin{equation}
    y_i| x_i, \beta_0,\beta, \sigma^2 \sim Normal(\beta_0 + \sum^p_{j=1} x_{ij} \beta_j, \sigma^2),
    \label{eq:lh}
\end{equation}

\noindent
where $y_i$ represents the outcome for an observation $i$, $\beta_0$ the intercept, $x_i$ the covariates for observation $i$, $\beta = (\beta_1, ..., \beta_p)$ a vector of regression coefficients with a specific $\beta$ denoted by $j$, and $\sigma^2$ is the residual variance. The model assumes that the outcome is normally distributed around the predicted values, with some noise around it. All the $X$ variables are standardized to be on the same scale, which is important when using shrinkage priors to ensure the shrinkage has the same effect on each parameter.

\subsection{Prior}
Selecting an appropriate prior for the regression coefficients can be a challenge; depending on the setting, there might be different optimal priors \cite{van_de_schoot_bayesian_2021}. Choosing a good prior for the problem at hand is very important. The choice of prior can improve the inference and the predictive performance of the model \cite{fortuin_priors_2022, van_erp_shrinkage_2019}. Specifically, the prior can have a strong influence over the posterior by concentrating the probability mass into a subspace of the parameter space \cite{gelman_prior_2017}.

\begin{figure}
    \centering
    \includegraphics[width=0.95\linewidth]{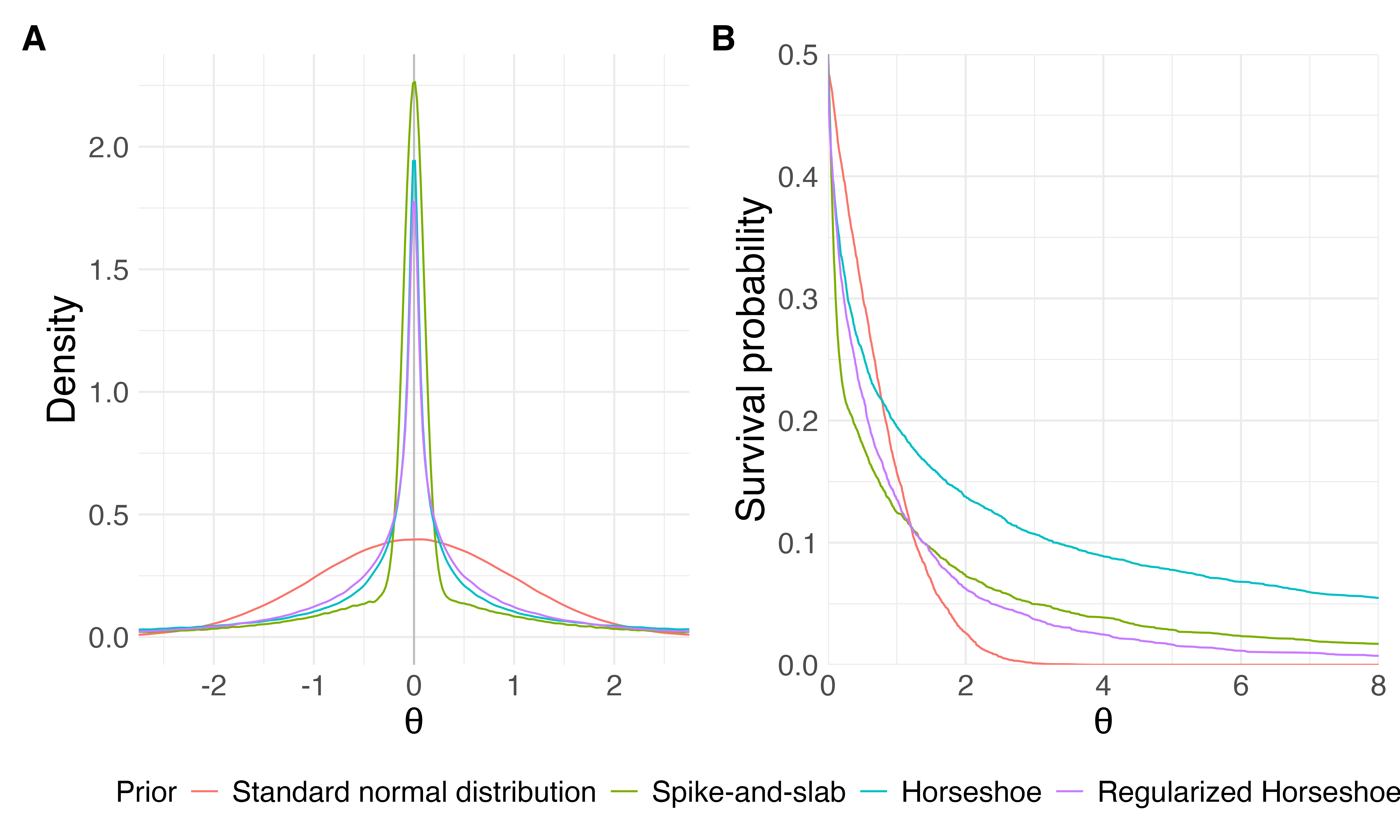}
    \caption{A visualization of the Spike-and-slab and (regularized) Horseshoe priors, adapted from \cite{van_erp_tutorial_2020}. Plot \textbf{A} shows the density of the different priors. There is a lot of density around zero for the shrinkage priors compared to a standard normal. Plot \textbf{B} illustrates the survival function, which is the probability (y-axis) that the parameter has a value greater than the values on the x-axis. The survival function gives insights into the behavior of the priors in the tails; the slower the function moves towards zero, the heavier the tails. }
    \label{fig:prior}
\end{figure}

Constraining the parameters to a specific space can be beneficial in the case of high-dimensional data. All variables should have a fair probability of being important in the prediction task, while pruning out non-important variables. This can be done by using a shrinkage prior. These shrinkage priors use distributions that tend to have high density at zero and wide tails, indicating that we expect sparsity with a few relevant variables \cite{polson_shrink_2010}. There are many different shrinkage priors (see \cite{van_erp_shrinkage_2019} for an overview). In this paper, we focus on two commonly used ones: the spike-and-slab \cite{mitchell_bayesian_1988, george_variable_1993}, and the (regularized) horseshoe \cite{carvalho_handling_2009, piironen_sparsity_2017}. A visualization of the shrinkage priors given in Figure \ref{fig:prior}; further explanation is given in the next sections.

\subsubsection{Spike-and-slab}

The idea of the spike-and-slab prior is to have a coefficient either be zero (spike) or non-zero (slab). The prior can be written as a two-component mixture of Gaussians:

\begin{equation}
    \beta_j | \lambda_j, c, \epsilon \sim \lambda_j Normal(0, c^2) + (1- \lambda_j) Normal(0, \epsilon^2),
    \label{eq:sl1}
\end{equation}

\begin{equation}
    \lambda_j \sim Ber(\pi),    j = 1,...,p ,
    \label{eq:sl2}
\end{equation}

\noindent
where both $ \epsilon$ and $c$ are variance terms with $\epsilon < c$, the assignment of the variable is performed by the indicator variable $\lambda_j \in {0,1}$ and $\pi$ indicates the prior inclusion probability, which represents the assumed sparsity of the coefficient vector. If the coefficient $\beta_j$ is close to zero then $\lambda_j = 0$, and $\lambda_j = 1$ otherwise.

When using the spike-and-slab prior, values for $\epsilon$, the slab width c and the prior inclusion probability $\pi$ need to be chosen. For the indicator variable $\lambda_j$, only two values are allowed, making it a discrete parameter.

\subsubsection{(regularized) Horseshoe}

An continuous alternative to the spike-and-slab is the horseshoe prior \cite{carvalho_handling_2009}:

\begin{equation}
    \beta_j | \lambda_j, \tau \sim Normal(0, \tau^2 \lambda^2_j),
    \label{eq:hs1}
\end{equation}

\begin{equation}
    \lambda_j \sim C^+ (0,1), j = 1,...,p , 
    \label{eq:hs2}
\end{equation}

\begin{equation}
    \tau \sim C^+ (0,\sigma_\tau),
    \label{eq:hs3}
\end{equation}

\noindent
where $\tau$ is a global hyperparameter that shrinks all parameters towards zero, and it has a half-Cauchy hyperprior with variance $\sigma_\tau$, which is often set to 1. Here, $C^+$ denotes the positive half-Cauchy distribution. The local hyperparameter $\lambda_j$ allows some $\beta_j$ to escape the shrinkage through the heavy-tailed half-Cauchy prior \cite{carvalho_handling_2009}. More recently, the regularized horseshoe was introduced \cite{piironen_sparsity_2017}:

\begin{equation}
    \beta_j | \lambda_j, \tau, c \sim Normal(0, \tau^2  \Tilde{\lambda^2_j}),
    \label{eq:hs_r1}
\end{equation}

\begin{equation}
    \Tilde{\lambda^2_j} = \frac{c^2 \lambda^2_j}{c^2 + \tau^2 \lambda^2_j}
    \label{eq:hs_r2}
\end{equation}

\begin{equation}
    \lambda_j \sim C^+ (0,1), j = 1,...,p.
    \label{eq:hs_r3}
\end{equation}

\noindent
The idea is that not only small coefficients, but also large ones are shrunken through the $c$ parameter \cite{piironen_sparsity_2017}. Instead of choosing a single value, it is common to put a prior on the value of $c$:

\begin{equation}
    c^2 = \text{Inv-Gamma}(\frac{v}{2}, \frac{vs^2}{2}) 
\end{equation}

where $v$ indicates the degrees of freedom and $s$ being a scale parameter. 

Prior information about the sparsity in the system can be included in the prior through the parameter $\tau$:

\begin{equation}
    \tau_0 = \frac{p_0}{p - p_0} \frac{\sigma}{\sqrt{n}},
\end{equation}

\noindent
with $p_0$ being the number of non-zero parameters, $p$ the total number of predictors, $\sigma$ the residual standard deviation and $n$ the number of samples. 

The main advantage of the regularized horseshoe prior is that it can regularize parameters that are far from zero, which is helpful if there is weak evidence in the data. Furthermore, setting an informative value for the global shrinkage prior $\tau$ can improve the parameter estimates and predictive performance \cite{piironen_sparsity_2017}.

\subsection{Posterior Predictive distribution}

The posterior for the latent variables can be used to make predictions for future events ($\hat{y}$), given new data ($D_{new}$). The combination of the posterior distribution with the new data is called the posterior predictive distribution \cite[p.~7]{gelman_bayesian_2013}:

\begin{equation}
    P(D_{new}|D) = \int P(D_{new}|\theta)P(\theta|D)d\theta
\end{equation}

When using MCMC, we obtain M draws from the posterior distribution of the model parameters. In the context of linear regression (Equation \ref{eq:lh}), this equation can be written as:

\begin{equation}
    P(y_{new}|x_{new}, x, y) = \int P(y_{new}|x_{new}, \theta)P(\theta|y, x)d\theta
\end{equation} 

These draws can then be used to obtain the Monte Carlo estimate of the posterior predictive distribution:

\begin{equation}
    P(y_{new}|x_{new}, x, y)  = \frac{1}{M} \sum_{m=1}^{M}  P(y_{new}|x_{new}, \theta^{(m)})
\end{equation}

where $\theta^{(m)} \sim P(\theta|y, x)$. $P(y_{new}|x_{new}, x, y)$ is the distribution of the predicted values, and can be used to make inferences about the uncertainty of future events. Each new sample thus obtains an individual PPD, since its features $x_{new}$ are multiplied by all $m$ draws. There are different types of PPD, described in [p.~115] \cite{gelman_regression_2021}, based on the level of uncertainty. In this paper, we use the predictive distribution for a new observation, which is defined as follows for the linear regression model:

\begin{equation}
     P(y_{new}|x_{new}, x, y) = \frac{1}{M} \sum_{m=1}^{M} Normal(\beta_0^{(m)} + \sum^p_{j=1} x_{new_{j}} \beta_j^{(m)}, \sigma^{2(m)})
\end{equation}

\section{Markov Chain Monte Carlo (MCMC) sampling}

Markov Chain Monte Carlo (MCMC) sampling provides a method to obtain samples from the posterior through simulations, despite the posterior being only known up to a constant ($P(D)$, Equation \ref{eq:bayes2}). The method combines the concepts of Markov chains with Monte Carlo simulations. 

The general idea of Markov chains is to have a sequence of possible events, where the next event only depends on the current event. When looking for the posterior, the Markov chain can be seen as a movement of points through the parameter space by sequentially applying Markov transitions. A Markov transition entails moving from an initial state to a new state based solely on the initial state. The Markov chain is useful if the Markov transition preserves the target distribution (posterior), meaning that it eventually reaches and explores the target distribution. As the Markov chain depends on the previous values, the entire chain is conditional on the starting value. In practice, multiple Markov chains are used to ensure that a specific starting value does not send the Markov chain into an area where it will get stuck.  

The Monte Carlo aspect refers to the fact that the algorithm uses repeated sampling to obtain the posterior. Subsequently, these samples can be used to represent aspects of the posterior, such as the mean, by simply adding all samples and dividing them by the total number of samples. Furthermore, credibility intervals can be taken by selecting a region containing 1-alpha \% of the samples. 

The exploration of the target distribution by the Markov chains can be specified in three phases: 1) looking for the high density area; 2) finding the high density area; 3) exploring the high density area. Phases 1 and 2 are part of the warm-up period, and the draws obtained in these phases are not used in the final calculation of the statistics. Given sufficient steps, the Markov chain will eventually explore our desired posterior \cite{betancourt_conceptual_2018}. In practice, the computational costs might be very high for a specific problem. 

Many different MCMC algorithms exist that differ in how proposals in the Markov chain are chosen and assessed. In this paper, the focus will be on three commonly used methods: Metropolis-Hastings, Gibbs sampling and Hamiltonian Monte Carlo. A high-level overview of the algorithms is given in Table \ref{tab:mcmc}, the next sections describe the algorithms in more depth. 

\begin{table}[htbp]
\centering
\resizebox{\textwidth}{!}{%
\begin{tabular}{|l|l|l|l|l|}
\hline
\textbf{Method} & \textbf{\begin{tabular}[c]{@{}l@{}}Acceptance \\ rule required\end{tabular}} & \textbf{\begin{tabular}[c]{@{}l@{}}Conjugate priors \\ required\end{tabular}} & \textbf{\begin{tabular}[c]{@{}l@{}}Gradient of \\ posterior required\end{tabular}} & \textbf{\begin{tabular}[c]{@{}l@{}}Discrete parameters \\  allowed\end{tabular}} \\ \hline
Metropolis-Hastings & Yes & No & No & Yes \\ \hline
Gibbs sampling & No & Yes & No & Yes \\ \hline
Hamiltonian Monte Carlo & Yes & No & Yes & No \\ \hline
\end{tabular}%
}
\caption{Summary of the MCMC methods discussed in this paper.}
\label{tab:mcmc}
\end{table}

\subsection{Metropolis-Hastings}
The Metropolis-Hastings (MH) algorithm is an often used method to obtain posterior samples using MCMC \cite{metropolis_equation_1953,hastings_monte_1970}. In the MH algorithm, there are two steps: a proposal of a new step and an assessment (either an acceptance or rejection) of this proposal. The proposals can be seen as a random walk through the parameter space. The acceptance criterion is based on how likely the proposal is compared to the sample from the previous step. The probability of accepting a proposal is given by:

\begin{equation}
   r = \frac{p(\theta_{t+1}|D)}{p(\theta_{t}|D)} \frac{J(\theta_{t}|\theta_{t+1})}{J(\theta_{t+1}|\theta_{t})},
    \label{eq:HM}
\end{equation}

\noindent
with $t$ indicating the current position and $t+1$ the proposal,  $p(\theta|D)$ being the unnormalized density \footnote{The ratio between the unnormalized posterior at time $t$ and time $t+1$ is the same as the ratio between the normalized posterior at the same points. This is due to the fact that the normalizing constant $P(D)$ scales the both the numerator and the denominator ar the same rate$ \frac{\frac{p(D| \theta_{t+1}) \cdot p(\theta_{t+1})}{p(D)}}{\frac{p(D| \theta_{t}) \cdot p(\theta_{t})}{p(D)}} = \frac{p(D| \theta_{t+1}) \cdot p(\theta_{t+1})}{p(D| \theta_{t}) \cdot p(\theta_{t})} = \frac{p(\theta_{t+1}|D)}{p(\theta_{t}|D)} $
.} and $J(\theta_{t}|\theta_{t}) $ is the proposal density at $\theta_t$ if the current position is $\theta_{t+1}$ \cite[p.~303]{lambert_students_2018}. The first fraction compares the density of the new point to the old point. The second fraction ensures that the proposal density can be asymmetric, which is sometimes necessary for constrained values \cite[p.~303]{lambert_students_2018}. In the case that $ J(\theta_{t+1}|\theta_{t}) = J(\theta_{t}|\theta_{t+1})$, which happens if the proposal distribution is Gaussian, Equation \ref{eq:HM} simplifies to:

\begin{equation}
   r = \frac{p(\theta_{t+1}|D)}{p(\theta_{t}|D)}.
    \label{eq:RWHM}
\end{equation}

\noindent
A value is then drawn from a uniform distribution with bounds at 0 and 1; if $r> u \sim U(0,1)$ then the proposal is accepted ($\theta_{t+1}\rightarrow \theta_{t}$). 

\begin{figure}[H]
    \centering
    \includegraphics[width=0.95\linewidth]{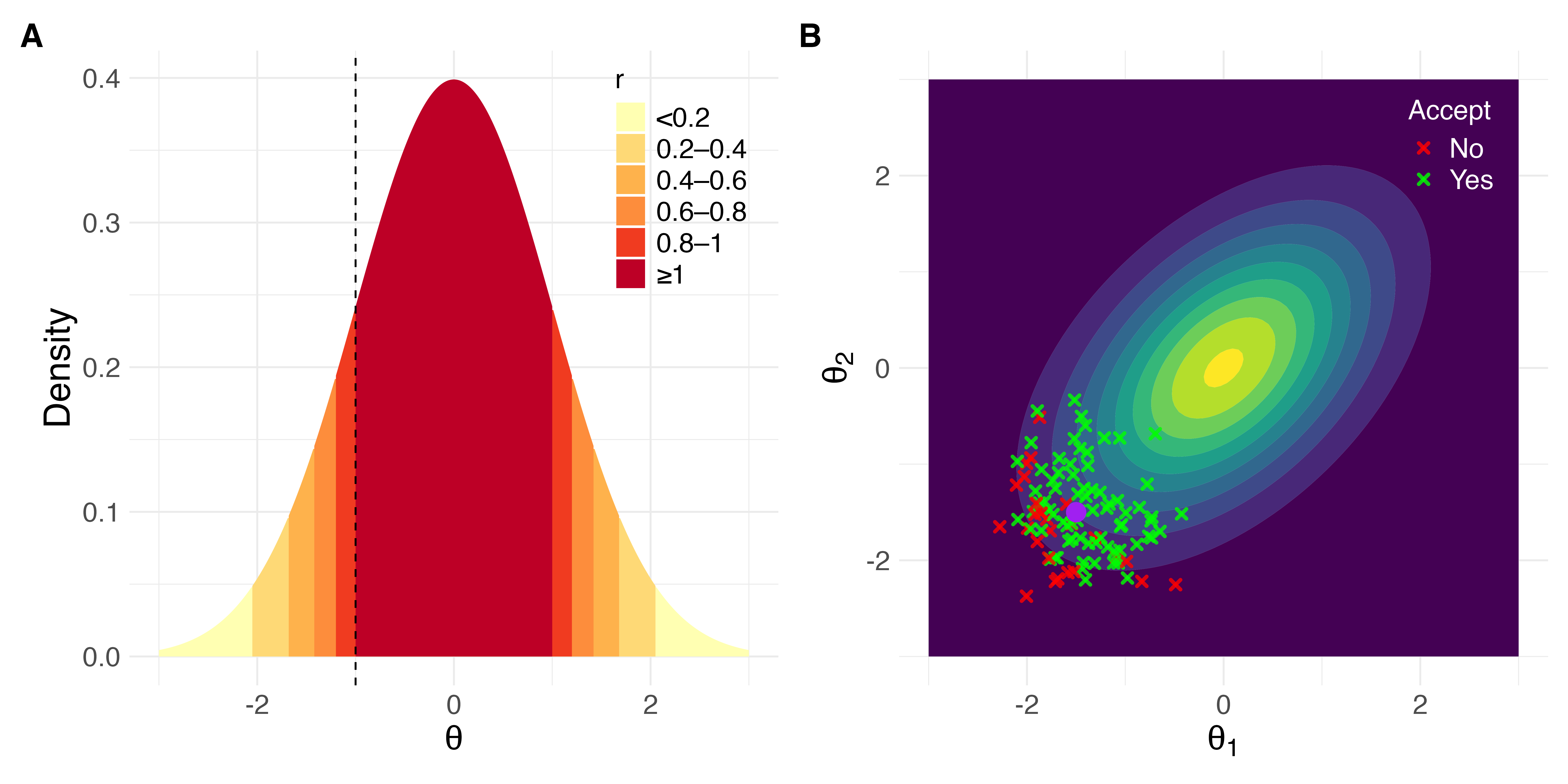}
    \caption{Illustration of the Metropolis-Hastings algorithm. In plot \textbf{A}, a standard normal distribution is used as the target distribution. The position of the current parameter value is indicated with a dashed line. In the MH algorithm, a proposal value is drawn from the proposal distribution and this new value is compared to the current value based on the ratio (r) between the current and the proposal sample. In the plot, the r values are shown based on the region where the new sample will be. All areas with higher density compared to the initial point have a r $>$ 1, so their values will be accepted. The further we move towards the tails, the lower the r values, and these samples will thus be more often rejected. In plot \textbf{B}, the target distribution is a bivariate normal distribution with the means set to 0 and the covariance to 0.5. The initial point is indicated in purple. 100 data points have been sampled from a proposal distribution, and the r value has been calculated and evaluated again $U(0,1)$. As expected, most newly accepted values have higher density, although some have lower density.}
    \label{fig:HM}
\end{figure}

As a result, a proposal sample with a higher posterior density value than the current sample is always accepted, and a proposal with a lower density value compared to the current sample is only sometimes accepted, as can be seen in Figure \ref{fig:HM}. All parameters are updated simultaneously, and if the proposal is rejected, the values in the Markov chain for the new step are the same as the previous step. The proposal distribution is not fixed, as the variance term of the proposal distribution needs to be tuned for optimal performance \cite{roberts_optimal_2001}.  A downside of the Metropolis Hastings algorithm is that the proposals are random around the current value, 
leading to the rejection of a lot of samples. This causes the performance to scale poorly with an increasing dimension and complexity of the target distribution \cite{betancourt_conceptual_2018}.

\subsection{Gibbs sampler}
The Gibbs sampler is a special version of the MH algorithm, where the proposal value is always accepted \cite{geman_stochastic_1984, carlo_markov_2004}. The key difference is that in the Gibbs sampler, the conditional distributions are used instead of a joint distribution in the MH algorithm. So not all parameters are updated at once; instead, this happens sequentially. The Gibbs sampler is much quicker in obtaining samples compared to MH, however, the sampler can only be used when the conditional distributions can be derived \cite[p.~333]{lambert_students_2018}.

The algorithm works in the following way: a vector of random values is generated for the parameters in the model $(\theta_1^0, \theta_2^0, \theta_3^0, ..., \theta_n^0)$. In an iterative process, new values are sampled using the conditional distributions, e.g., for $\theta_1^0$:

\begin{equation}
    \theta_1^1 \sim\ p(\theta_1 | \theta_2^0, \theta_3^0, ..., \theta_p^0).
\end{equation}

\begin{figure}[h]
    \centering
    \includegraphics[width=0.9\linewidth]{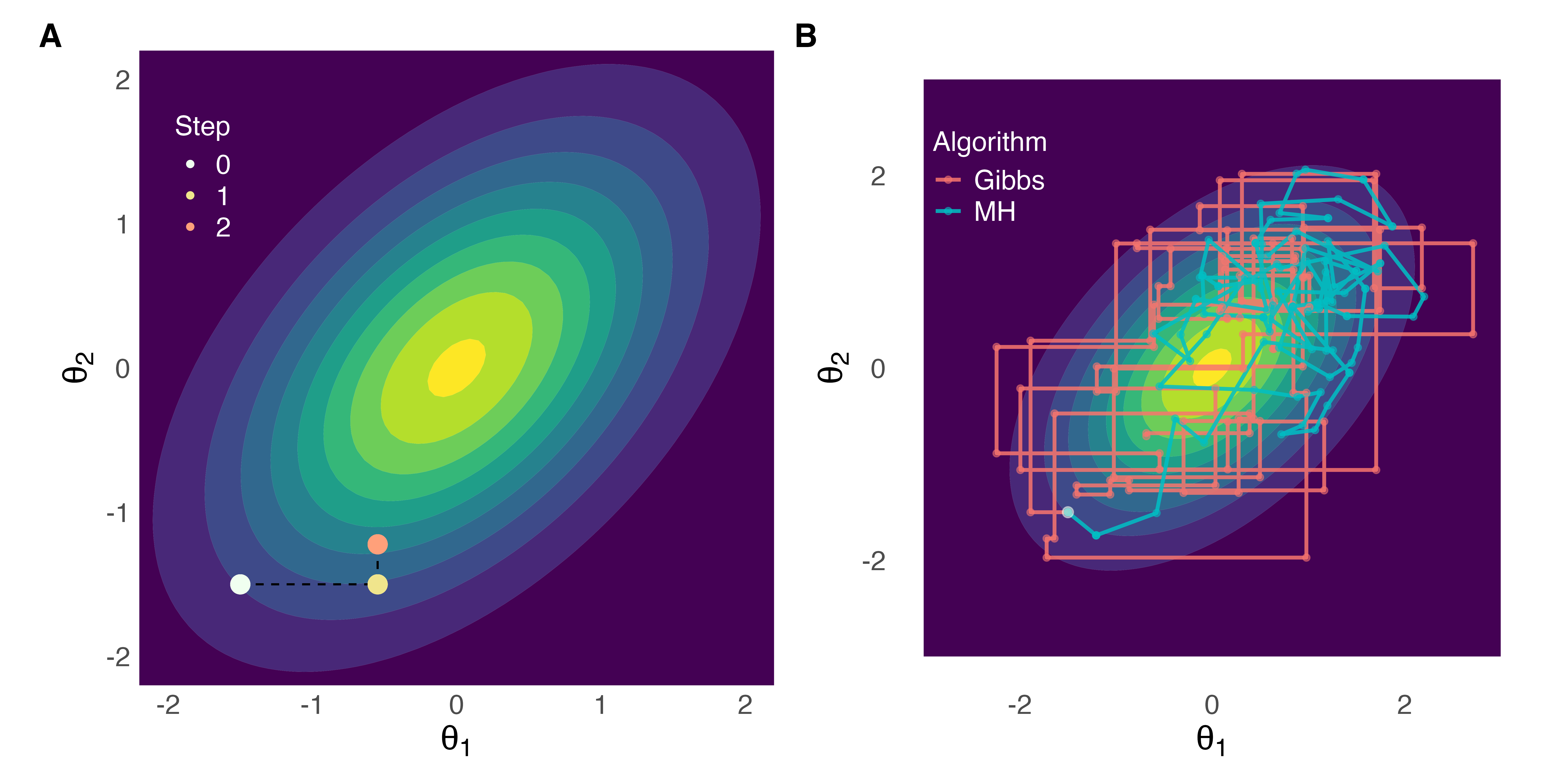}
    \caption{Illustration of the Gibbs sampler. The same bivariate normal distribution is used as in Figure \ref{fig:HM}B. In plot \textbf{A} it is shown how the Gibbs sampler iteratively updates the parameters, starting from the initial point in step 0. In step 1, the value for $\theta_1$ is updated based on the conditional distribution of $\theta_1$ given $\theta_2$; similarly, this happens for $\theta_2$ in step 2. Parameters are thus updated sequentially. In plot \textbf{B}, the samples from the Gibbs sampler are compared to those of the MH algorithm, when starting at the same initial points as in plot \textbf{A}.}
    \label{fig:GIBBS}
\end{figure}

\noindent
This process continues for each $\theta$ variable, until at some point all random values are replaced (see Figure \ref{fig:GIBBS} for an example). On the kth iteration, we will have the following:

\begin{equation}
    \theta_k^m \sim\ p(\theta_m | \theta_1^k, \theta_2^k, ..., \theta_p^{k-1}).
\end{equation}

\noindent
This results in a Gibbs sequence of length k. Similarly to the Random Walk Metropolis algorithm, the Gibbs sequence converges to a stationary distribution that is independent of the starting values. By design, this stationary distribution is the same as our target distribution \cite{tierney_markov_1994, carlo_markov_2004}.

Derivation of the conditional distributions requires the use of conjugate prior distributions, meaning that the posterior distribution will be in the same probability distribution family as the prior distribution. For example, a normal likelihood function paired with a normal prior for the mean and an inverse gamma prior for the variance leads to a normal-inverse gamma posterior distribution \cite[p. ~67]{gelman_bayesian_2013}. Based on this information, the conditional distributions can be derived for each parameter and used in the Gibbs sampler. In more complex cases, it might be possible to derive part of the conditional distributions but not all, in which case a combination of the Gibbs algorithm with a Metropolis-Hastings step can be utilized, leading to Metropolis-within-Gibbs algorithms \cite{gilks_adaptive_1995}. There are, however, many problems in which the posterior will have a shape that cannot be deduced using known probability functions. The Gibbs sampler cannot then be used. 

\subsection{Hamiltonian Monte Carlo}
Hamiltonian Monte Carlo (HMC) is a second adaptation of MH, which uses information about the shape of the posterior, specifically the gradient, to guide the proposals on a path towards high density parts of the posterior. The acceptance rate can be much higher for HMC than for MH \cite{roberts_weak_1997}, making it an attractive alternative. HMC can be used more generally compared to the Gibbs sampler since it does not require the derivation of conditional distributions, only the existence and derivation (or approximation) of the gradient. This does come at the cost that discrete parameters cannot be used, as the gradient would then not be computable. A downside of HMC is that for every latent variable (i.e., parameter) an additional momentum variable is estimated, thereby increasing the complexity of the model. 

The idea of guiding the Markov chain uses concepts from physics, namely Hamiltonian dynamics. To define the movement in the Hamiltonian system, there are two forces: potential and kinetic energy. Furthermore, the system has a rule: total energy (potential + kinetic) is always constant. However, the division of the energy does not need to be constant:

\begin{equation}
    H(\theta, m) = U(\theta) + K(m),
    \label{eq:ham}
\end{equation}

\noindent
with $U(\theta)$ being the potential energy, $K(m)$ being the kinetic energy and $H(\theta, m)$ being the Hamiltonian (total energy). The potential energy, $U(\theta)$, is defined as the negative log of the posterior probability density. The kinetic energy, $K(m)$, is introduced to ensure that spaces in which the posterior density is lower are also sampled, and its state is described by the value of the momentum variable ($m$). The Hamiltonian is the joint density between our parameters $\theta$ and the momentum variables \cite{hoffman_no-u-turn_2014}.

\begin{figure}[H]
    \centering
    \includegraphics[width=0.85\linewidth]{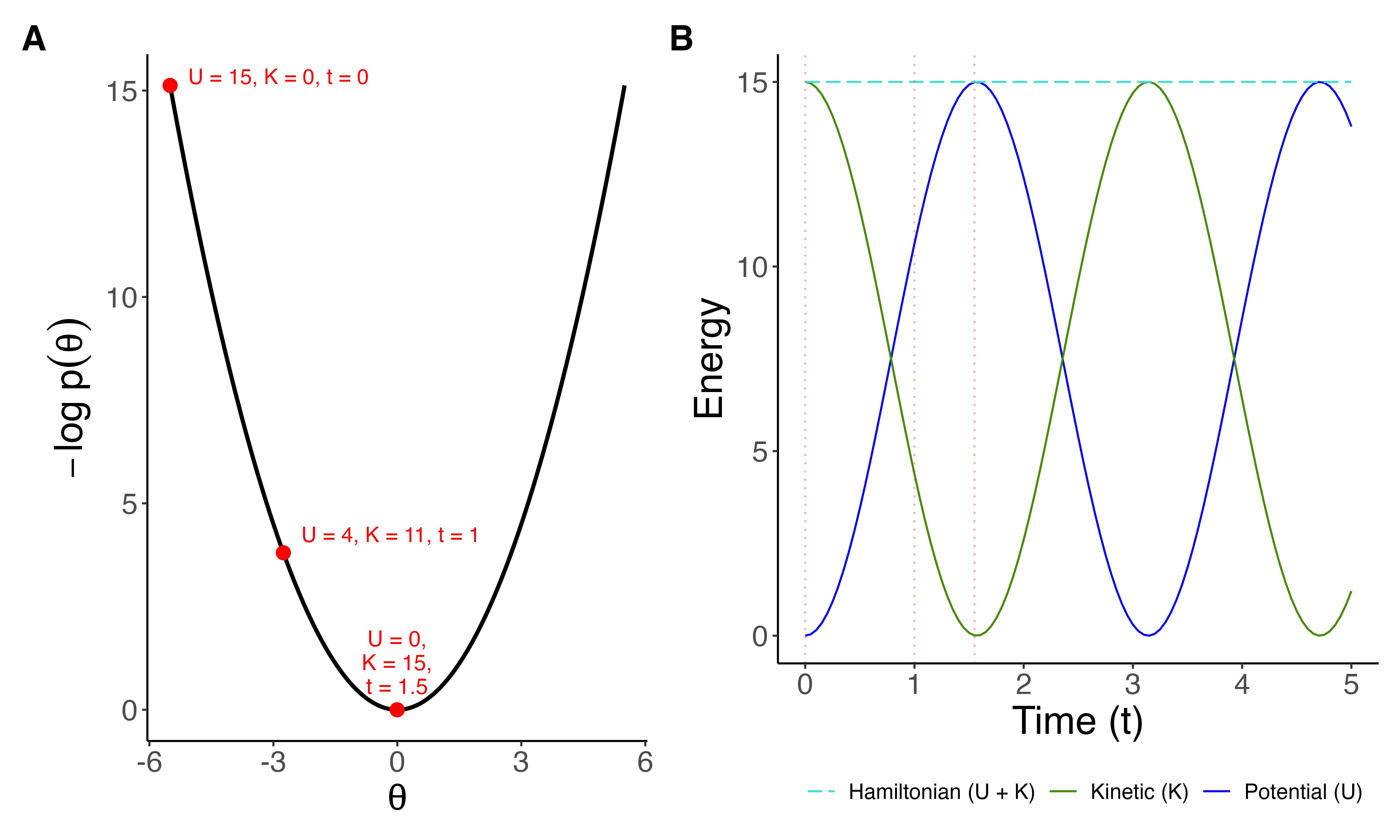}
    \caption{Illustration of Hamiltonian dynamics over time. In plot \textbf{A}, a ball is rolled from the top of the parabola. At three time points (t = 0, 1, 1.5), the values for the kinetic (K) and potential (U) energy are displayed. The values of the potential energy decrease as the ball rolls over time, while the kinetic energy increases until it reaches the bottom. Note that the valleys of -log p(x) relate to peaks of p(x). In the plot \textbf{B}, the values of the kinetic and potential energy are shown over time. The dotted lines correspond to the three time points in plot \textbf{A}. Since the system conserves energy (the Hamiltonian is unchanged over time), the ball will keep rolling down and up.}
    \label{fig:parabole}
\end{figure}

To obtain new samples, movement is needed in the parameter space. The momentum variable serves exactly this purpose; it creates the ability to guide samples through the parameter space while using information about the shape of the posterior. To visualize the Hamiltonian system, we introduce an often used example \cite{monnahan_faster_2017,brooks_mcmc_2011}. Imagine there is a ball rolling frictionless from one of the slopes of a parabola (top point, left side Figure \ref{fig:parabole}). The potential energy, which can be seen as the distance to the bottom of the parabola, decreases as the kinetic energy, which can be seen as the speed of the ball, increases when the ball moves to the second point. When the ball reaches the third point, there is no more potential energy, only kinetic energy. In all phases, the Hamiltonian, or the total energy, is constant. The goal of HMC is to explore the posterior in high and low density areas, and the right side of Figure \ref{fig:parabole} shows that as time progresses, the potential energy passes through both high and low value states. By working with the negative of the log posterior, areas with the highest density become valleys with high kinetic energy and low potential energy. 

\begin{figure}[H]
    \centering
    \includegraphics[width=.8\linewidth]{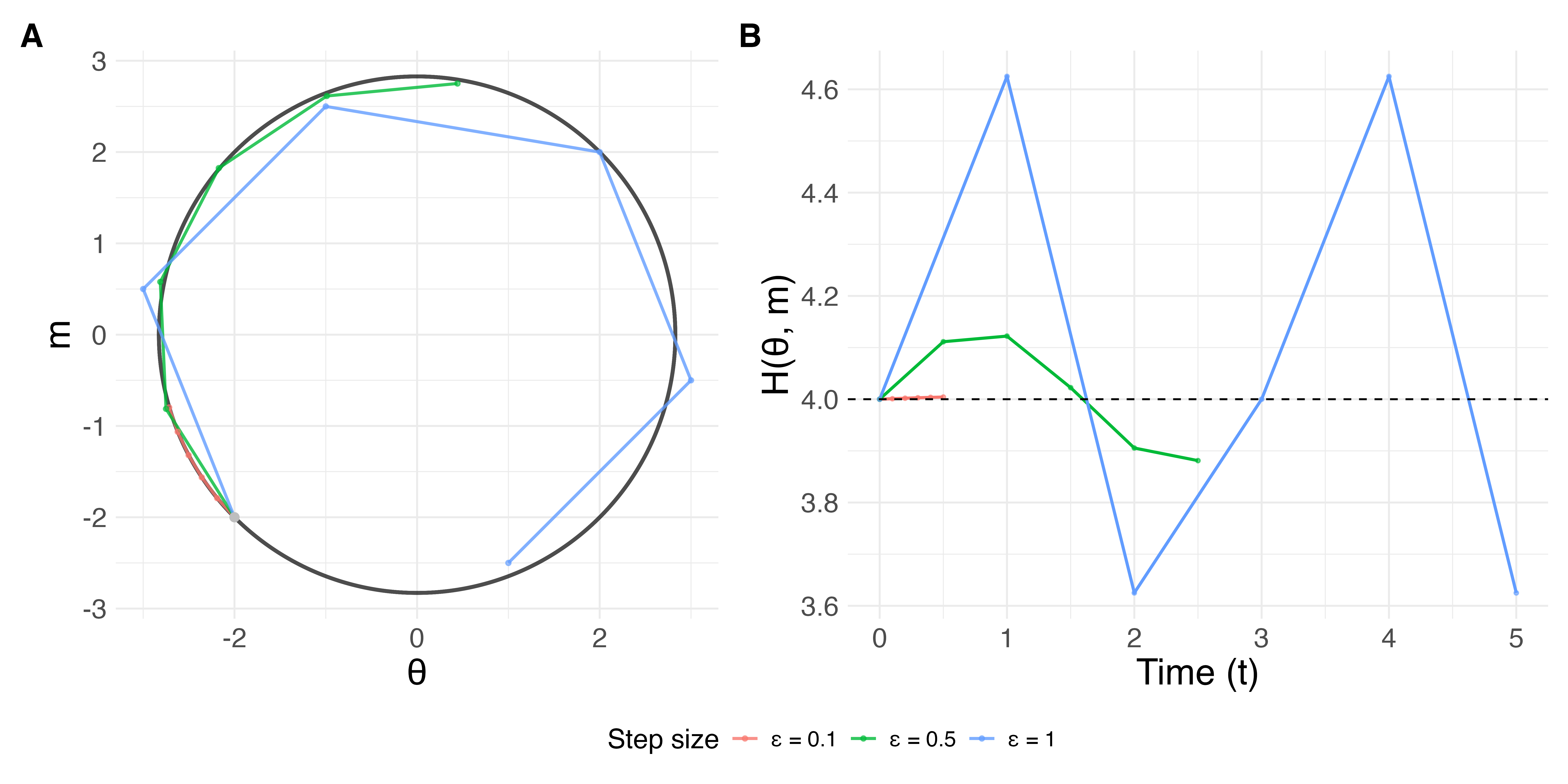}
    \caption{An example of the Leapfrog approximation of the Hamiltonian based on different values for the step size, where $m$ is the momentum variable and $\theta$ the parameter of interest. In plot \textbf{A}, the leapfrog algorithm is used to estimate the path (in black) for 5 steps, given different step sizes. Using a small step size causes the path to be walked slowly, but it helps keep the Hamiltonian conserved. In plot \textbf{B} the Hamiltonian is shown to not always be conserved as the step size increases, due to the local approximation of the gradients. }
    \label{fig:leap}
\end{figure}

The exact movement of the ball over the parabola, or any other function, can be calculated using the partial derivatives over the Hamiltonian \cite{brooks_mcmc_2011}: 

\begin{equation}
    \frac{d\theta}{dt} = \frac{dH}{dm},
    \label{eq:ham_d1}
\end{equation}

\begin{equation}
    \frac{dm}{dt} = -\frac{dH}{d\theta},
    \label{eq:ham_d2}
\end{equation}

\noindent
where $\frac{d\theta}{dt}$ is the gradient of the negative log of the posterior density. These equations allow us to map the potential and kinetic energies over time $t$. Unfortunately, the differential equations (Equation \ref{eq:ham_d1} and \ref{eq:ham_d2}) have no analytical solution in most cases. To implement HMC in practice, the new position (denoted by the energy states) needs to be approximated for a specific time point after $t$, which is commonly done using the leapfrog integrator (See Appendix A for the full algorithm). The idea is to obtain tangent lines to the posterior at predefined time points to approximate the movement for a set step size. Instead of making one big step, it is more common to make many small steps in one iteration. The step size and number of steps are important tuning parameters in HMC, as can be seen in Figure \ref{fig:leap}. There are algorithms to automatically choose these parameters, such as the No U-Turn Sampler \cite{hoffman_no-u-turn_2014}. 


Due to the approximation of the leapfrog algorithm, the energy is not always conserved in the system. There is thus still a need for an acceptance criterion to ensure that the Markov chain is not sent to an area with limited density due to the bad approximation of the leapfrog:

\begin{equation}
    r = \exp (H(\theta_{t}, m_{t}) -H(\theta_{t+1}, m_{t+1}))  = \exp(U(\theta_{t}) - U(\theta_{t+1}) + K(m_{t}) - K(m_{t+1})).
    \label{eq:ham_accept}
\end{equation}

Similar to MH, a value is accepted if $r> u \sim U(0,1)$. When the Hamiltonian is constant between the two states, the difference is zero and $r = 1$, resulting in the acceptance of the new value. Proposals are thus always accepted if the total energy stays the same or decreases \cite{monnahan_faster_2017}. This is important, as a lower value for the total energy indicates a higher density area in the joint distribution of $H(\theta, m)$.

\begin{figure}[H]
    \centering
    \includegraphics[width=0.75\linewidth]{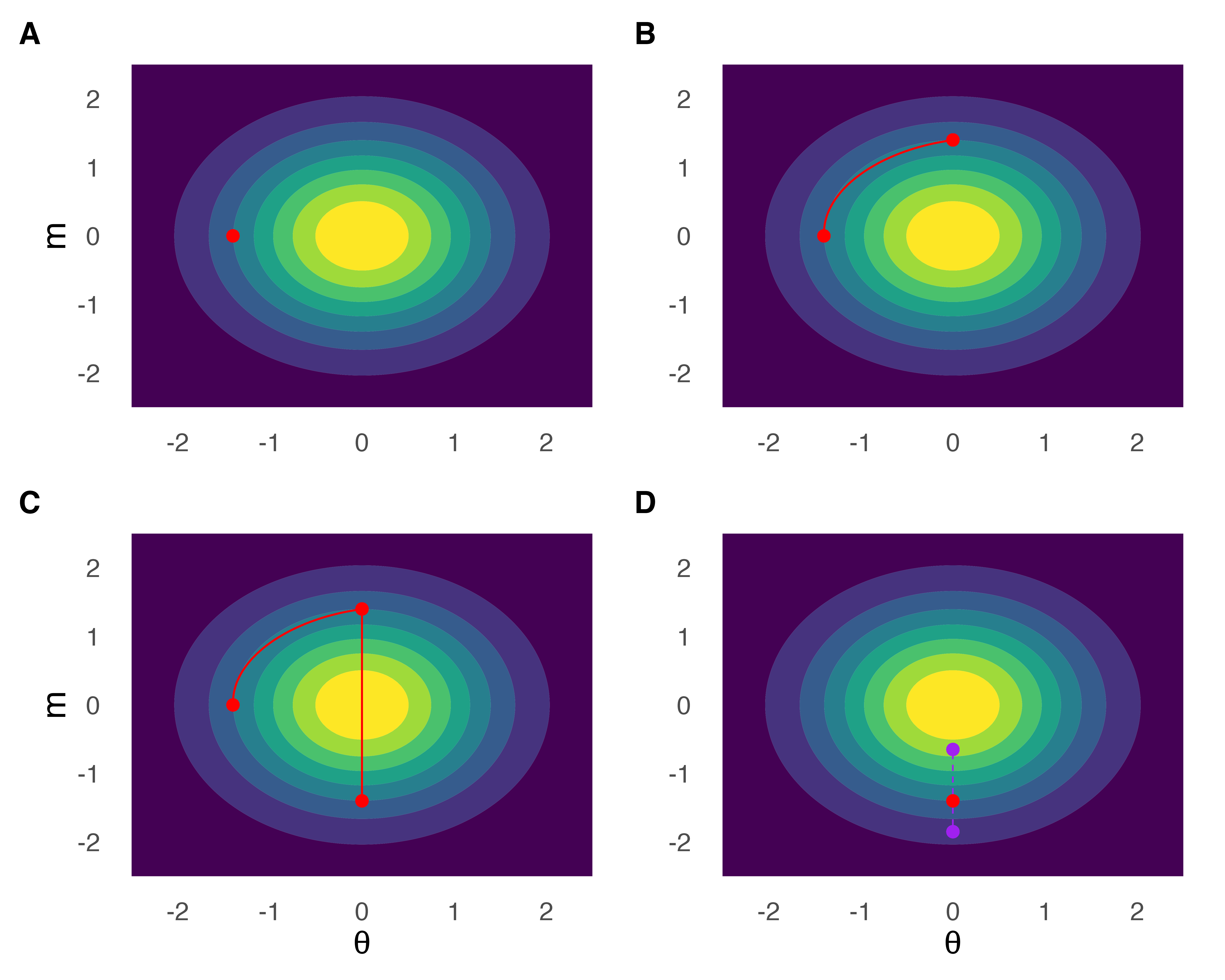}
    \caption{An example of how a single new parameter value is obtained using HMC. The joint distribution between the latent variable of interest ($\theta$) and the momentum variable (m) is depicted. Plot \textbf{A} shows an initial point on the Hamiltonian. In plot \textbf{B} we move to a new point using the leapfrog algorithm. In plot \textbf{C} the sign of the momentum variable is switched, after which the point is evaluated based on  Equation \ref{eq:ham_accept}. Plot \textbf{D} the new value is accepted, and a new value for the momentum is sampled to start the process over again. }
    \label{fig:HMC}
\end{figure}

A high-level overview of one iteration of the HMC algorithm can be seen in Figure \ref{fig:HMC}, where there is one latent variable ($\theta$) and thus one momentum variable (m). The joint probability density contours are depicted for $\theta$ and m, indicating where the values of $H(\theta, m)$ are constant. Just as in MH, a starting value is needed for $\theta$, and a value of m is sampled usually from a standard Gaussian. Combining these values gives an initial point in the joint distribution of $\theta$ and m (Figure \ref{fig:HMC}, A). The next step is to move along the Hamiltonian using the leapfrog integrator for several steps with a specific step size, returning a proposal point (Figure \ref{fig:HMC}, B). To ensure that the Hamiltonian transition is reversible, the negative of the momentum is taken for our final point \cite{betancourt_conceptual_2018} (Figure \ref{fig:HMC}, C). In practice, this step can be omitted if the goal is to sample from the posterior \cite{brooks_mcmc_2011, hoffman_no-u-turn_2014}. Once there is a proposal sample, it is assessed using the acceptance rule in Equation \ref{eq:ham_accept}. If the proposal is accepted, the value of $\theta$ is kept and a new value for the momentum is drawn, which can go both up and down, and the whole process can start over again (Figure \ref{fig:HMC}, D). For more elaborate examples of HMC in practical settings, see \cite{monnahan_faster_2017} and for more details on the HMC algorithm, see \cite{betancourt_conceptual_2018, brooks_mcmc_2011, neal_bayesian_2012}.

\subsection{Assessing Convergence}
When using any MCMC algorithm, the goal is to obtain samples from the posterior. All algorithms start with random values, relating to a random place in the multi dimensional parameter space. It is likely that this random place is not within the high density area of the posterior. To ensure that only samples are used from the posterior, and not the path from the initial value to the posterior there is often a warmup or burn-in phase  in the sampling proces to learn the tuning parameters (e.g. the covariance matrix of the proposal distribution of MH) \cite{andrieu_tutorial_2008}, and to move towards the posterior. After the posterior is "found", many samples are taken to reduce the variance of our parameters of interest \cite{margossian_for_2024}. Often multiple random starting values are used, to create independent Markov chains. Using samples from these (Markov) chains, convergence to the target posterior distribution can be investigated. In addition to visual diagnostics such as traceplots, often used metrics that are available for MCMC-methods are the $\hat{R}$ and ESS. The focus in this section is convergence diagnostic based on multiple chains, but there are also metrics for single chain convergence. An example is the Geweke diagnostic which compares the first part of the chain to a latter part using a statistical test to investigate if the samples stem from a different distribution \cite{geweke_evaluating_1991}. 

\begin{figure}[H]
    \centering
    \includegraphics[width=0.95\linewidth]{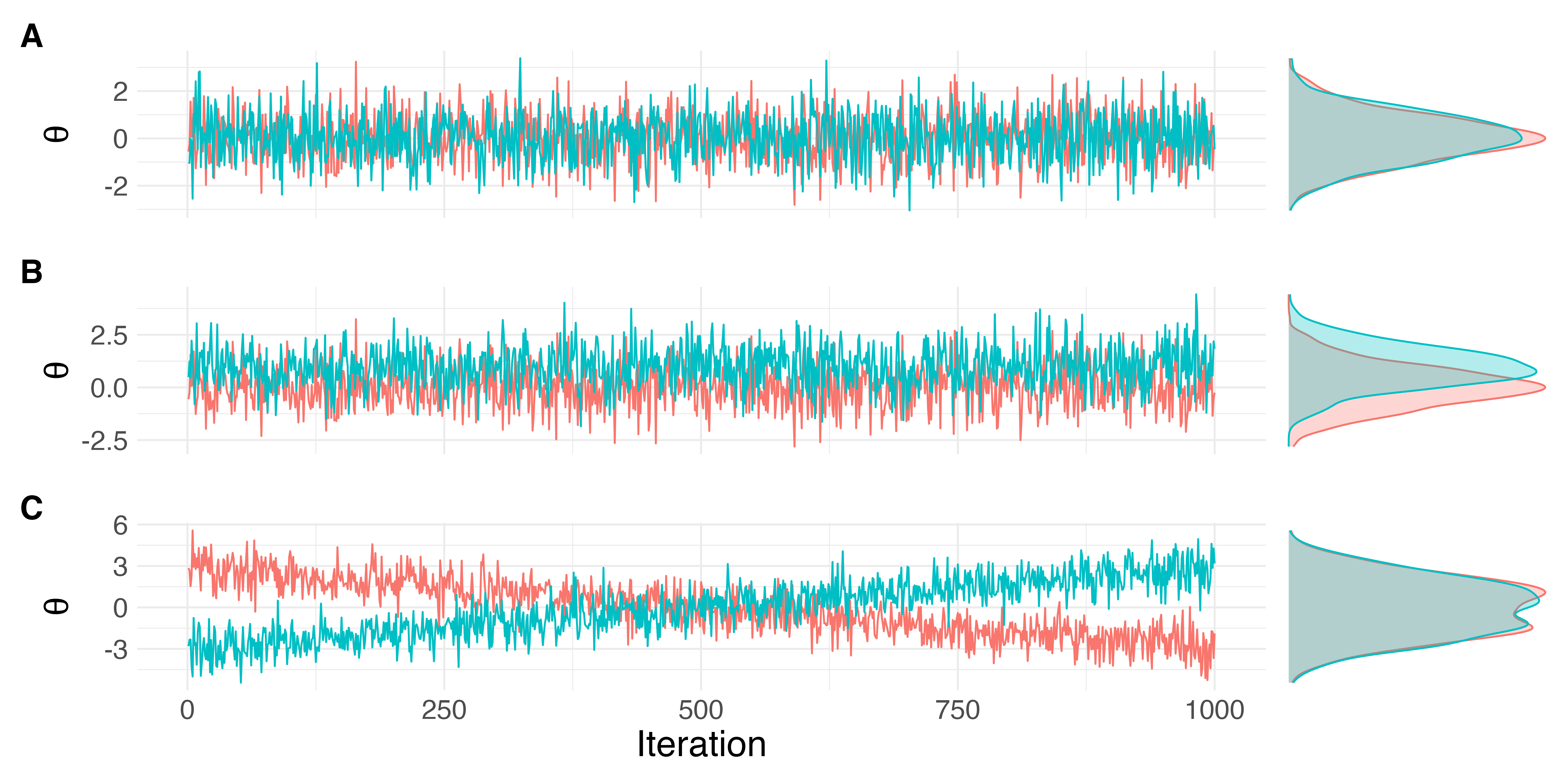}
    \caption{Assessing the convergence of two chains (after burn-in) based on the sample values from the chains (left) and the marginal distributions (right); example based on Figure 1 from \cite{lambert_r_2022} and Figure 1 from \cite{vehtari_rank-normalization_2021}. Plot \textbf{A} shows that the chains have samples that cover the same marginal distribution, there seems to be convergence as the $\hat{R}$ and split-$\hat{R}$ values are both 1. In Plot \textbf{B} the chains seems to sample from different distributions, the $\hat{R}$ is 1.16 (split-$\hat{R}$ 1.16) and this would thus not be considered as converged.  In Plot \textbf{C} the marginal distributions of the chains overlap, but the chains are non-stationary. Here the $\hat{R}$ value is 1, while the split-$\hat{R}$ is 1.56. It is thus evident that the split-$\hat{R}$ can capture non-convergence, when the classical $\hat{R}$ fails. }
    \label{fig:rhat}
\end{figure}

To assess the convergence of the Markov chains the classical $\hat{R}$ statistic \cite{gelman_inference_1992} measures the ratio of the variance within a Markov chain compared to the variance between Markov chains that were initialized at different values:

\begin{equation}
    \hat{R} = \sqrt{\frac{\frac{n-1}{n}W + \frac{1}{n}B}{W}},
\end{equation}

\noindent
where $W$ is the within-chain variance and B is the between chain variance. The numerator is an estimate of the population variance $\hat{\sigma^2}$, while the denominator is the sample variance $s^2$. We assume that $\hat{\sigma^2}$ is an overestimation of $\sigma^2$, and that $s^2$ is a underestimation of $\sigma^2$. Both are consistent for $\sigma^2$, so the $\hat{R}$ decreases to 1 as n increases \cite{gelman_inference_1992, vats_revisiting_2021}. The statistic indicates if the chains mix well, and are converged to a common distribution (Figure \ref{fig:rhat}. An adaptation is the split-$\hat{R}$ that divides the chains into two and pretends like there are now twice as many chains \cite[p.~284]{gelman_bayesian_2013}. This can capture non convergence even if there are mirrored trends in the chains. The split-$\hat{R}$ is only defined if the marginal posteriors have a finite mean and variance, to alleviate this assumption the original draws can be replaces by the rank normalized values \cite{vehtari_rank-normalization_2021}. In this paper we will use the rank-normalized split-$\hat{R}$, and from this point this will be referred to as the $\hat{R}$. If the $\hat{R}$ is larger than 1 one might be concerned; a common threshold for approximate convergence is $\hat{R}<1.01$ \cite{vehtari_rank-normalization_2021}.

The samples obtained through MCMC are inherently depended on each other, due to the use of Markov chains. This means that N draws from the posterior does not equal N independent draws. To ensure that the sample sizes in our Markov chains are large enough to obtain stable estimates (of uncertainty), another metric is used called the effective sample size (ESS): 

\begin{equation}
    \hat{n}_{eff} = \frac{mn}{1 + 2 \sum_{t=1}^T \hat{p}_t},
    \label{eq:ess}
\end{equation}

\noindent 
where $\hat{p}_t$ is the estimate for the autocorrelation (correlation between a sample and the sample before) at time $t$, m is the number of chains and n the number of samples. The idea behind the ESS is to calculate the number of independent samples needed to obtain the information in our dependent samples. If the autocorrelation becomes zero then the ESS equals the total number of draws. The larger the autocorrelation, the lower the ESS \cite[p.~286]{gelman_bayesian_2013}. An often used threshold for the ESS to be larger than 400 \cite{vehtari_rank-normalization_2021}. For further information on ESS see \cite{vehtari_rank-normalization_2021}.

In addition, for HMC there is another metric that indicates how often the total energy in the Hamiltonian is not conserved between steps (due to the leapfrog approximation). If the discrepancy is large (Equation \ref{eq:ham_accept}), a step is called a divergent transition. In practice, these steps are often never accepted and this can lead to an under exploration of an area of the posterior. 

\section{Non-MCMC-based methods: Variational inference, Laplace approximation and Pathfinder}
MCMC methods attempt to obtain an "exact" approximation of the posterior, which is possible when there are many samples drawn ($n \rightarrow \infty)$, resulting in a large computational burden. When complex models are fitted with many parameters, e.g. hierarchical models or neural networks, Bayesian inference relying on MCMC becomes infeasible. One solution is to use cruder approximations that restrict the shape of the posterior and replace the sampling scheme with an optimization problem. Popular non-MCMC methods include variational inference and the Laplace approximation. These and a new method named Pathfinder, which can be used to obtain good initial values for HMC, will be discussed in the following sections. However, many other new algorithms exist, such as Amortized approximators \cite{radev_jana_2023} and Prior-Data Fitted Networks \cite{muller_transformers_2024}. 

\subsection{Variational inference}
Variational Inference (VI) is a method that allows us to directly model the posterior of latent variables using gradient descent, which is much faster than MCMC. The goal is to obtain a simple distribution for latent variables, such as regression coefficients, that approximates the posterior as closely as possible. The loss in the optimization problem is obtained using the Kullback-Leibler (KL) divergence \cite{blei_variational_2017}, although many other divergences can be used \cite{margossian_variational_2025}. The KL divergence indicates the similarity between the posterior, ($p(\theta | D) $), and the approximate distribution. The approximate distribution is commonly denoted by $q(\theta)$. The KL divergence can be defined as:

\begin{equation}
    \text{KL}(q(\theta) || p(\theta | D)) = \int q(\theta) \log \frac{q(\theta)}{p(\theta|D)} d\theta,
\end{equation}

\noindent
where $q(\theta)$ and this is a "simple" distribution \cite{zhang_advances_2019}. Note that the KL divergence is not symmetric: $\text{KL}(q(\theta) || p(\theta | D)) \neq \text{KL}(p(\theta | D) || q(\theta))$. The choice between these two, sometimes called exclusive and inclusive (or reverse and forwards) KL, will influence the approximation $q(\theta)$ \cite{dhaka_challenges_2021, margossian_variational_2025}. Minimizing $\text{KL}(p(\theta | D) || q(\theta))$ results in the correct marginal variances, which is often of interest in a Bayesian analysis, but it involves an expectation over the posterior $p(\theta | D)$ that is hard to compute \cite{margossian_variational_2025}. For this reason, the exclusive/reverse KL is often chosen, although it leads to an underestimation of the marginal variances \cite{blei_variational_2017}. Either way, the KL divergence should be minimized to find the optimal choice for $q$: 

\begin{equation}
    q^*(\theta) = \argmin_{q(\theta) \in Q} \text{KL}(q(\theta) || p(\theta|D)),
    \label{eq:kl_min}
\end{equation}

where $Q$ indicates a family of possible distributions. To evaluate and subsequently minimize the KL divergence, the posterior is needed (Equation \ref{eq:bayes}). 
In practice, the posterior distribution is not known. However, by rewriting the KL divergence, we obtain the following expression:

\begin{equation}
    \begin{aligned}
        \text{KL}(q(\theta) || p(\theta | D)) &= E_{q(\theta)}\left[\log \frac{q(\theta)}{p(\theta|D)}\right] \\
        &= E_{q(\theta)}[\log q(\theta)] -E_{q(\theta)}[\frac{\log p(\theta, D)}{\log p(D)}] \\
        &= E_{q(\theta)}[\log q(\theta)] - E_{q(\theta)}[\log p(\theta, D)] + \log p(D).
    \end{aligned}
    \label{eq:kl}
\end{equation}

From this, we can obtain the evidence lower bound (ELBO) by dropping the reliance on the constant $\log p(D)$:

\begin{equation}
    \text{ELBO}(q) = E_{q(\theta)}[\log p(\theta, D)] - E_{q(\theta)}[\log q(\theta)]
\end{equation}

As such, the ELBO is equivalent to the KL divergence up to a constant. To understand the parts of the ELBO, it can be rewritten as follows:

\begin{equation}
    \begin{aligned}
        \text{ELBO}(q) &=  E_{q(\theta)}[\log p(\theta)] + E_{q(\theta)}[\log p(D|\theta)] - E_{q(\theta)}[\log q(\theta)] \\
        &=  E_{q(\theta)}[\log p(D|\theta)] - KL(q(\theta) || p(\theta))\\
    \end{aligned}
    \label{eq:elbo}
\end{equation}

The first part of the ELBO evaluates how much density of the approximation is in high density areas of the posterior defined by the expected likelihood, while the second part is the divergence between the approximation and the prior \cite{blei_variational_2017}. Maximizing the ELBO minimizes the KL divergence \cite{jordan_introduction_1999}:

 \begin{equation}
    q^*(\theta) = \argmax_{q \in Q} \text{ELBO}(q)
\end{equation}

In theory, the KL divergence, by means of the ELBO, could be minimized to zero. In that case, the variational distribution perfectly matches the posterior; however, often this is not possible \cite{zhang_advances_2019}. In practice, the variational distribution will not be flexible enough to capture the true nature of the posterior exactly. 

To estimate the parameters of the variational distribution, multiple methods can be used \cite{kucukelbir_automatic_2017, kingma_auto-encoding_2022, rezende_variational_2015}. A popular method relies on gradient descent using the reparameterization trick. Gradient descent requires setting a learning rate for how quickly the parameters are adjusted. The learning rate is a hyperparameter that can be hard to tune. For work on automatically adjusting the learning rate for VI, see \cite{agrawal_advances_2020, welandawe_framework_2024}. Note that for restricted parameters, such as variance terms, a reparameterization often needs to occur before VI can be used \cite{kucukelbir_automatic_2017}. 






In this paper two common approaches of VI, mean-field and full-rank, are discussed. The mean-field method uses independent distributions for the latent variables, while the full-rank method constructs one joint distribution. 

\subsubsection{Mean-field}
In the mean-field approach, the latent variables are modeled as being mutually independent, each having its own distribution \cite{blei_variational_2017}:

\begin{equation}
    q(\theta) = \prod_{p = 0}^{P-1} q_p(\theta_p)
\end{equation}

where $\theta \in R^P$. 



\begin{figure}[H]
    \centering
    \includegraphics[width=0.95\linewidth]{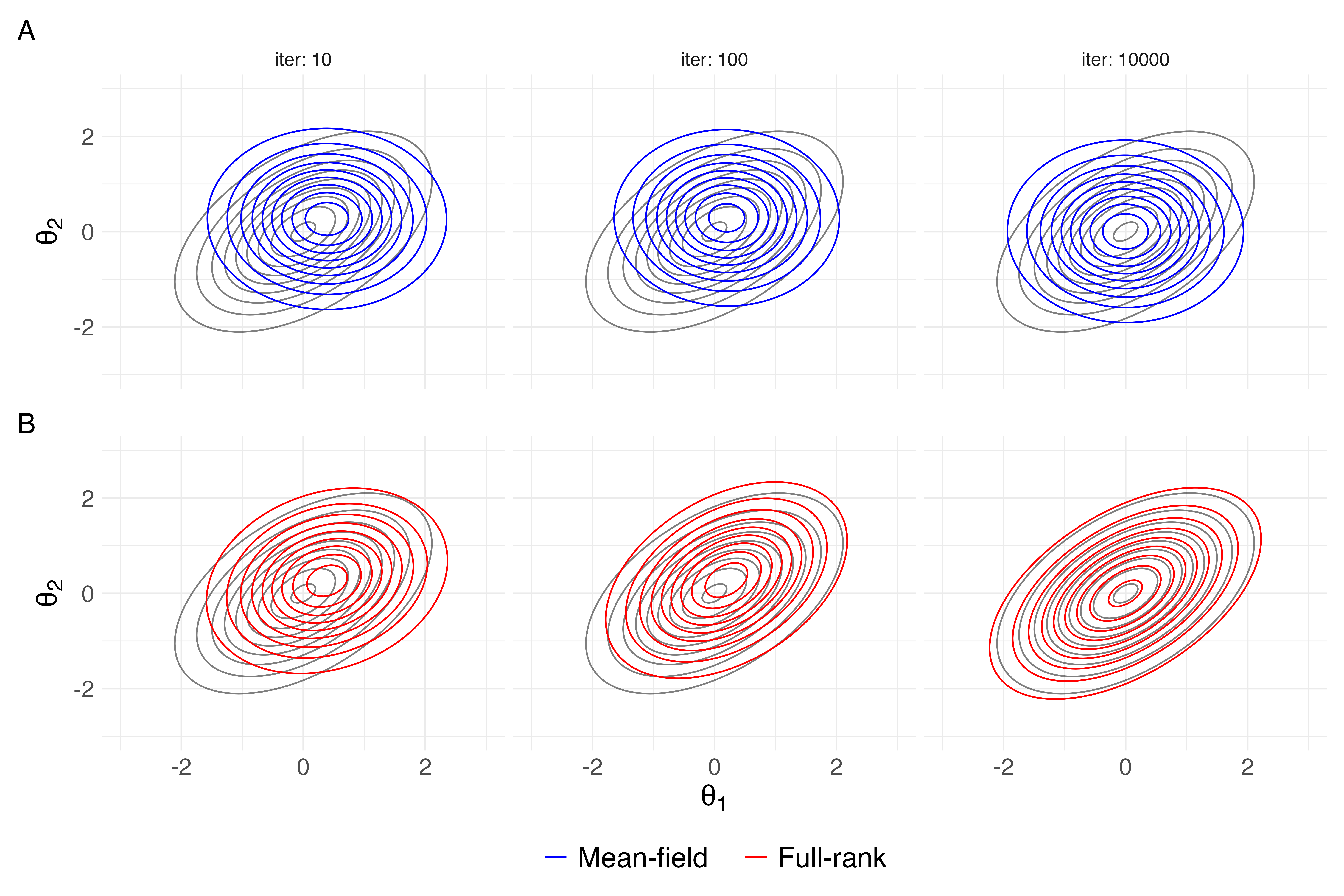}
    \caption{Visualization of the optimization paths for the mean-field and full-rank algorithms. Plot \textbf{A} depicts the posterior used in Figures \ref{fig:HM}B/\ref{fig:GIBBS} in black contour lines. The blue contour lines show the approximations after a set number of iterations optimizing the ELBO. The approximation becomes better as the number of iterations increases, but the mean-field fails to capture the correlation structure in the data. Plot \textbf{B} shows the same for the full-rank approximation. After 10000 iterations, the algorithm's approximation comes very close to the posterior, and is much better compared to the mean-field. }
    \label{fig:VI}
\end{figure}

The mean-field family can capture the marginal density of the latent variables. However, it does not model correlations between them \cite{blei_variational_2017} and thus cannot model all measures of uncertainty around the posterior well.  In general, the mean-field VI methods tend to approximate posterior distributions that are too compact, when using the exclusive KL \cite{bishop_pattern_2006}. In Figure \ref{fig:VI}, an example is shown of a mean-field approximation after a certain number of iterations minimizing the ELBO. It is clear that with a limited number of iterations, the mean-field approximation is far off. As the number of iterations increases, the approximations become better. In the example, there is a correlation structure between the parameters, something the mean-field approximation cannot capture. Not all density in the tails of the posterior will be covered by the mean-field approximation in this case, and this behaviour will get worse as the correlations between parameters increase. 

\subsubsection{Full-rank}
The latent variables can also be modeled in a single multivariate distribution, which is known as the full-rank approach. It is a generalization of the mean-field approximation. In case of a multivariate Gaussian, the off-diagonal terms in the covariance matrix would include posterior correlations among the latent random variables \cite{kucukelbir_automatic_2017}. This can improve the posterior approximation, but comes at a computational cost as the number of parameters in the covariance matrix scales quadratically with the number of latent variables in the model \cite{opper_variational_2009}. 

The form of the joint distribution depends on the choice of $q$; if a Gaussian is used, then we have:

\begin{equation}
    q(\theta) = \text{MVN} (\boldsymbol{\mu}, \Sigma)
\end{equation}

where $\boldsymbol{\mu}$ is a vector with the means, and $\Sigma$ is the covariance matrix. 

In Figure \ref{fig:VI}, the full-rank approximation performs much better than the mean-field, given the maximum number of iterations. This relatively simple model can easily be estimated with the full-rank approach, as it only requires one extra off-diagonal covariance matrix entry. 

\subsection{Laplace}
A method that also attempts to fit a simpler function to the posterior is Laplace approximation. The general idea is to approximate the posterior using a (multivariate) Gaussian distribution. The mean of the approximation is set to the mode of the posterior, and the curvature at the mode is used to estimate the covariance matrix \cite{rue_bayesian_2017}. The posterior is expected to become increasingly better estimated by a Gaussian as the number of data points increases \cite{bishop_pattern_2006, gelman_bayesian_2013}.

The parameters of the Gaussian are estimated in a two-step procedure. First, the mode of the posterior is found using some optimization method. Then a second-order Taylor expansion is computed around the mode \cite{rue_bayesian_2017}; the second-order derivatives at the mode describe the curvature at the mode \cite{opper_variational_2009}. The higher order derivatives can be added, but fade in importance to the second derivative \cite[p.~83]{gelman_bayesian_1995}. The normal approximation around the mode ($\hat{\theta}$) is given as:

\begin{equation}
    p(\theta|D) \approx N(\hat{\theta}, [I(\hat{\theta)}]^{-1}).
\end{equation}

where $I(\theta) = -\frac{d^2 }{d\theta^2}p(\theta|D)$.

\begin{figure}[H]
    \centering
    \includegraphics[width=0.8\linewidth]{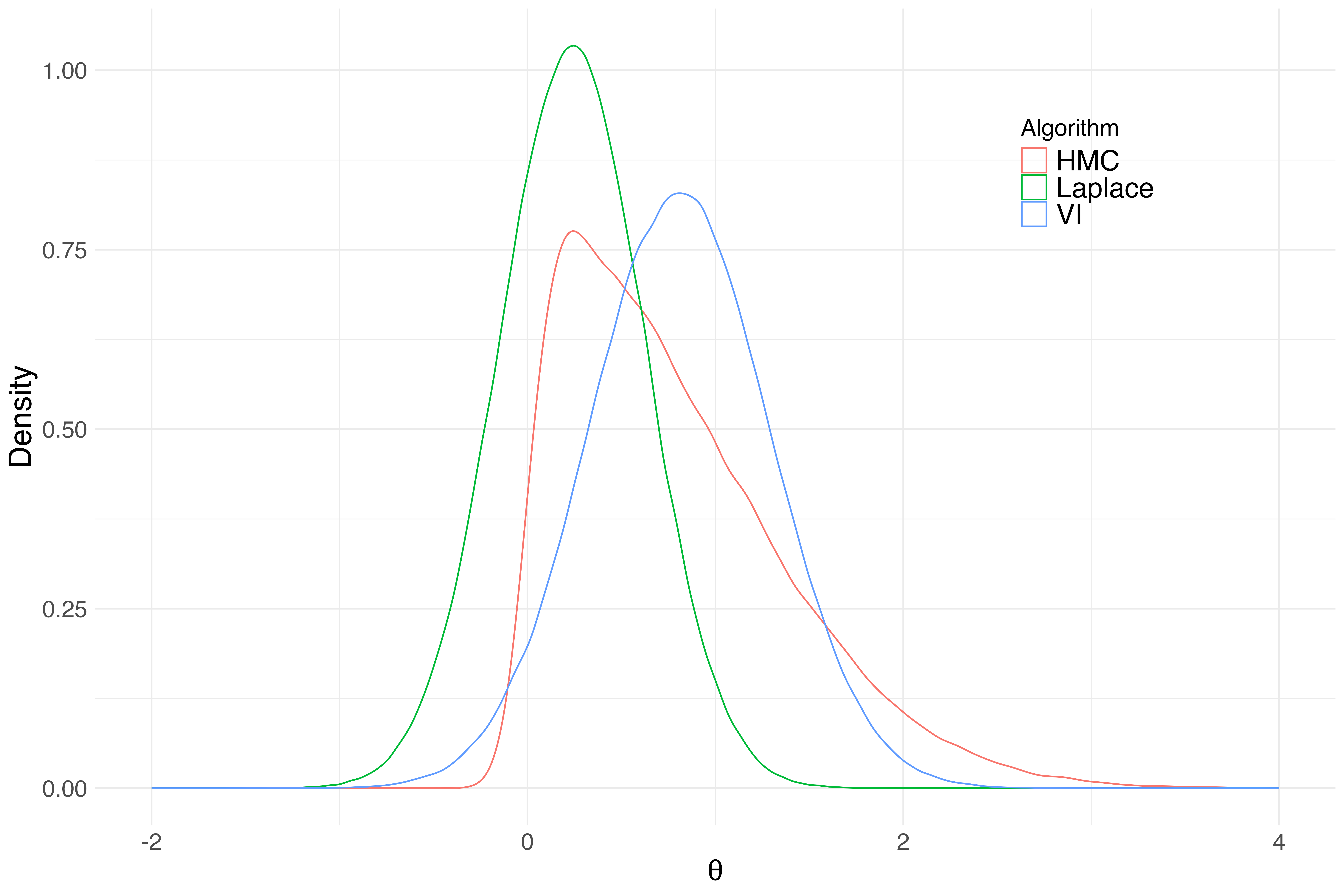}
    \caption{Visualization of different non-MCMC methods that approximate a skewed univariate posterior, inspired by Figure 10.14 \cite[p.~508]{bishop_pattern_2006}. In this model, a skew-normal posterior is estimated using HMC, Gaussian VI and Laplace. Note that since there is only a single parameter, the mean-field and full-rank VI methods are the same.}
    \label{fig:approx_exp}
\end{figure}

Variational inference methods using Gaussian distributions are similar to the Laplace approximation; the difference lies in how the parameters of the Gaussian are estimated. In Figure \ref{fig:approx_exp}, posterior samples from HMC are compared to VI and Laplace approximations of the same posterior. It is clear that the mean of Laplace is not the same as that for the VI method; this is due to the restriction that the mean of the Laplace approximation is set to the mode of the posterior. 

\subsection{Pathfinder}

A third method, the Pathfinder algorithm \cite{zhang_pathfinder_2022}, uses a variational algorithm to quickly obtain samples in areas of high density and subsequently uses these samples as initial values for an MCMC algorithm. As such, this option should be seen as a hybrid method.

\begin{figure}[H]
    \centering
    \includegraphics[width=1\linewidth]{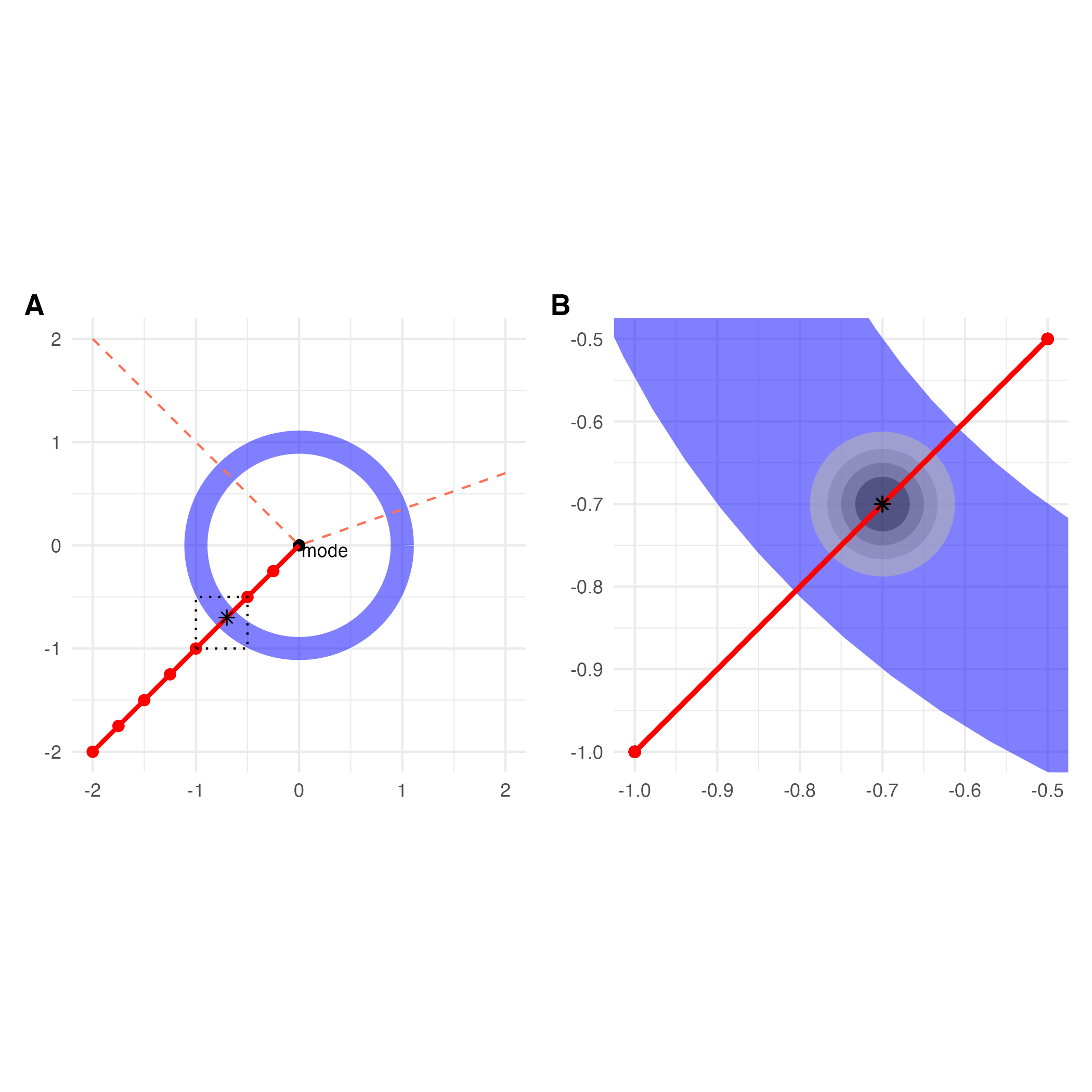}
    \caption{A low dimensionality example of the optimization path of the Pathfinder algorithm. A posterior density is shown where the regions of high density are indicated by the blue circle around the mode. The Pathfinder algorithm finds the path to the mode using the L-BFGS algorithm. At each step, the KL divergence is evaluated, and at the point with the lowest KL divergence, samples are drawn. Plot \textbf{A} indicates a path that Pathfinder finds toward the mode, where the dashed lines indicate other paths that the algorithm could have found. Plot \textbf{B} is a zoom of the black dotted box in Plot \textbf{A} showing the point with the lowest KL. This point is used as the mean of a Gaussian sampling distribution, where the covariance can be constructed using terms in the L-BFGS path. The samples drawn around this point are then in areas of high posterior density. }
    \label{fig:Pathfinder}
\end{figure}

Pathfinder uses a quasi-newton method named L-BFGS \cite{byrd_limited_1995, zhu_algorithm_1997} to move from an initial value on a path towards the mode. It can be shown in high dimensions, high density areas of a distribution are located in a sphere around the point which maximizes the probability density \cite{vowels_typical_2024}. It would be beneficial to find the points along the path that are in the sphere. After the path is created, a Gaussian approximation is fit at each step along the path. The covariance matrix can be obtained from the L-BFGS path. The KL divergence from these approximations to the target (posterior) is then evaluated at each step, and the step with the lowest KL is chosen to sample from (Figure \ref{fig:Pathfinder}). The evaluations can be performed in parallel, increasing the speed of Pathfinder. Also, multimodal distributions can be approximated by running Pathfinder from different initializations \cite{zhang_pathfinder_2022}.

The goal of Pathfinder is to find good initial values for HMC. In the example of Figure \ref{fig:Pathfinder}, it is evident that we end up sampling values in the high density areas. Still, the samples only cover a small area of the entire density, and thus MCMC is needed to ensure that samples are obtained from the entire distribution. The phase of finding the area with high density, the burn-in period, can thus be reduced when using Pathfinder. 

Pathfinder had been tested on various posteriors available in the posteriordb database \cite{zhang_pathfinder_2022, magnusson_posteriordb_2025}, but not explicitly in high-dimensional settings so it remains unclear what the benefits in these scenarios might be.

\subsection{Assessing Convergence}
\begin{figure}[H]
    \centering
    \includegraphics[width=0.75\linewidth]{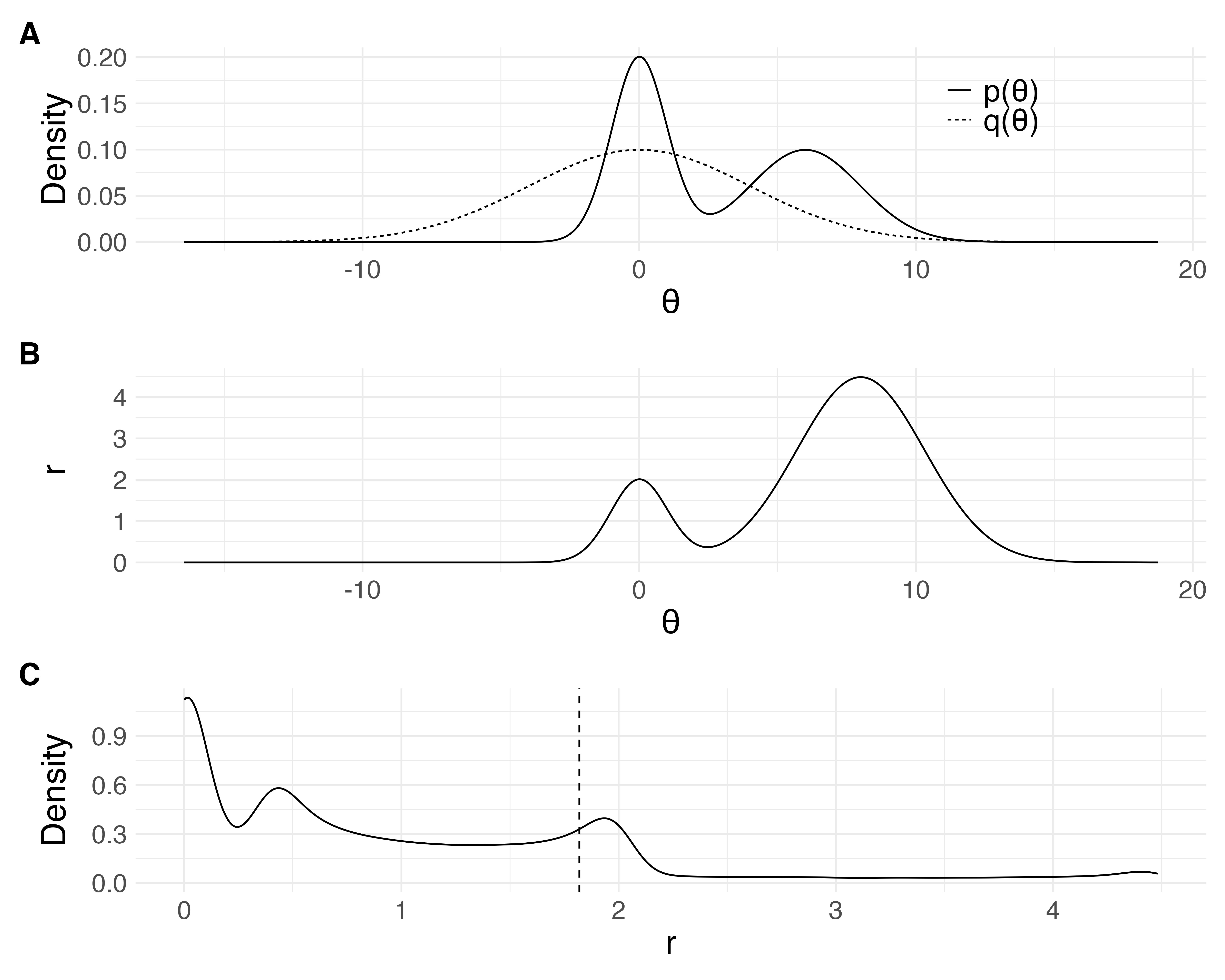}
    \caption{Visualization of using importance weights for calculating the Pareto $\hat{k}$ value. Plot \textbf{A} is an example of a posterior (mixture of two Gaussians), $p(\theta)$, with an ( Gaussian) approximation, $q(\theta)$. Plot \textbf{B} shows the IS weights, r, at each point of $\theta$. Plot \textbf{C} displays the distribution of the unnormalized IS weights and the cutoff value for the top 20\% of the weights.}
    \label{fig:IS}
\end{figure}

There are two main types of convergence assessments for non-MCMC-based methods: 1) has the algorithm converged to the optimal parameters for the approximation; 2) is the final approximation a close fit to the posterior \cite{zhang_advances_2019}. The first criterion is assessed via a stopping rule for the optimization threshold. Most implementations have automatic stopping rules (e.g., \href{https://mc-stan.org/docs/cmdstan-guide}{CmdStan User's guide}). For the second criterion, there are many reasons why the approximation might not be correct, e.g., the posterior is asymmetric or the tails are too light due to the direction of the KL divergence (inclusive versus exclusive) \cite{yao_yes_2018}. One method to test the fit of the approximation to the posterior is based on Pareto Smoothing Importance Sampling (PSIS) \cite{vehtari_pareto_2024}. Importance sampling is a method that can be used to obtain samples from a "complicated" posterior using a simpler proposal distribution. This aligns well with the idea behind the non-MCMC methods. The workflow can be seen in Figure \ref{fig:IS}. 

The first step is to run a model and obtain samples from an approximator $q(\theta)$ (Figure \ref{fig:IS}A). Then the importance ratios are calculated between this proposal and the unnormalised posterior (Figure \ref{fig:IS}B), using the following formula:

\begin{equation}
    r_s = \frac{p(\theta_s | D)}{q(\theta_s)}
\end{equation}

It is evident that the weights are high where the density of the posterior is high, but the density of the approximation is low. The same holds the other way around, as can be seen in Figure \ref{fig:IS}. The importance weights can then be used to improve the draws by importance sampling. Here, a new set of samples is created by drawing from the original samples using the importance weights \cite[p. ~266]{gelman_bayesian_2013}. Importance resampling decreases the bias of the samples. In practice, the importance ratios can vary tremendously in their value, causing some draws to become very dominant in the estimate. This can result in very high variance in the resampled draws \cite{huggins_validated_2020,vehtari_pareto_2024}. To improve this, the M largest weights are selected (Figure \ref{fig:IS}C), based on this formula:

\begin{equation}
    M = \min (0.2S, 3\sqrt{S}),
\end{equation}

with S the number of samples. 

Lastly, a generalized Pareto distribution, GPD$(\mu, \sigma, k)$, is fit on a subset of the normalized weights (Figure \ref{fig:IS}C) by a computationally cheap method (e.g., \cite{zhang_new_2009}). The number of moments of the GDP is defined as 1/k, so if $k < 1$, the mean can be estimated, and if $k < 0.5$, the variance can also be estimated. The shape parameter k of this distribution provides a diagnostic measurement between the true posterior and the approximation \cite{yao_yes_2018}; the larger the shape parameter, the wider the distribution. If the distribution of the importance ratios has a heavy tail, then this can indicate a poor approximation, and that will be captured in the shape parameter \cite{vehtari_pareto_2024}. A threshold for the Pareto $\hat{k}$ value that is used in practice is 0.7 or $min (1-  1/ log10(S), 0.7)$ \cite{yao_yes_2018, vehtari_pareto_2024}; values within this range would indicate that the approximation $q(\theta)$ is still useful. 

Finding an approximation that matches the high density areas of the posterior tends to become more difficult in high dimensions; it is thus expected that the Pareto $\hat{k}$ value increases as the dimensions increase \cite{dhaka_challenges_2021}. A large Pareto $\hat{k}$ value
does not necessarily mean that the approximations are not useful 
\footnote{Samples can still be useful even with high Pareto $\hat{k}$ values, as can be seen in this \href{https://users.aalto.fi/~ave/casestudies/Birthdays/birthdays.html}{example}.}.

\section{Simulation}
The goal of the simulation study is to evaluate the performance of non-MCMC methods, compared to HMC, for prediction tasks with high-dimensional tabular data with Bayesian regularized regression. We limit the scope of our investigation to the algorithms available in Stan, a popular probabilistic programming language for Bayesian analysis \cite{gelman_stan_2015}.

\subsection{Data generating mechanism}
In prediction (supervised learning) tasks, the goal is to estimate a model that relates observations, \textit{X}, to an outcome, \textit{Y}. For the simulation, multivariate normal data are generated for the \textit{X} values:

\begin{equation}
    X \sim N(\boldsymbol{\mu}, \Sigma),
    \label{eq:sim_x}
\end{equation}

with $\boldsymbol{\mu}$ being a vector of zeros equal to the number of parameters, and $\Sigma$ the chosen covariance matrix. Using the \textit{X} values, the \textit{Y} values are generated:

\begin{equation}
    Y = X \cdot B \cdot  \sqrt{\frac{R^2}{B \cdot \Sigma \cdot B}} + N(0, \sqrt{1-R^2)},
    \label{eq:sim_Y}
\end{equation}

where $B$ denotes a vector of the chosen coefficients, and $R^2$ equals a chosen value for the explained variance. The coefficients are thus scaled to fix the explained variance. The parameters $\Sigma$, $B$ and $R^2$ will vary in our simulation study. First, the values for \textit{X} are sampled, and then for \textit{Y} according to equation \ref{eq:sim_Y}. Using this data generating model, different settings are explored (Table \ref{tab:sim_set}). The scenarios become increasingly higher dimensional, and thus more computationally intensive. For each scenario number, there is an \textit{a} and \textit{b} variant, where all parameters are the same except the structure of the covariance matrix and the effect sizes of the predictors. Specifically, in the \textit{a} scenario, there is only a covariance structure between the relevant predictors, where in scenario \textit{b} the covariance is between all predictors. Furthermore, in \textit{a} all parameters have the same effect size, while in scenario \textit{b} the effect sizes differ. Due to computational constraints, a full factorial design was not used. The simulation settings are:

\begin{table}[http]
\begin{tabular}{|p{1.5cm}|p{1.5cm}|p{2cm}|p{1.5cm}|p{2cm}|p{1.5cm}|p{2cm}|p{2cm}|}
\hline
\textbf{Scenario} & \textbf{Training set}  & \textbf{Number of predictors (p/n ratio)} & \textbf{Sparsity ratio} & \textbf{Correlations} & $\boldsymbol{R^2}$ & \textbf{Covariance structure} & \textbf{Coefficients} \\ \hline
1a & 100 & 20 (0.2) & 0.2 & 0.5 & 0.5 & Partial & Equal \\
\hline
1b & 100 & 20 (0.2) & 0.2 & 0.5 & 0.5 & Full & Different \\
\hline
2a & 100 & 100 (1) & 0.05 & 0.5 & 0.5 & Partial & Equal \\
\hline
2b & 100 & 100 (1) & 0.05 & 0.5 & 0.5 & Full & Different \\
\hline
3a & 100 & 500 (5) & 0.05 & 0.9 & 0.5 &  Partial & Equal\\
\hline
3b & 100 & 500 (5) & 0.05 & 0.9 & 0.5 &  Full & Different\\
\hline
4a & 100 & 1000 (10) & 0.05 & 0.9 & 0.2 &  Partial & Equal\\
\hline
4b & 100 & 1000 (10) & 0.05 & 0.9 & 0.2 &  Full & Different\\
\hline
\end{tabular}
\caption{Simulation scenarios by their characteristics. }
\label{tab:sim_set}
\end{table}

\noindent
\textbf{Training set}: In all scenarios, the models are trained using a sample size of 100. The models are compared to a test set (size 1000), which is generated in the same way as the training set.

\noindent
\textbf{Number of predictors}: The training set stays the same in all scenarios, but the number of predictors increases, from a non-high-dimensional setting example, \textit{p} to \textit{n} ratio of 0.2, to a very high-dimensional setting, \textit{p} to \textit{n} ratio of 10.

\noindent
\textbf{Sparsity ratio}: The sparsity ratio indicates how many of the predictors are directly correlated with the outcome variable. In the first scenario (1a\&b), the sparsity ratio is 0.2 and there are 20 predictors leading to 4 (0.2 *20) predictors that are directly related to the outcome, the vector $B$ from Equation \ref{eq:sim_Y} will then be $(1,1,1,1,0,0,..,0)$. In all other scenarios (2/3/4) the sparsity ratio drops to 0.05.

\noindent
\textbf{Correlations}: The correlations are between the predictor variables. The correlations become increasingly large, leading to a structure where it is harder to distinguish between a predictor that directly affects \textit{Y} and one that only indirectly affects \textit{Y}. 

\noindent
$\boldsymbol{R^2}$: The proportion of explained variance is fixed in all scenarios to 0.5, except the last two, where the value is 0.2. Lower values for the $R^2$ will lead to lower effect sizes of the coefficients.

\noindent
\textbf{Covariance structure}: Two covariance structures were considered for $\Sigma$ in Equation \ref{eq:sim_x}. In the partial case, only the correlations between \textit{X} values that are directly related to \textit{Y} are non-zero. In the full scenario, all correlations are non-zero, regardless of their effect on \textit{Y}. 

\noindent
\textbf{Coefficients}: The values of the coefficients were split into a scenario where all coefficients had the same value (set to 1 before scaling). Alternatively, the second predictor is twice as important as the first, the third three times as important as the first, etc. 

\noindent
\textbf{Total number of simulated datasets}: This is set to 500 for the first 4 scenarios from Table \ref{tab:sim_set} and to 250 for the last 4 scenarios due to computational constraints. 

\subsection{Methods}
A quick outline of the methods (from Section 3/4) used in the simulation study is given in Table \ref{tab:sim_methods}. In all analyses, a regularized horseshoe prior (section 2.2.2) with 3 degrees of freedom is used for the regression parameters. Other priors were checked in a sensitivity analysis, which indicated the best predictive performance for the HS prior (Appendix D).

\begin{table}[H]
\centering
\small
\begin{tabular}{|p{2cm}|p{2.5cm}|p{6cm}|p{3cm}|}
\hline
\textbf{Method} & \textbf{Name} & \textbf{Description} & \textbf{Settings} \\ \hline
MCMC & Hamiltonian Monte Carlo \newline (section 3.3) & Sampling method that uses gradients of log posterior through Hamiltonian dynamics.  & \begin{tabular}[c]{@{}l@{}}Chains = 4\\ N warmup = 1000\\ N draws = 2000\end{tabular} \\ \hline
\multirow{3}{*}{Non-MCMC} & Mean-field VI \newline (section 4.1.1) & Assumes independent known distributions for latent variables. & \begin{tabular}[c]{@{}l@{}}  $Q$  = Gaussian \\ Max N iter = 10000\\ N draws = 2000\end{tabular} \\ \cline{2-4} 
 & Full-rank VI \newline (section 4.1.2) & Assumes dependent known distributions for latent variables. & \begin{tabular}[c]{@{}l@{}} $Q$  = Gaussian \\ Max N iter = 10000\\ N draws = 2000\end{tabular} \\ \cline{2-4} 
 & Laplace \newline (section 4.2) & Two-step Gaussian approximation: 1) Find mode, 2) Estimate covariance using second-order Taylor approximation around mode. & N draws = 2000 \\ \hline
Hybrid & Pathfinder → HMC \newline (section 4.3) & Finds path to mode and evaluates KL divergence at each step. Draws from a Gaussian at the step with the lowest KL value, using the resulting samples as initial values for HMC. & \begin{tabular}[c]{@{}l@{}}N paths = 4\\ N iter = 1000\\  PSIS = FALSE \\ \hdashline Chains = 4 \\ N warmup = 100 \\ N draws = 2000\end{tabular} \\ \hline
\end{tabular}
\caption{Overview of methods and settings used in the simulation study. For the full details on the default settings, see the \href{https://mc-stan.org/docs/cmdstan-guide}{CmdStan user guide}.}
\label{tab:sim_methods}
\end{table}

All analyses were performed in R \cite{r_core_team_r_2024}. The methods were implemented using Stan \cite{carpenter_stan_2017}; specifically \texttt{cmdstanr} \cite{garby_cmdstanr_2024}. For HMC, the No U-Turn Sampler (NUTS) algorithm is implemented to help select the moment at which the leapfrog integrator should stop \cite{hoffman_no-u-turn_2014}. Variational inference is implemented using Automatic Differentiation Variational Inference (ADVI) \cite{kucukelbir_automatic_2017}; the variational family consists of Gaussians, which are fitted in the unconstrained space. The Gaussians are univariate in the case of mean-field VI and multivariate for full-rank VI. 

The hyperparameters of the different algorithms used in this study can be seen in Table \ref{tab:sim_methods}. The warmup phase for HMC might be considered low for the high-dimensional settings, but it has been shown that fewer than 1000 iterations can be enough to reach stationarity for high-dimensional penalized regression \cite{biswas_coupling-based_2022}. When using Pathfinder to initialize the values for HMC, the warm up phase of HMC is decreased from 1000 to 100. 


\subsection{Outcomes}

To appraise the performance of the algorithms, the following metrics are considered: convergence ratio, quality of estimates, running time, bias in the parameters, predictive performance and uncertainty estimation \cite{burkner_models_2023}.

The convergence ratio is the percentage of models that return estimates after running. The quality of the estimates in this study is based on three categories (Table \ref{tab:quality}). The convergence of the MCMC algorithms is assessed based on the $\hat{R}$, the bulk effective sample size (ESS) and the number of divergent transitions \cite{vehtari_rank-normalization_2021}. For the non-MCMC methods, the Pareto $\hat{k}$ value is used \cite{yao_yes_2018}. Originally, a cutoff of 0.7 is recommended, so if Pareto $\hat{k}$ values exceed 0.7, we label them as “questionable”. However, since it is known that $\hat{k}$  values can increase as the dimensions increase, we further investigate a more lenient cutoff of 5 and label $\hat{k}$ values exceeding 5 as “bad”.

\begin{table}[]
\centering
\begin{tabular}{|l|l|l|l|}
\hline
Quality & Good & Questionable & Bad \\ \hline
MCMC & 
\begin{tabular}[c]{@{}l@{}} 
$\hat{R} \leq 1.05$ \\ 
$\text{ESS}/N \geq 0.5$ \\ 
$N_{\text{div}}/N \leq 0.1$ 
\end{tabular} & 
\begin{tabular}[c]{@{}l@{}} 
$1.05 < {\hat{R}} \leq 1.1$ \\ 
$0.1 \leq \text{ESS}/N < 0.5$ \\ 
$N_{\text{div}}/N \leq 0.1$ 
\end{tabular} & 
\begin{tabular}[c]{@{}l@{}} 
$\hat{R} > 1.1$ \\ 
$\text{ESS}/N < 0.1$ \\ 
$N_{\text{div}}/N > 0.1$ 
\end{tabular} \\ \hline
non-MCMC & 
${\hat{k}} \leq 0.7$ & 
$0.7 < \hat{k} \leq 5$ & 
${\hat{k}} > 5$ \\ \hline
\end{tabular}
\caption{Quality measures for models. The overall quality of a model corresponds to the lowest category among its individual diagnostic metrics.}
\label{tab:quality}
\end{table}

Simulation results will be used if the quality was not Bad (Table \ref{tab:quality}). Furthermore, for every scenario, a method should have estimates for at least 25 datasets; otherwise, the results are not used due to the small sample size. The running time is defined as the difference between calling the algorithm and obtaining results. The bias in the parameters is estimated separately for the intercept, non-zero parameters and zero parameters as follows:

\begin{equation}
    Bias_\theta =  \frac{1}{N} \sum^N_{i = 1}  |\hat{\theta}_i - \theta| 
\end{equation}

where $\theta$ denotes the true value and $\hat{\theta}_i$ the estimated parameter in simulation $i$. The average is then taken of the bias for each type of parameter. 

The predictive performance is calculated on the test set. Mean squared error (MSE) is used to assess the predictive performance:

\begin{equation}
    MSE = \frac{1}{N} \sum^N_{i = 1} (\hat{y}_i - y_i)^2,
    \label{eq:mse}
\end{equation}

with $\hat{y}_i$ being the mean of the PPD for observation $i$. The uncertainty estimation is assessed by looking at the coverage of the prediction intervals. If there is a 95\% prediction interval, then 95\% of the time the prediction should fall within this interval. The coverage can be calculated using the following formula:

\begin{equation}
    \text{coverage} = \frac{1}{N} \sum^N_{i = 1} (L_{\hat{y}_i} \leq  y_i \leq U_{\hat{y}_i}),
\end{equation}

with $L_{\hat{y}_i}$ being the $\alpha/2$ quintile and $U_{\hat{y}_i}$ the $1-\alpha/2$ quintile of the PPD.

The simulation was run on an Intel(R) Xeon(R) LATINUM 8580 CPU. The simulation iterations were ran in parallel, not the chains of MCMC methods. All R \cite{r_core_team_r_2024} code for the simulation and the following analyses are available on \href{https://osf.io/mdk9z}{OSF}. The main packages used are: \texttt{future} \cite{RJ-2021-048}, \texttt{furrr}  \cite{davis_vaughan_and_matt_dancho_furrr_2022}, \texttt{cmdstanr} \cite{Gabry_cmdstanr_2024}, \texttt{brms} \cite{Burkner_bmrs_2021}, \texttt{posterior} \cite{Bürkner_posterior_2025}, \texttt{loo} \cite{Vehtari_loo_2024}, \texttt{dplyr} \cite{Wickham_dplyr_2023} and \texttt{ggplot} \cite{Wickham_ggplot2_2016}. 

\subsection{Results}

The results section focuses on the main results, primarily the convergence of the algorithms, the bias in the coefficients, the MSE and the coverage of the prediction interval. Additional results can be found in Appendix B. 

\begin{figure}[H]
    \centering
    \includegraphics[width=0.95\linewidth]{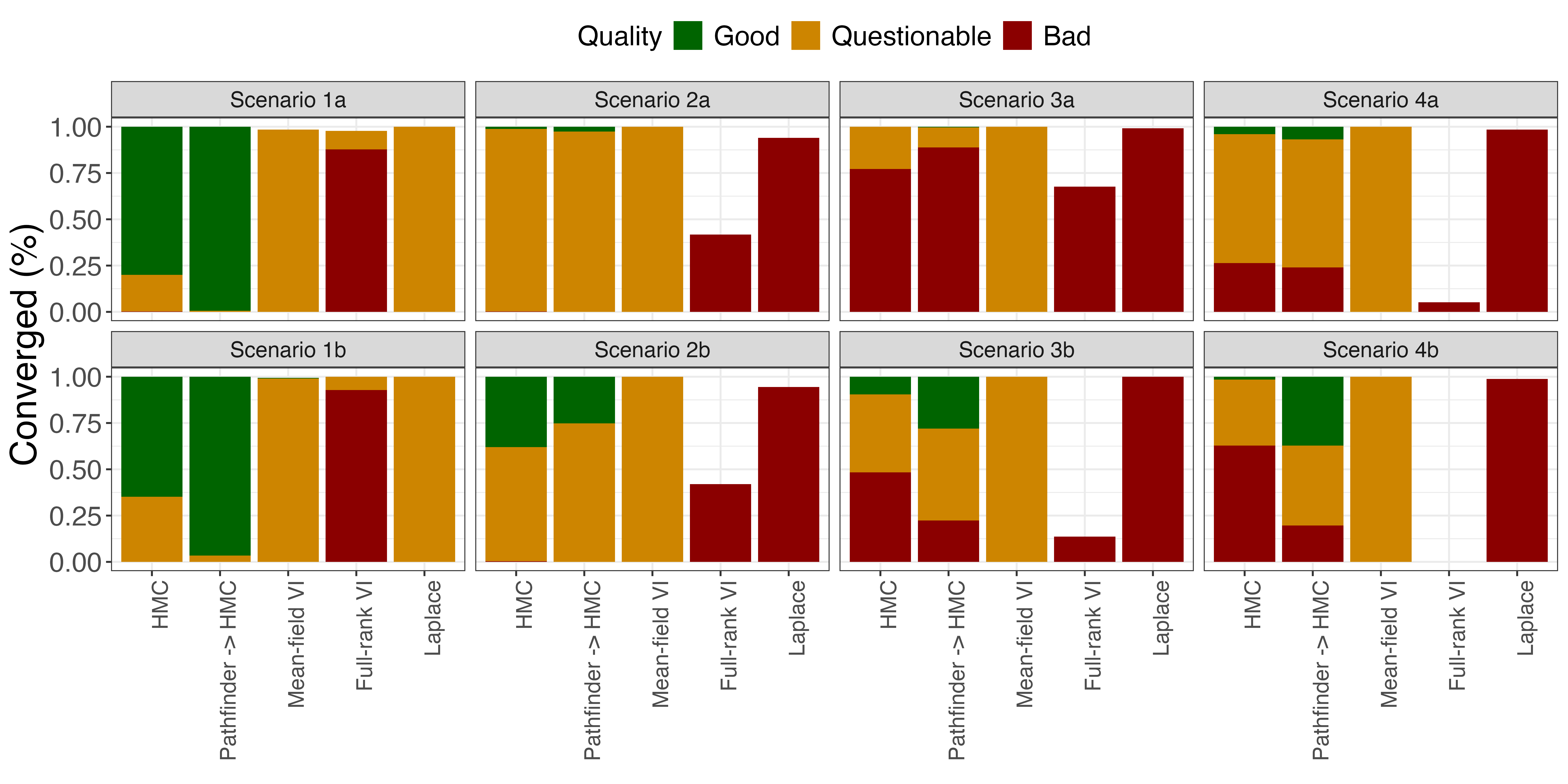}
    \caption{The percentage of models that gave estimates after running for the different approximators, split by the quality of the estimates. }
    \label{fig:conv3}
\end{figure}

\begin{figure}[H]
    \centering
    \includegraphics[width=0.4\linewidth]{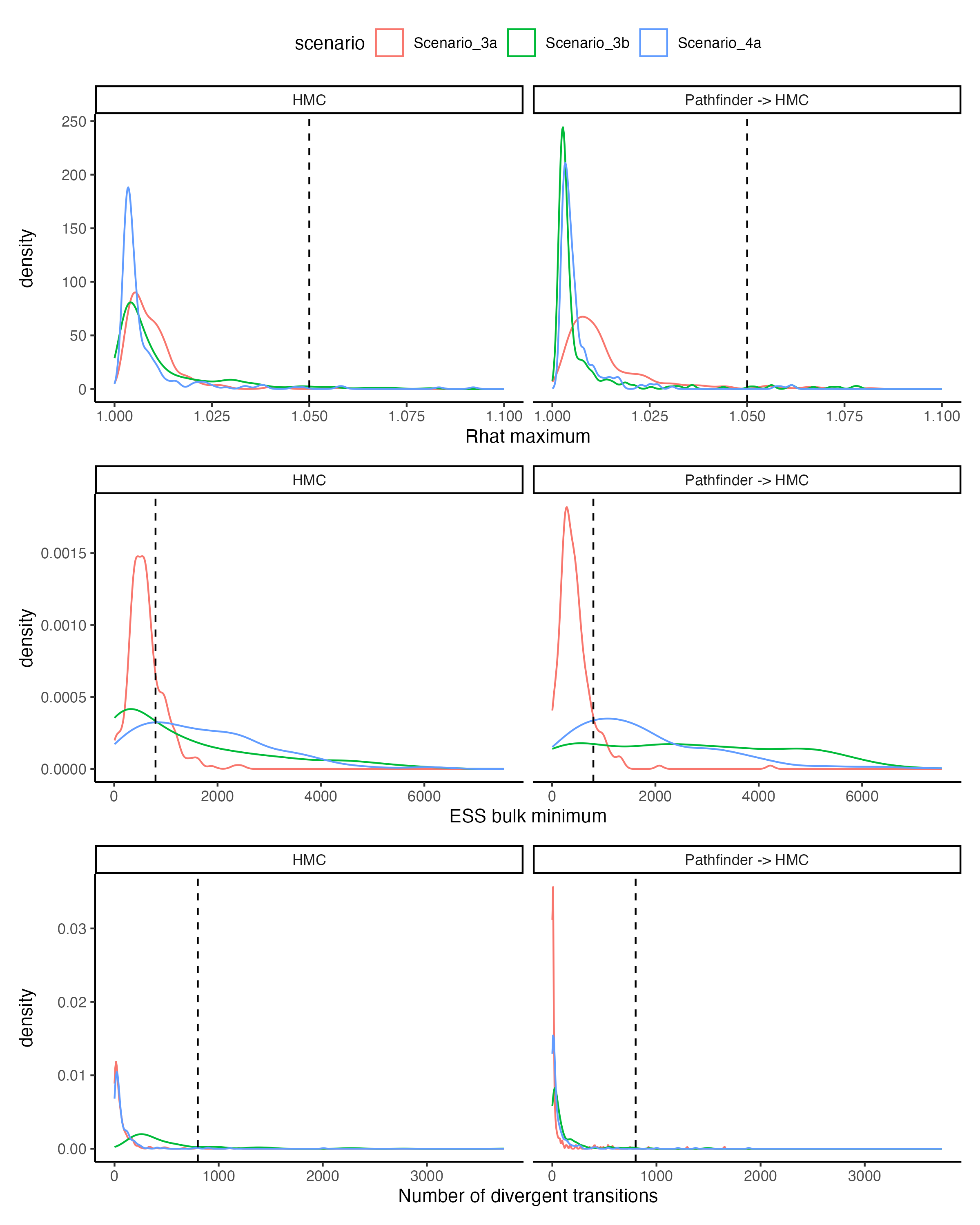}
    \caption{A closer look at the convergence criteria for the MCMC methods in scenario 3a compared to scenario 3b and 4b, based on the ESS, $\hat{R}$ and the number of divergent transitions. IT seems that the lowest ESS for the resgression parameters is very low in scenario 3b leading to many runs labels as bad. }
    \label{fig:conv_3a}
\end{figure}

The percentage of models that provided estimates after running was close to 100\% for all algorithms except full-rank, which had fewer returned estimates as the scenarios became more difficult (Figure \ref{fig:conv3}). Based on the criteria defined in this paper (Table \ref{tab:quality}) the quality of the estimates became increasingly worse when the \textit{n} to \textit{p} ratio increased. Laplace, for example, did not provide a single good estimate when $p >= n$. Mean-field was the only method that return estimates which were not labeled as bad quality estimates in the high-dimensional scenarios (Table \ref{tab:n_conv}, Appendix B). HMC did not perform well in scenario 3a; this was due to a low ESS estimate (Figure \ref{fig:conv_3a}, Appendix B).  The full results before filtering the bad quality estimates are shown in Table \ref{tab:full_res1_all}/\ref{tab:full_res2_all}, and the results after filtering are shown in Table \ref{tab:full_res1}/\ref{tab:full_res2} (Appendix B).

\begin{figure}[H]
    \centering
    \includegraphics[width=0.8\linewidth]{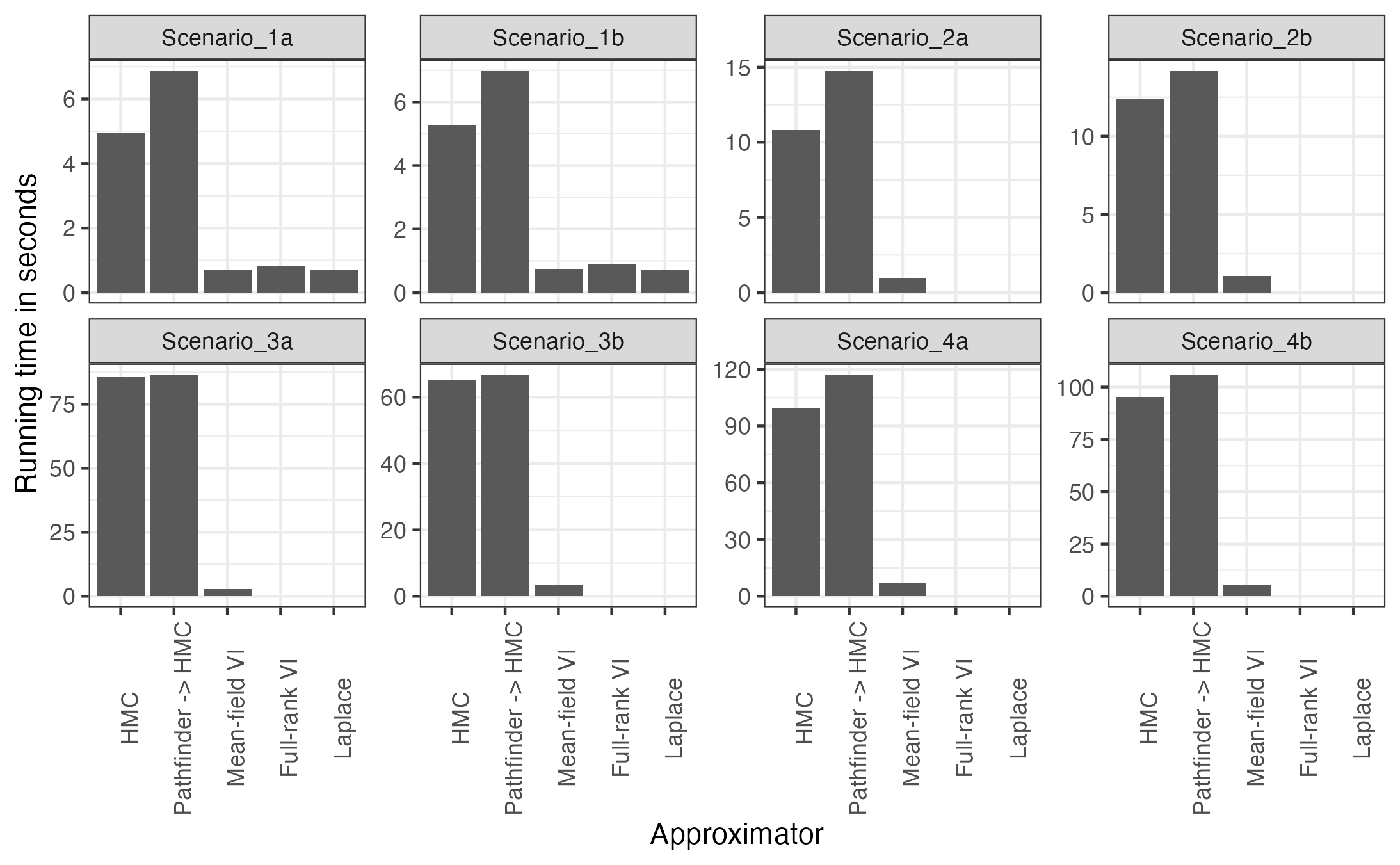}
    \caption{Run time of models.}
    \label{fig:run_time}
\end{figure}

The running time is the lowest for the non-MCMC algorithms. The ratio between the running time of mean-field and HMC becomes larger as the scenarios become more complicated, from a 7-10 times speedup in scenarios $n \geq p$ to 15-30 for $n < p$ (Figure \ref{fig:run_time}, Appendix B.). Pathfinder as initialization takes a bit longer than HMC, even though the burn-in time is cut from 1000 to 100. 

\begin{figure}[H]
    \centering
    \includegraphics[width=0.95\linewidth]{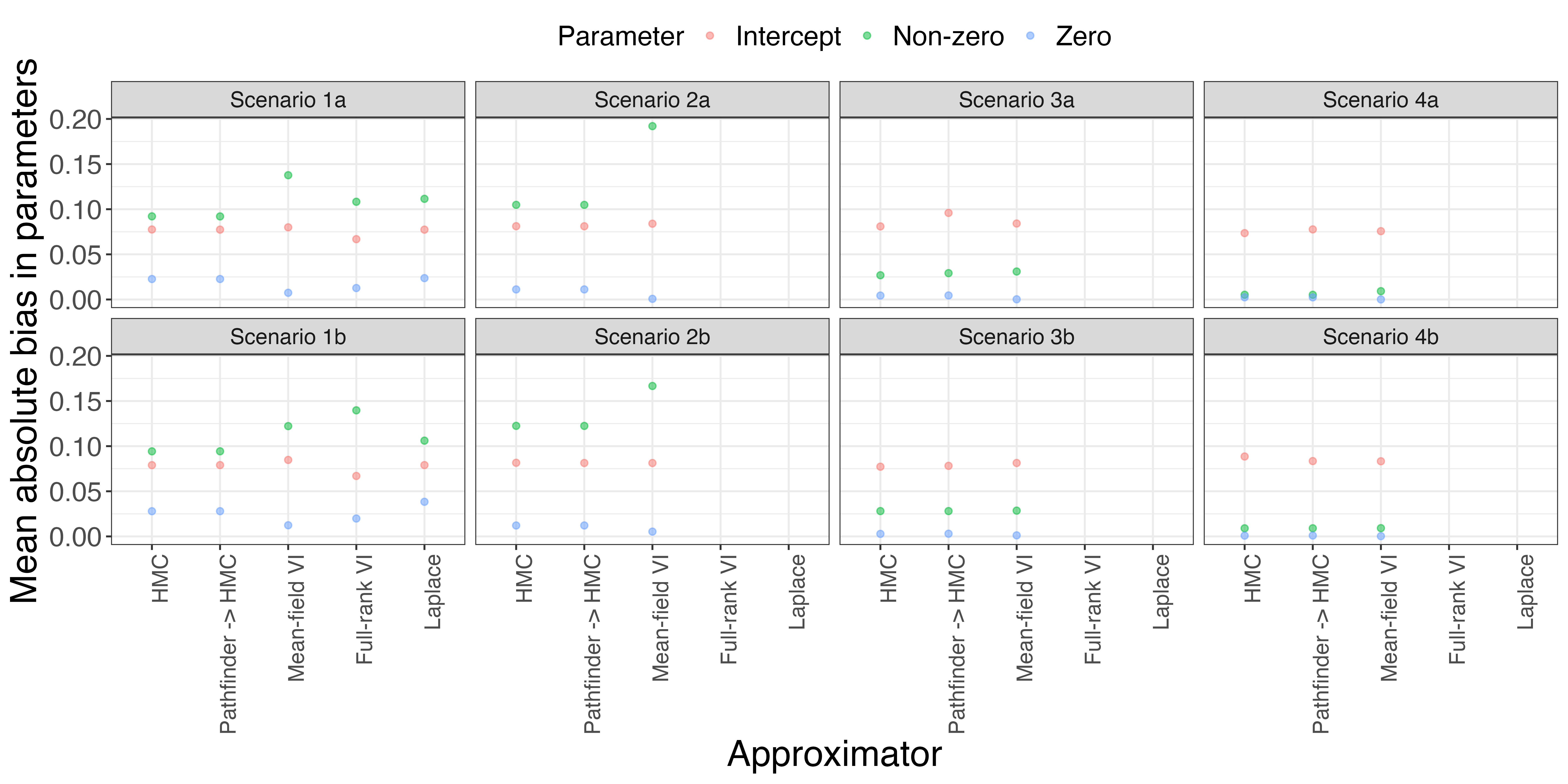}
    \caption{Absolute difference between estimated regression parameters and real parameters, split by type of parameters. For all scenarios except 1a and 1b, there were no results of full-rank and Laplace with good or questionable quality, so these are missing from the figure. }
    \label{fig:bias_params}
\end{figure}

Using a shrinkage prior will, by design, bias the parameter estimates. To investigate how this bias differs based on the algorithm and type of parameters, the absolute error is inspected. Pathfinder $\rightarrow$ HMC performs very similarly to HMC, while mean-field has less bias in the zero parameters but often more in the non-zero parameters (Figure \ref{fig:bias_params}). All methods have similar patterns across the scenarios. 

\begin{figure}[H]
    \centering
    \includegraphics[width=0.95\linewidth]{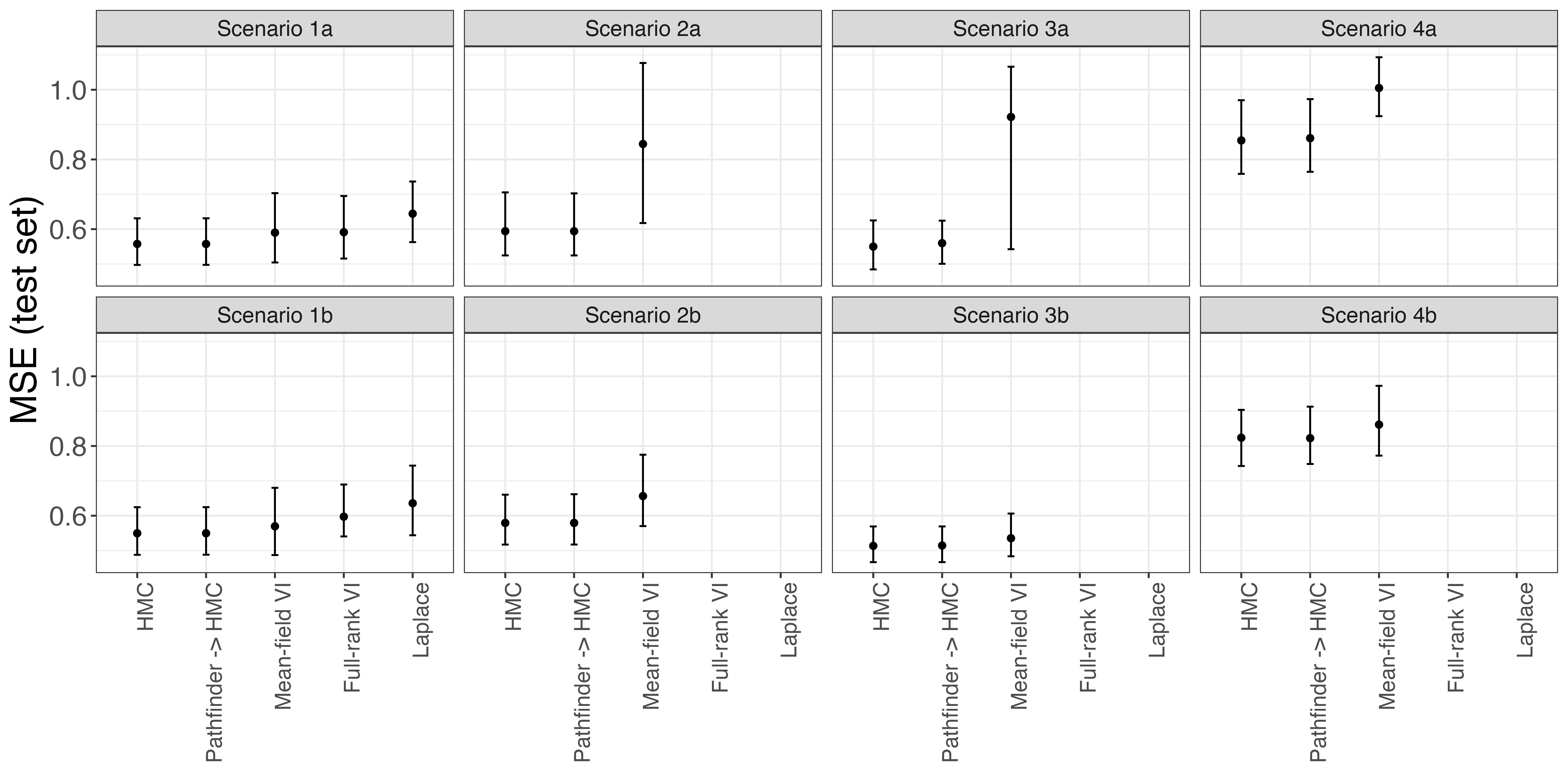}
    \caption{The MSE on the test set with the 95\% empirical interval, for the different methods split by scenario.  Some methods are missing, due to non-convergence. Note that scenarios 4a/4b have a lower $R^2$, so a higher MSE can be expected.}
    \label{fig:mse}
\end{figure}

The predictive performance, defined as MSE, is overall the lowest for HMC and Pathfinder as initialization for HMC (Figure \ref{fig:mse}). Mean-field can get close to the predictive performance of the MCMC methods. This indicates that after the MCMC methods, mean-field VI can also be a contender if the goal is to obtain a prediction model. The Pareto $\hat{k}$ values for the mean-field algorithm were not an indicator of predictive performance based on the MSE (Appendix E).

\begin{figure}[H]
    \centering
    \includegraphics[width=0.8\linewidth]{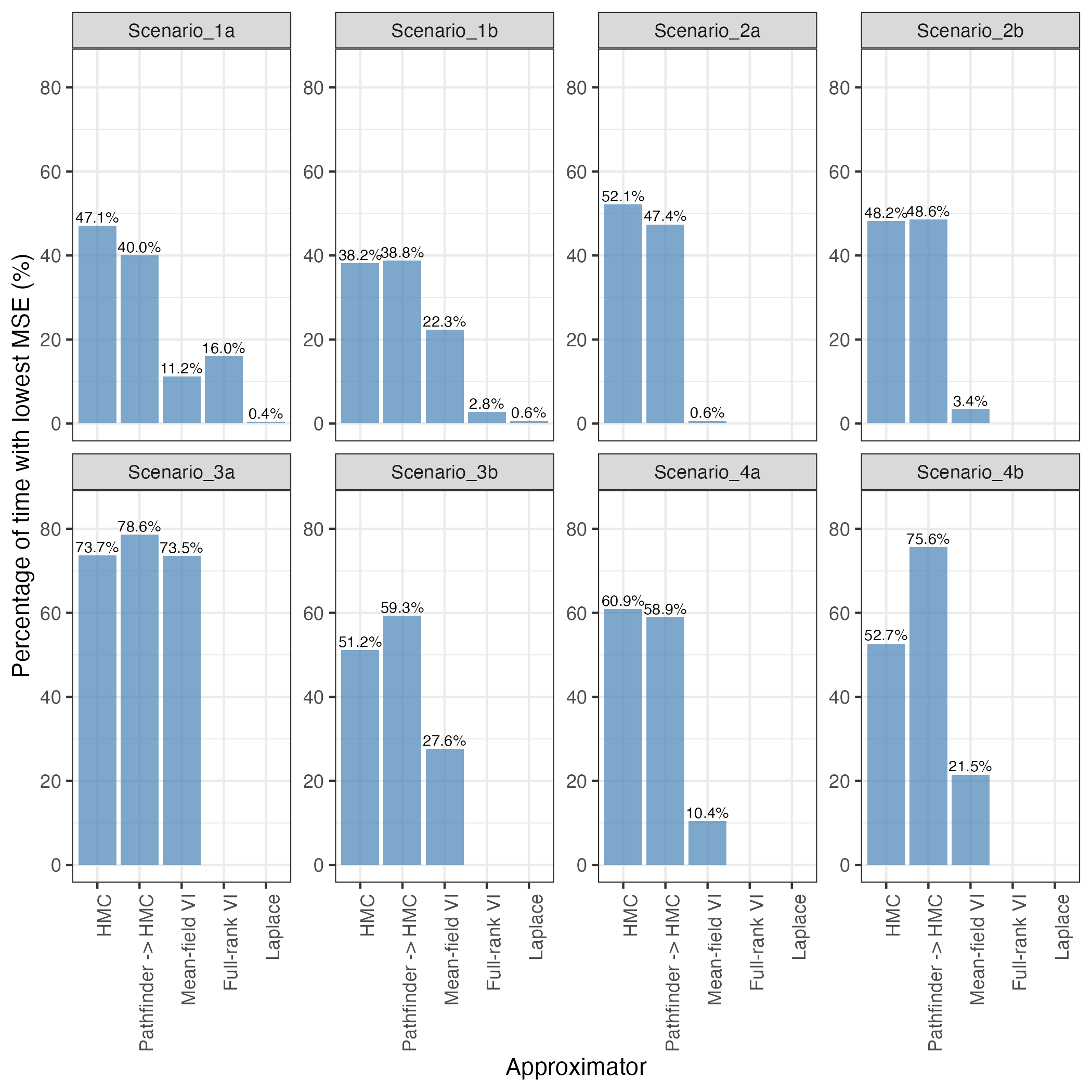}
    \caption{The percentage of models that had the lowest MSE in the runs that converged. Note that the number of converged datasets differed, so the percentages do not add up to 100\%. }
    \label{fig:mse_lowest}
\end{figure}

By averaging over the simulations, it cannot be seen how often a method had the best performance in a specific dataset. In Figure \ref{fig:mse_lowest} (Appendix B.), the percentage of times is shown that the method had the best performance in the simulations that converged. HMC and Pathfinder as initialization perform the best, with HMC seemingly performing better in the lower dimensional setting and Pathfinder $->$ HMC in the higher dimensional settings. Mean-field VI is also a strong contender, with the model being the best 10-30\% of the time. 

\begin{figure}[H]
    \centering
    \includegraphics[width=0.95\linewidth]{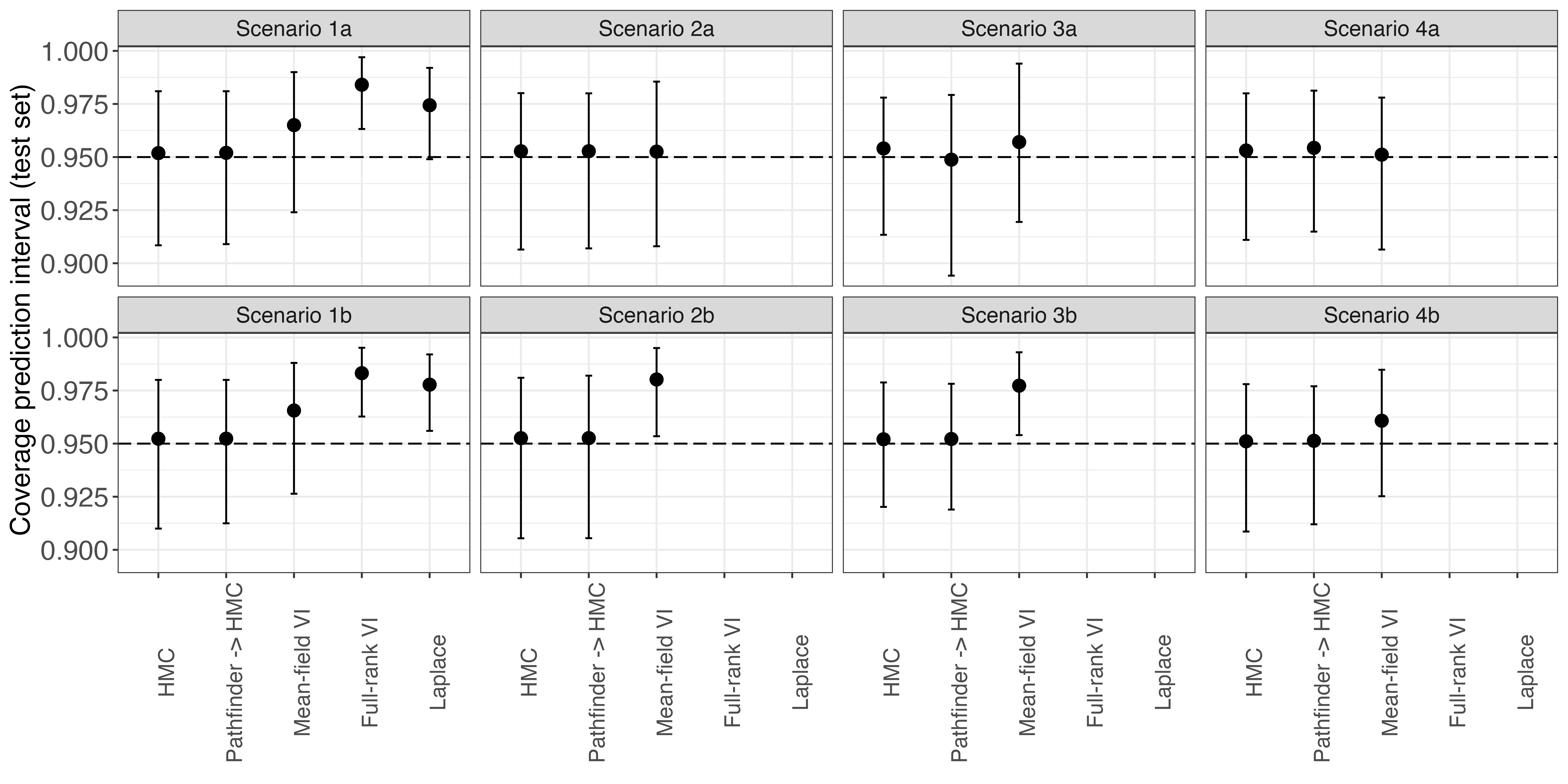}
    \caption{The coverage of the 95\% prediction interval on the test set for the different methods, with the corresponding 95\% empirical interval. The dashed line indicates the correct coverage.}
    \label{fig:cov}
\end{figure}

The coverage of the prediction interval is nominal for HMC and Pathfinder to HMC, and too high for mean-field VI, full-rank VI and Laplace, as can be seen in Figure \ref{fig:cov}. Often, undercoverage is seen when using variational inference methods with the exclusive (reversed) KL \cite{dhaka_challenges_2021, blei_variational_2017}. The overcoverage is a result of using the regularized horseshoe prior, which causes long tails towards zero in the non-zero parameter posteriors (Appendix C). Using a different prior can correct the overcoverage, at the cost of much higher predictive error (Appendix D).

\section{Real data}
To investigate if the findings from the simulation study have external validity, empirical datasets are used to examine the performance of the non-MCMC methods. The datasets have different ratios of predictors to samples, and the outcome variables are continuous and binary. A summary of the datasets used can be found in Table \ref{tab:datasets}.

\begin{table}[H]
\centering
\begin{tabular}{|l|l|r|r|r|l|}
\hline
\textbf{Dataset} & \textbf{Outcome type} & \textbf{Sample size} & \textbf{\# Predictors} & \textbf{Ratio (p/n)} & \textbf{Source (id)} \\ \hline
Math & Continuous & 395 & 42 & 0.11 & UCL ML repository (320) \\ \hline
Loan & Continuous & 1042 & 5742 & 5.51 & OpenML  (42572) \\ \hline
Topo & Continuous & 8885 & 266 & 0.03 & OpenML  (422) \\ \hline
QSAR & Continuous & 5742 & 1024 & 0.18 & OpenML  (44988) \\ \hline
Cristalli & Continuous & 32 & 1143 & 35.72 & OpenML  (420) \\ \hline
ALLAML & Binary & 72 & 7129 & 99.01 & ASU FS datasets\textsuperscript{*} \\ \hline
Parkinson & Binary & 756 & 753 & 1.00  & ASU FS datasets\textsuperscript{*} \\ \hline
Prostate & Binary & 102 & 5966 & 58.49 & ASU FS datasets\textsuperscript{*} \\ \hline
Anomalies & Binary & 1763 & 1558 & 0.88 & Kaggle\textsuperscript{†} \\ \hline
\end{tabular}
\vspace{0.5em}

{\raggedright\footnotesize
\textsuperscript{*} Under the Biological Data section at \texttt{https://jundongl.github.io/scikit-feature/datasets.html} \\
\textsuperscript{†} Can be found under the name: \emph{Detecting Anomalies in Wafer Manufacturing}
\par}
\caption{Summary of the empirical datasets used.}
\label{tab:datasets}
\end{table}

The methods that are used are the same as in the simulation (Table \ref{tab:sim_methods}), with one addition: ridge regression \cite{hoerl_ridge_1970}. This non-Bayesian method offers a way to obtain quick point estimates. The uncertainty in the predictions is not directly captured, as it lacks a posterior predictive distribution. Uncertainty estimates are thus not considered.

For all methods, 5-fold cross-validation was used to calculate the predictive performance of the models. The outcome measures are averaged over the folds where the models converged. All rows were randomly shuffled before the cross-validation, to ensure that the ordering of the data had no influence on the outcome. For the \textit{math} dataset, there is an online appendix showcasing how to implement the various models in R using stan. A deeper dive is taken in an online appendix to explain the effect of PSIS on the samples. 

The empirical datasets were run on a MacBook with an M2 chip and 16GB of RAM and all R-code \cite{r_core_team_r_2024} is available at \href{https://osf.io/mdk9z}{OSF}. 

The Ethics Review Board of the Faculty of Social and Behavioural Sciences from Utrecht University has provided Ethical approval for the use of the empirical datasets, under application number 25-0365.

\subsection{Continuous outcome}

The outcome metrics used are the same as in the simulation study, and are explained in Section 5.3.

\subsubsection{Results}

The main results for the data with a continuous outcome can be seen in Table \ref{tab:emp_res_cont}, further results are in Table \ref{tab:emp_res_cont3} (Appendix F). Similar to the simulation study, full-rank did not converge often, only in the low dimensional \textit{math} dataset. Except in the loan dataset and for the full-rank approximation, all methods converged. The running time was significantly higher than in the simulation study, HMC took 100 minutes in the \textit{loan} dataset, while mean-field was about 3 and a half minutes. In the \textit{qsar} dataset Pathfinder as initialization of HMC took 4x longer than HMC. The Pathfinder part only took 2 minutes; thus, it seems that HMC has some difficulty sampling with the starting values from Pathfinder. The number of divergent transitions was below 10\% for all datasets, apart from Pathfinder as initialization to HMC in the \textit{cristalli} data, where it was 24 \% (Table \ref{tab:emp_res_cont3}, Appendix F). The Pareto $\hat{k}$ value was above 0.7 for all the non-MCMC methods in all datasets. 

For the predictive performance, HMC performed the best in 4 out of the 5 datasets. This can be expected as it is our gold standard. Of the Bayesian models, mean-field gets the closest to HMC in terms of MSE, it even outperformed HMC in the \textit{cristalli} dataset while being 90x quicker than HMC. Ridge regression also has competitive performance with HMC in terms of MSE, but at the cost of not providing uncertainty estimates. 


\begin{table}[H]
\centering
\begin{tabular}{llrlll}
  \hline
\textbf{Dataset} & \textbf{Method} & 
\begin{tabular}[c]{@{}l@{}}\textbf{Converged\textsuperscript{*}} \\ \textbf{(\%)} \end{tabular}  & 
\begin{tabular}[c]{@{}l@{}}\textbf{Runtime} \\ \textbf{(seconds)} \end{tabular} & \textbf{MSE} & 
\begin{tabular}[c]{@{}l@{}}\textbf{Coverage} \\ \textbf{(predictions)} \end{tabular}\\ 
  \hline
\multirow{6}{*}{\textit{math}}
  & HMC & 100 & 14 & \textbf{3.81202} & 0.94 \\ 
  & Pathfinder $\rightarrow$ HMC & 100 & 19 (1)\textsuperscript{†} & 3.82381 & 0.94 \\ 
  & VI (mean-field) & 100 & 0.39 & 4.01795 & 0.95 \\
  & VI (full-rank) & 40 & 1 & 101.08411 & 0.18 \\  
  & Laplace & 100 & 0.37 & 5.21049 & 0.94 \\ 
  & Classical Ridge & 100 & 0.03 & 3.82507 &  \\ 
  \hline
\multirow{6}{*}{\textit{loan}}
  & HMC & 100 & 6095 & \textbf{2.72473} & 0.96 \\ 
  & Pathfinder $\rightarrow$ HMC & 0 &  &  &  \\ 
  & VI (mean-field) & 100 & 202 & 330.92177 & 1.00 \\ 
  & VI (full-rank) & 0 &  &  &  \\ 
  & Laplace & 0 &  &  &  \\ 
  & Classical Ridge & 100 & 21 & 2.80739 &  \\ 
  \hline
\multirow{6}{*}{\textit{topo}}
  & HMC & 100 & 3826 & \textbf{0.00084} & 0.92 \\ 
  & Pathfinder $\rightarrow$ HMC & 100 & 5317 (44)\textsuperscript{†} & \textbf{0.00084} & 0.92 \\ 
  & VI (mean-field) & 100 & 9.9 & 0.00087 & 0.92 \\ 
  & VI (full-rank) & 0 &  &  &  \\ 
  & Laplace & 40 & 9.52 & 269.36271 & 0.51 \\ 
  & Classical Ridge & 100 & 2.36 & 0.00085 &  \\ 
  \hline
\multirow{6}{*}{\textit{qsar}}
  & HMC & 100 & 4236 & \textbf{0.02029} & 0.95 \\ 
  & Pathfinder $\rightarrow$ HMC & 100 & 17378 (142)\textsuperscript{†} & 0.0203 & 0.95 \\ 
  & VI (mean-field) & 100 & 44.76 & 0.03867 & 0.96 \\ 
  & VI (full-rank) & 0 &  &  &  \\ 
  & Laplace & 100 & 31.91 & 0.02708 & 0.99 \\ 
  & Classical Ridge & 100 & 10.56 & 0.02081 &  \\ 
  \hline
\multirow{6}{*}{\textit{cristalli}}
  & HMC & 100 & 423 & 0.10374 & 0.81 \\ 
  & Pathfinder $\rightarrow$ HMC & 100 & 577 (12)\textsuperscript{†} & 0.10686 & 0.81 \\ 
  & VI (mean-field) & 100 & 4.75 & \textbf{0.08205} & 0.91 \\ 
  & VI (full-rank) & 0 &  &  &  \\ 
  & Laplace & 100 & 8.72 & $>$10\textasciicircum4 & 1.00 \\ 
  & Classical Ridge & 100 & 0.05 & 0.10311 &  \\ 
  \hline
\end{tabular}

\vspace{0.5em}

{\raggedright\footnotesize
\textsuperscript{*} Percentage of folds in the cross-validation that provided results.\\

\textsuperscript{†} The values in brackets show the runtime of Pathfinder.
\par}

\caption{Results from empirical datasets with a continuous outcome. The bold MSE values are the lowest within each dataset.}
\label{tab:emp_res_cont}
\end{table}

\subsection{Binary outcome}

To ensure that our results generalize to other types of outcomes, the methods are also compared on datasets with binary outcomes. Logistic regression is used to model the binary outcomes. Many metrics could be used to evaluate the predictive performance of models with binary outcomes \cite{calster_performance_2024}. In this work, we focus on three metrics: 1) accuracy; 2) Area under the curve (AUC); and 3) the Brier score. The accuracy is the percentage of predictions that have the same label as the observed outcome. For calculating the accuracy a threshold is needed to transform the risk prediction into a label; in our case, a threshold of 50\% is used meaning a risk prediction of above 0.5 resulted in a predicted label of 1 and 0 otherwise. The AUC quantifies the predictive power across all thresholds. For both the accuracy and AUC, higher values are better, with 1 being the maximum. The Brier score considers both predictive performance and calibration (how well the predicted risks are in line with the observed risks). It can be calculated as the MSE between the observed (binary) value and the risk prediction, where lower values are better. 

\subsubsection{Results}

The full-rank method only converged in one fold of one dataset, and Laplace and Pathfinder did not converge once in the \textit{Parkinson} dataset. The runtime of mean-field VI was 45-200 times quicker than HMC. The number of divergent transitions was below 10\%, while the Pareto $\hat{k}$ value was above 0.7 for all datasets (Table \ref{tab:emp_res_bin2}). The MCMC methods performed the best in one of the three performance measures in all but one dataset. Ridge performed much better in the \textit{Parkinson} dataset, while surprisingly Laplace had good performance in both the \textit{ALLAML} and \textit{prostate} datasets. 


\begin{table}[H]
\centering
\begin{tabular}{llrlrrr}
  \hline
\textbf{Dataset} & \textbf{Method} & 
\begin{tabular}[c]{@{}l@{}}\textbf{Converged\textsuperscript{*}} \\ \textbf{(\%)} \end{tabular} & 
\begin{tabular}[c]{@{}l@{}}\textbf{Runtime} \\ \textbf{(seconds)} \end{tabular} & \textbf{AUC} & \textbf{Accuracy} & \textbf{Brier score} \\ 
  \hline
\multirow{6}{*}{\textit{ALLAML}}
  & HMC & 100 & 1597 & 0.973 & \textbf{0.986} & 0.071 \\ 
  & Pathfinder $\rightarrow$ HMC & 100 & 921 (32)\textsuperscript{†}  & 0.978 & \textbf{0.986} & 0.072 \\ 
  & VI (mean-field) & 100 & 18 & 0.856 & 0.652 & 0.218 \\ 
  & VI (full-rank) & 0 &  &  &  &  \\ 
  & Laplace & 100 & 475 & \textbf{0.982} & 0.957 & 0.217 \\ 
  & Classical Ridge & 100 & 0.2 & 0.973 & 0.930 & \textbf{0.055} \\ 
  \hline
\multirow{6}{*}{\textit{parkinson}}
  & HMC & 100 & 652 & 0.567 & 0.557 & 0.323 \\ 
  & Pathfinder $\rightarrow$ HMC & 0 &  &  &  &  \\ 
  & VI (mean-field) & 100 & 7 & 0.800 & 0.802 & 0.144 \\ 
  & VI (full-rank) & 0 &  &  &  &  \\ 
  & Laplace & 0 &  &  &  &  \\ 
  & Classical Ridge & 100 & 1 & \textbf{0.887} & \textbf{0.862} & \textbf{0.108} \\ 
  \hline
\multirow{6}{*}{\textit{prostate}}
  & HMC & 100 & 1757 & 0.945 & \textbf{0.913} & \textbf{0.085} \\ 
  & Pathfinder $\rightarrow$ HMC & 100 & 943 (28)\textsuperscript{†}  & 0.945 & \textbf{0.913} & \textbf{0.085} \\ 
  & VI (mean-field) & 100 & 34 & 0.752 & 0.678 & 0.233 \\ 
  & VI (full-rank) & 0 &  &  &  &  \\ 
  & Laplace & 100 & 367 & \textbf{0.951} & \textbf{0.913} & 0.183 \\ 
  & Classical Ridge & 100 & 0.2 & 0.931 & 0.911 & 0.088 \\ 
  \hline
\multirow{6}{*}{\textit{anomalies}}
  & HMC & 100 & 2905 & \textbf{0.892} & 0.914 & 0.059 \\ 
  & Pathfinder $\rightarrow$ HMC & 100 & 4156 (61)\textsuperscript{†}  & \textbf{0.892} & 0.915 & 0.059 \\ 
  & VI (mean-field) & 100 & 14 & 0.868 & \textbf{0.921} & \textbf{0.057} \\ 
  & VI (full-rank) & 20 & 2064 & 0.743 & 0.881 & 0.106 \\ 
  & Laplace & 80 & 27 & 0.641 & 0.836 & 0.150 \\ 
  & Classical Ridge & 100 & 12 & 0.500 & 0.919 & 0.250 \\ 
  \hline
\end{tabular}

\vspace{0.5em}

{\raggedright\footnotesize
\textsuperscript{*} Percentage of folds in the cross-validation that provided results. \\

\textsuperscript{†} The values in brackets show the runtime of Pathfinder.
\par}

\caption{Results from empirical datasets with a binary outcome. For the AUC, Accuracy and the Brier score, the best performing methods in each dataset are in bold.}
\label{tab:emp_res_bin}
\end{table}

\section{Discussion}
In this paper, we have given an extensive overview of MCMC and non-MCMC methods that can be used to obtain posterior samples. More specifically, the focus was on the non-MCMC methods that are available in Stan \cite{carpenter_stan_2017}, although some of these algorithms are also available in other software programs such as PYMC \cite{abril-pla_pymc_2023} or INLA \cite{martins_bayesian_2013}. The MCMC and non-MCMC algorithms were compared in a predictive regression modeling setting using a simulation and empirical data sets. 

Our results showed that for high-dimensional problems, HMC, Pathfinder as an initialization for HMC and mean-field VI generally returned reliable results. Laplace would often produce unreliable results, whereas full-rank VI returned fewer results as the dimensions increased. The MCMC methods have a number of convergence criteria, of which $\hat{R}$, ESS and divergent transitions were used in this study. The simulation highlighted that with a burn-in period of 1000 draws and 2000 thereafter, HMC would still be able to obtain good quality estimates even in the scenarios with a \textit{p} to \textit{n} ratio of 10. For the non-MCMC algorithms, there are less well known metrics to assess how close the approximation comes to the posterior. The Pareto $\hat{k}$ value was used, which stems from Pareto smoothed importance sampling, where the goal is to resample the draws based on the Pareto-smoothed importance weights. Unfortunately, our experiments found that almost all models had a Pareto $\hat{k}$ above the recommended 0.7 threshold \cite{yao_yes_2018}, which makes it hard to distinguish good from bad approximations. Especially the full-rank and Laplace methods resulted in many models with high Pareto $\hat{k}$ values ($>5$). A potential reason is that these methods construct a full covariance matrix, compared to the diagonal covariance matrix of the mean-field. The mean-field algorithm thus scales linearly with regard to the number of regression parameters, while full-rank and Laplace algorithms scale quadratically. When having thousands of predictors, this can lead to millions of parameters, with sometimes only hundreds of observations. One can imagine this might lead to unstable results. More research is needed to evaluate the usefulness of the Pareto $\hat{k}$ value for assessing the quality of (high-dimensional) posterior approximations.  

For the methods that provided good-quality results, we generally see the best predictive performance for (Pathfinder $->$) HMC. This is expected. In the simulation study, the only reliable non-MCMC method was mean-field, which had an increase of 3-90\% in MSE over the different scenarios while reducing the running time by 7-30 times. To investigate if the Pareto $\hat{k}$ value can be used to assess the predictive performance of the approximation, we ran some experiments (Appendix E.). A higher Pareto $\hat{k}$ value did not result in a higher MSE compared to HMC, while there was a relation between the coverage and the Pareto $\hat{k}$ (Appendix E.). The uncertainty around the predictions was evaluated by the coverage of the predictive interval. The non-MCMC methods obtained small overcoverage and thus posterior predictive distributions that were a bit too wide. This was quite a surprising result, as undercoverage is expected when using the exclusive $KL(q||p)$. A possible explanation is that the regularized horseshoe prior paired with the non-MCMC methods causes the posterior for non-zero parameters to have a long tail towards zero, leading to overcoverage (Appendix C.).  

There were some exceptions to the results of the simulation study in the empirical datasets. First, the mean-field approximation had the best performance in the \textit{cristali} dataset. Second, Laplace approximation seems to perform better modeling a binary outcome as it had good performance in both the \textit{ALLAML} and \textit{prostate} datasets. In general, there was a tremendous speed-up of the non-MCMC algorithms compared to HMC. An example is in the \textit{topo} dataset, where the mean-field approximation was almost 400 times quicker than HMC while having an MSE only 4\% higher, and in the \textit{cristalli} dataset, mean-field VI even outperformed HMC with a 90 times speed-up. This result did not hold across all empirical datasets, as in the \textit{topo} dataset, the MSE increased over 100x with only a speed-up of 30x, compared to HMC. Notably, the running time of Pathfinder as initialization of HMC was occasionally much higher than for HMC. For example, in the \textit{qsar} dataset, Pathfinder $->$ HMC resulted in a 4x longer running time compared to HMC; the increase in running time came from the HMC part as the Pathfinder algorithm is relatively quick. 
 
Our simulations were limited in some aspect. By fixing the R2 value, the coefficients of the actual predictors became very small. The effect sizes of the predictors were also rather similar, which could be different in the merpical datasets if there were only a couple of relevant predictors. Furthermore, there was always a non-diagonal covariance structure, either between the covariates with a non-zero effect size or between all covariates, when generating the data. This could harm the performance of the full-rank and Laplace algorithms if such a structure is not present in empirical datasets. This might explain the good performance of Laplace in the \textit{ALLAML} and \textit{prostate} datasets. Moreover, in our analysis, the settings were somewhat high-dimensional, but not extremely so. This was a practical choice in order to be able to run MCMC as a comparison. However, in reality, researchers might simply run MCMC in these settings anyway. Further research could look into more high-dimensional settings.

The focus in this paper was on algorithms that are implemented in Stan, although there are many other algorithms that could also be considered. For example, the family of approximate distributions for VI could be extended beyond Gaussians (e.g., normalizing flows \cite{agrawal_advances_2020, papamakarios_normalizing_2021}). Alternatively, instead of using gradient descent to minimize the KL loss, one can adopt a more stable approach, such as score matching \cite{modi_variational_2023}. Using score functions opens the door for new approximate distributions for VI \cite{cai_eigenvi_2024}. For Laplace approximation, there are also other popular implementations such as INLA \cite{rue_approximate_2009} or methods that approximate the covariance matrix by a less computationally expensive method \cite{daxberger_laplace_2021, miani_laplacian_2022}.  An important thing that is still missing is good indicators of the quality of non-MCMC algorithms, and until we have these, it becomes hard to trust the results. There is recent work on providing error bounds on specific quantities, such as the mean or standard deviation, of the approximation \cite{huggins_validated_2020, kasprzak_how_2025}. Furthermore, more research is needed on why the initial values based on the Pathfinder algorithm lead to more computing time for HMC in high-dimensional settings. 

In general, non-MCMC methods should be used carefully. The methods might be quicker, but it can be difficult to assess the quality of the estimates, especially in high-dimensional settings. For model building, the non-MCMC algorithms can be extremely helpful to decrease the computational burden of MCMC. However, more research into non-MCMC methods is needed before we can replace the final MCMC run with an approximation.




\section{Acknowledgements}
We would like to thank Gerko Vink and Matthijs Vákár for valuable discussions, Alex Carriero for comments on paper and Aki Vehtari and Bob Carpenter for answering questions relating to the Stan implementation of the algorithms on the Stan forum. 
This research was supported by a NWO Veni Grant (Vl.Veni.221G.005) to SvE.

\newpage

\bibliographystyle{apalike}
\bibliography{Bayes}

\begin{thebibliography}{}

\bibitem[Abril-Pla et~al., 2023]{abril-pla_pymc_2023}
Abril-Pla, O., Andreani, V., Carroll, C., Dong, L., Fonnesbeck, C.~J., Kochurov, M., Kumar, R., Lao, J., Luhmann, C.~C., Martin, O.~A., Osthege, M., Vieira, R., Wiecki, T., and Zinkov, R. (2023).
\newblock {PyMC}: a modern, and comprehensive probabilistic programming framework in {Python}.
\newblock {\em PeerJ Computer Science}, 9:e1516.
\newblock Publisher: PeerJ Inc.

\bibitem[Agrawal et~al., 2020]{agrawal_advances_2020}
Agrawal, A., Sheldon, D.~R., and Domke, J. (2020).
\newblock Advances in {Black}-{Box} {VI}: {Normalizing} {Flows}, {Importance} {Weighting}, and {Optimization}.
\newblock In {\em Advances in {Neural} {Information} {Processing} {Systems}}, volume~33, pages 17358--17369. Curran Associates, Inc.

\bibitem[Andrieu and Thoms, 2008]{andrieu_tutorial_2008}
Andrieu, C. and Thoms, J. (2008).
\newblock A tutorial on adaptive {MCMC}.
\newblock {\em Statistics and computing}, 18:343--373.
\newblock Publisher: Springer.

\bibitem[Ballnus et~al., 2017]{ballnus_comprehensive_2017}
Ballnus, B., Hug, S., Hatz, K., Görlitz, L., Hasenauer, J., and Theis, F.~J. (2017).
\newblock Comprehensive benchmarking of {Markov} chain {Monte} {Carlo} methods for dynamical systems.
\newblock {\em BMC Systems Biology}, 11(1):63.

\bibitem[Bengtsson, 2021]{RJ-2021-048}
Bengtsson, H. (2021).
\newblock A unifying framework for parallel and distributed processing in r using futures.
\newblock {\em The R Journal}, 13(2):208--227.

\bibitem[Betancourt, 2018]{betancourt_conceptual_2018}
Betancourt, M. (2018).
\newblock A {Conceptual} {Introduction} to {Hamiltonian} {Monte} {Carlo}.
\newblock arXiv:1701.02434 [stat].

\bibitem[Biondi and De~Luca, 2012]{biondi_bayesian_2012}
Biondi, D. and De~Luca, D.~L. (2012).
\newblock A {Bayesian} approach for real-time flood forecasting.
\newblock {\em Physics and Chemistry of the Earth, Parts A/B/C}, 42-44:91--97.

\bibitem[Bishop, 2006]{bishop_pattern_2006}
Bishop, C.~M. (2006).
\newblock {\em Pattern recognition and machine learning}.
\newblock Information science and statistics. Springer, New York.

\bibitem[Biswas et~al., 2022]{biswas_coupling-based_2022}
Biswas, N., Bhattacharya, A., Jacob, P.~E., and Johndrow, J.~E. (2022).
\newblock Coupling-based {Convergence} {Assessment} of some {Gibbs} {Samplers} for {High}-{Dimensional} {Bayesian} {Regression} with {Shrinkage} {Priors}.
\newblock {\em Journal of the Royal Statistical Society Series B: Statistical Methodology}, 84(3):973--996.

\bibitem[Blei et~al., 2017]{blei_variational_2017}
Blei, D.~M., Kucukelbir, A., and McAuliffe, J.~D. (2017).
\newblock Variational {Inference}: {A} {Review} for {Statisticians}.
\newblock {\em Journal of the American Statistical Association}, 112(518):859--877.

\bibitem[Byrd et~al., 1995]{byrd_limited_1995}
Byrd, R.~H., Lu, P., Nocedal, J., and Zhu, C. (1995).
\newblock A {Limited} {Memory} {Algorithm} for {Bound} {Constrained} {Optimization}.
\newblock {\em SIAM Journal on Scientific Computing}, 16(5):1190--1208.

\bibitem[Bürkner, 2021]{Burkner_bmrs_2021}
Bürkner, P.-C. (2021).
\newblock Bayesian item response modeling in {R} with {brms} and {Stan}.
\newblock {\em Journal of Statistical Software}, 100(5):1--54.

\bibitem[Bürkner et~al., 2025]{Bürkner_posterior_2025}
Bürkner, P.-C., Gabry, J., Kay, M., and Vehtari, A. (2025).
\newblock posterior: Tools for working with posterior distributions.
\newblock R package version 1.6.1.

\bibitem[Bürkner et~al., 2023]{burkner_models_2023}
Bürkner, P.-C., Scholz, M., and Radev, S.~T. (2023).
\newblock Some models are useful, but how do we know which ones? {Towards} a unified {Bayesian} model taxonomy.
\newblock {\em Statistics Surveys}, 17(none):216--310.
\newblock Publisher: Amer. Statist. Assoc., the Bernoulli Soc., the Inst. Math. Statist., and the Statist. Soc. Canada.

\bibitem[Cai et~al., 2024]{cai_eigenvi_2024}
Cai, D., Modi, C., Margossian, C.~C., Gower, R.~M., Blei, D.~M., and Saul, L.~K. (2024).
\newblock {EigenVI}: score-based variational inference with orthogonal function expansions.
\newblock {\em Advances in Neural Information Processing Systems}, 37:132691--132721.

\bibitem[Calster et~al., 2024]{calster_performance_2024}
Calster, B.~V., Collins, G.~S., Vickers, A.~J., Wynants, L., Kerr, K.~F., Barreñada, L., Varoquaux, G., Singh, K., Moons, K. G.~M., Hernandez-boussard, T., Timmerman, D., Mclernon, D.~J., Smeden, M.~V., and Steyerberg, E.~W. (2024).
\newblock Performance evaluation of predictive {AI} models to support medical decisions: {Overview} and guidance.
\newblock arXiv:2412.10288 [cs].

\bibitem[Carlo, 2004]{carlo_markov_2004}
Carlo, C.~M. (2004).
\newblock Markov chain monte carlo and gibbs sampling.
\newblock {\em Lecture notes for EEB}, 581(540):3.
\newblock Publisher: Citeseer.

\bibitem[Carpenter et~al., 2017]{carpenter_stan_2017}
Carpenter, B., Gelman, A., Hoffman, M., Lee, D., Goodrich, B., Betancourt, M., Brubaker, M., Guo, J., Li, P., and Riddell, A. (2017).
\newblock Stan : {A} {Probabilistic} {Programming} {Language}.
\newblock {\em Journal of Statistical Software}, 76.

\bibitem[Carvalho et~al., 2009]{carvalho_handling_2009}
Carvalho, C.~M., Polson, N.~G., and Scott, J.~G. (2009).
\newblock Handling sparsity via the horseshoe.
\newblock In {\em Artificial intelligence and statistics}, pages 73--80. PMLR.

\bibitem[Dang and Kishino, 2019]{dang_stochastic_2019}
Dang, T. and Kishino, H. (2019).
\newblock Stochastic {Variational} {Inference} for {Bayesian} {Phylogenetics}: {A} {Case} of {CAT} {Model}.
\newblock {\em Molecular Biology and Evolution}, 36(4):825--833.

\bibitem[Daxberger et~al., 2021]{daxberger_laplace_2021}
Daxberger, E., Kristiadi, A., Immer, A., Eschenhagen, R., Bauer, M., and Hennig, P. (2021).
\newblock Laplace redux-effortless bayesian deep learning.
\newblock {\em Advances in neural information processing systems}, 34:20089--20103.

\bibitem[Dhaka et~al., 2021]{dhaka_challenges_2021}
Dhaka, A.~K., Catalina, A., Welandawe, M., Andersen, M.~R., Huggins, J., and Vehtari, A. (2021).
\newblock Challenges and opportunities in high dimensional variational inference.
\newblock {\em Advances in Neural Information Processing Systems}, 34:7787--7798.

\bibitem[Fortuin, 2022]{fortuin_priors_2022}
Fortuin, V. (2022).
\newblock Priors in {Bayesian} {Deep} {Learning}: {A} {Review}.
\newblock {\em International Statistical Review}, 90(3):563--591.

\bibitem[Gabry et~al., 2024]{Gabry_cmdstanr_2024}
Gabry, J., Češnovar, R., Johnson, A., and Bronder, S. (2024).
\newblock {\em cmdstanr: R Interface to 'CmdStan'}.
\newblock R package version 0.8.1, https://discourse.mc-stan.org.

\bibitem[Garby et~al., 2024]{garby_cmdstanr_2024}
Garby, J., Cesnovar, R., Johnson, A., and Bronder, S. (2024).
\newblock cmdstanr: {R} {Interface} to '{CmdStan}'.

\bibitem[Gelman et~al., 2013]{gelman_bayesian_2013}
Gelman, A., Carlin, J.~B., Stern, H.~S., Dunson, D.~B., Vehtari, A., and Rubin, D.~B. (2013).
\newblock Bayesian {Data} {Analysis}.
\newblock {\em Bayesian Data Analysis}.
\newblock ISBN: 9781439840955.

\bibitem[Gelman et~al., 1995]{gelman_bayesian_1995}
Gelman, A., Carlin, J.~B., Stern, H.~S., and Rubin, D.~B. (1995).
\newblock {\em Bayesian data analysis}.
\newblock Chapman and Hall/CRC.

\bibitem[Gelman et~al., 2021]{gelman_regression_2021}
Gelman, A., Hill, J., and Vehtari, A. (2021).
\newblock {\em Regression and other stories}.
\newblock Cambridge University Press.

\bibitem[Gelman et~al., 2015]{gelman_stan_2015}
Gelman, A., Lee, D., and Guo, J. (2015).
\newblock Stan: {A} {Probabilistic} {Programming} {Language} for {Bayesian} {Inference} and {Optimization}.
\newblock {\em Journal of Educational and Behavioral Statistics}, 40(5):530--543.
\newblock Publisher: American Educational Research Association.

\bibitem[Gelman and Rubin, 1992]{gelman_inference_1992}
Gelman, A. and Rubin, D.~B. (1992).
\newblock Inference from iterative simulation using multiple sequences.
\newblock {\em Statistical science}, 7(4):457--472.
\newblock Publisher: Institute of Mathematical Statistics.

\bibitem[Gelman et~al., 2017]{gelman_prior_2017}
Gelman, A., Simpson, D., and Betancourt, M. (2017).
\newblock The prior can often only be understood in the context of the likelihood.
\newblock {\em Entropy}, 19(10):555.
\newblock Publisher: MDPI.

\bibitem[Geman and Geman, 1984]{geman_stochastic_1984}
Geman, S. and Geman, D. (1984).
\newblock Stochastic relaxation, {Gibbs} distributions, and the {Bayesian} restoration of images.
\newblock {\em IEEE Transactions on pattern analysis and machine intelligence}, (6):721--741.
\newblock Publisher: IEEE.

\bibitem[George and McCulloch, 1993]{george_variable_1993}
George, E.~I. and McCulloch, R.~E. (1993).
\newblock Variable {Selection} via {Gibbs} {Sampling}.
\newblock {\em Journal of the American Statistical Association}, 88(423):881--889.

\bibitem[Geweke, 1991]{geweke_evaluating_1991}
Geweke, J. (1991).
\newblock Evaluating the accuracy of sampling-based approaches to the calculation of posterior moments.
\newblock Technical report, Federal Reserve Bank of Minneapolis.

\bibitem[Gilks et~al., 1995]{gilks_adaptive_1995}
Gilks, W.~R., Best, N.~G., and Tan, K. K.~C. (1995).
\newblock Adaptive {Rejection} {Metropolis} {Sampling} within {Gibbs} {Sampling}.
\newblock {\em Journal of the Royal Statistical Society. Series C (Applied Statistics)}, 44(4):455--472.
\newblock Publisher: [Royal Statistical Society, Oxford University Press].

\bibitem[Gruber et~al., 2023]{gruber_sources_2023}
Gruber, C., Schenk, P.~O., Schierholz, M., Kreuter, F., and Kauermann, G. (2023).
\newblock Sources of {Uncertainty} in {Machine} {Learning} -- {A} {Statisticians}' {View}.
\newblock arXiv:2305.16703 [cs, stat].

\bibitem[Gunapati et~al., 2022]{gunapati_variational_2022}
Gunapati, G., Jain, A., Srijith, P.~K., and Desai, S. (2022).
\newblock Variational inference as an alternative to {MCMC} for parameter estimation and model selection.
\newblock {\em Publications of the Astronomical Society of Australia}, 39:e001.
\newblock Publisher: Cambridge University Press.

\bibitem[Hastings, 1970]{hastings_monte_1970}
Hastings, W.~K. (1970).
\newblock Monte {Carlo} sampling methods using {Markov} chains and their applications.
\newblock {\em Biometrika}, 57(1):97--109.

\bibitem[Hoerl and Kennard, 1970]{hoerl_ridge_1970}
Hoerl, A.~E. and Kennard, R.~W. (1970).
\newblock Ridge {Regression}: {Biased} {Estimation} for {Nonorthogonal} {Problems}.
\newblock {\em Technometrics}, 12(1):55--67.
\newblock Publisher: [Taylor \& Francis, Ltd., American Statistical Association, American Society for Quality].

\bibitem[Hoffman and Gelman, 2014]{hoffman_no-u-turn_2014}
Hoffman, M.~D. and Gelman, A. (2014).
\newblock The {No}-{U}-{Turn} sampler: adaptively setting path lengths in {Hamiltonian} {Monte} {Carlo}.
\newblock {\em J. Mach. Learn. Res.}, 15(1):1593--1623.

\bibitem[Holt and Laury, 2002]{holt_risk_2002}
Holt, C.~A. and Laury, S.~K. (2002).
\newblock Risk {Aversion} and {Incentive} {Effects}.
\newblock {\em The American Economic Review}, 92(5):1644--1655.
\newblock Publisher: American Economic Association.

\bibitem[Huggins et~al., 2020]{huggins_validated_2020}
Huggins, J., Kasprzak, M., Campbell, T., and Broderick, T. (2020).
\newblock Validated variational inference via practical posterior error bounds.
\newblock In {\em International {Conference} on {Artificial} {Intelligence} and {Statistics}}, pages 1792--1802. PMLR.

\bibitem[Izmailov et~al., 2021]{izmailov_what_2021}
Izmailov, P., Vikram, S., Hoffman, M.~D., and Wilson, A. G.~G. (2021).
\newblock What {Are} {Bayesian} {Neural} {Network} {Posteriors} {Really} {Like}?
\newblock In {\em Proceedings of the 38th {International} {Conference} on {Machine} {Learning}}, pages 4629--4640. PMLR.
\newblock ISSN: 2640-3498.

\bibitem[Johnstone and Titterington, 2009]{johnstone_statistical_2009}
Johnstone, I.~M. and Titterington, D.~M. (2009).
\newblock Statistical challenges of high-dimensional data.
\newblock {\em Philosophical Transactions of the Royal Society A: Mathematical, Physical and Engineering Sciences}, 367(1906):4237--4253.

\bibitem[Jordan et~al., 1999]{jordan_introduction_1999}
Jordan, M.~I., Ghahramani, Z., Jaakkola, T.~S., and Saul, L.~K. (1999).
\newblock An {Introduction} to {Variational} {Methods} for {Graphical} {Models}.
\newblock {\em Machine Learning}, 37(2):183--233.

\bibitem[Kasprzak et~al., 2025]{kasprzak_how_2025}
Kasprzak, M.~J., Giordano, R., and Broderick, T. (2025).
\newblock How good is your {Laplace} approximation of the {Bayesian} posterior? {Finite}-sample computable error bounds for a variety of useful divergences.
\newblock arXiv:2209.14992 [math].

\bibitem[Kingma and Welling, 2022]{kingma_auto-encoding_2022}
Kingma, D.~P. and Welling, M. (2022).
\newblock Auto-{Encoding} {Variational} {Bayes}.
\newblock arXiv:1312.6114 [stat].

\bibitem[Kucukelbir et~al., 2017]{kucukelbir_automatic_2017}
Kucukelbir, A., Tran, D., Ranganath, R., Gelman, A., and Blei, D.~M. (2017).
\newblock Automatic differentiation variational inference.
\newblock {\em Journal of machine learning research}, 18(14):1--45.

\bibitem[Lambert, 2018]{lambert_students_2018}
Lambert, B. (2018).
\newblock {\em A {Student}’s {Guide} to {Bayesian} {Statistics}}.
\newblock SAGE.
\newblock Google-Books-ID: ZvBUDwAAQBAJ.

\bibitem[Lambert and Vehtari, 2022]{lambert_r_2022}
Lambert, B. and Vehtari, A. (2022).
\newblock R*: {A} robust {MCMC} convergence diagnostic with uncertainty using decision tree classifiers.
\newblock {\em Bayesian Analysis}, 17(2):353--379.
\newblock Publisher: International Society for Bayesian Analysis.

\bibitem[Magnusson et~al., 2025]{magnusson_posteriordb_2025}
Magnusson, M., Torgander, J., Bürkner, P.-C., Zhang, L., Carpenter, B., and Vehtari, A. (2025).
\newblock posteriordb: {Testing}, {Benchmarking} and {Developing} {Bayesian} {Inference} {Algorithms}.
\newblock In {\em Proceedings of {The} 28th {International} {Conference} on {Artificial} {Intelligence} and {Statistics}}, pages 1198--1206. PMLR.
\newblock ISSN: 2640-3498.

\bibitem[Margossian and Gelman, 2024]{margossian_for_2024}
Margossian, C.~C. and Gelman, A. (2024).
\newblock For how many iterations should we run {Markov} chain {Monte} {Carlo}?
\newblock arXiv:2311.02726 [stat].

\bibitem[Margossian et~al., 2025]{margossian_variational_2025}
Margossian, C.~C., Pillaud-Vivien, L., and Saul, L.~K. (2025).
\newblock Variational {Inference} for {Uncertainty} {Quantification}: an {Analysis} of {Trade}-offs.
\newblock arXiv:2403.13748 [stat].

\bibitem[Martins et~al., 2013]{martins_bayesian_2013}
Martins, T.~G., Simpson, D., Lindgren, F., and Rue, H. (2013).
\newblock Bayesian computing with {INLA}: {New} features.
\newblock {\em Computational Statistics \& Data Analysis}, 67:68--83.

\bibitem[Metropolis et~al., 1953]{metropolis_equation_1953}
Metropolis, N., Rosenbluth, A.~W., Rosenbluth, M.~N., Teller, A.~H., and Teller, E. (1953).
\newblock Equation of state calculations by fast computing machines.
\newblock {\em The journal of chemical physics}, 21(6):1087--1092.
\newblock Publisher: American Institute of Physics.

\bibitem[Miani et~al., 2022]{miani_laplacian_2022}
Miani, M., Warburg, F., Moreno-Muñoz, P., Skafte, N., and Hauberg, S. (2022).
\newblock Laplacian autoencoders for learning stochastic representations.
\newblock {\em Advances in Neural Information Processing Systems}, 35:21059--21072.

\bibitem[Mitchell and Beauchamp, 1988]{mitchell_bayesian_1988}
Mitchell, T.~J. and Beauchamp, J.~J. (1988).
\newblock Bayesian {Variable} {Selection} in {Linear} {Regression}.
\newblock {\em Journal of the American Statistical Association}, 83(404):1023--1032.

\bibitem[Modi et~al., 2023]{modi_variational_2023}
Modi, C., Gower, R., Margossian, C., Yao, Y., Blei, D., and Saul, L. (2023).
\newblock Variational inference with {Gaussian} score matching.
\newblock {\em Advances in Neural Information Processing Systems}, 36:29935--29950.

\bibitem[Monnahan et~al., 2017]{monnahan_faster_2017}
Monnahan, C.~C., Thorson, J.~T., and Branch, T.~A. (2017).
\newblock Faster estimation of {Bayesian} models in ecology using {Hamiltonian} {Monte} {Carlo}.
\newblock {\em Methods in Ecology and Evolution}, 8(3):339--348.
\newblock \_eprint: https://onlinelibrary.wiley.com/doi/pdf/10.1111/2041-210X.12681.

\bibitem[Müller et~al., 2024]{muller_transformers_2024}
Müller, S., Hollmann, N., Arango, S.~P., Grabocka, J., and Hutter, F. (2024).
\newblock Transformers {Can} {Do} {Bayesian} {Inference}.
\newblock arXiv:2112.10510 [cs, stat].

\bibitem[Neal, 2011]{brooks_mcmc_2011}
Neal, R.~M. (2011).
\newblock {MCMC} {Using} {Hamiltonian} {Dynamics}.
\newblock In {\em Handbook of {Markov} {Chain} {Monte} {Carlo}}, pages 113--162. Chapman and Hall/CRC, New York, 1 edition.

\bibitem[Neal, 2012]{neal_bayesian_2012}
Neal, R.~M. (2012).
\newblock {\em Bayesian learning for neural networks}, volume 118.
\newblock Springer Science \& Business Media.

\bibitem[Opper and Archambeau, 2009]{opper_variational_2009}
Opper, M. and Archambeau, C. (2009).
\newblock The {Variational} {Gaussian} {Approximation} {Revisited}.
\newblock {\em Neural Computation}, 21(3):786--792.

\bibitem[Papamakarios et~al., 2021]{papamakarios_normalizing_2021}
Papamakarios, G., Nalisnick, E., Rezende, D.~J., Mohamed, S., and Lakshminarayanan, B. (2021).
\newblock Normalizing {Flows} for {Probabilistic} {Modeling} and {Inference}.
\newblock {\em Journal of Machine Learning Research}, 22(57):1--64.

\bibitem[Papamarkou et~al., 2024]{papamarkou_position_2024}
Papamarkou, T., Skoularidou, M., Palla, K., Aitchison, L., Arbel, J., Dunson, D., Filippone, M., Fortuin, V., Hennig, P., Hernández-Lobato, J.~M., Hubin, A., Immer, A., Karaletsos, T., Khan, M.~E., Kristiadi, A., Li, Y., Mandt, S., Nemeth, C., Osborne, M.~A., Rudner, T. G.~J., Rügamer, D., Teh, Y.~W., Welling, M., Wilson, A.~G., and Zhang, R. (2024).
\newblock Position: {Bayesian} {Deep} {Learning} is {Needed} in the {Age} of {Large}-{Scale} {AI}.
\newblock arXiv:2402.00809 [cs].

\bibitem[Piironen and Vehtari, 2017]{piironen_sparsity_2017}
Piironen, J. and Vehtari, A. (2017).
\newblock Sparsity information and regularization in the horseshoe and other shrinkage priors.
\newblock {\em Electronic Journal of Statistics}, 11(2):5018--5051.
\newblock Publisher: Institute of Mathematical Statistics and Bernoulli Society.

\bibitem[Polson and Scott, 2010]{polson_shrink_2010}
Polson, N.~G. and Scott, J.~G. (2010).
\newblock Shrink globally, act locally: {Sparse} {Bayesian} regularization and prediction.
\newblock {\em Bayesian statistics}, 9(501-538):105.
\newblock Publisher: Citeseer.

\bibitem[R~Core~Team, 2024]{r_core_team_r_2024}
R~Core~Team, R. (2024).
\newblock R: {A} language and environment for statistical computing.
\newblock Publisher: R foundation for statistical computing Vienna, Austria.

\bibitem[Radev et~al., 2023]{radev_jana_2023}
Radev, S.~T., Schmitt, M., Pratz, V., Picchini, U., Köthe, U., and Bürkner, P.-C. (2023).
\newblock Jana: {Jointly} amortized neural approximation of complex {Bayesian} models.
\newblock In {\em Proceedings of the {Thirty}-{Ninth} {Conference} on {Uncertainty} in {Artificial} {Intelligence}}, pages 1695--1706. PMLR.
\newblock ISSN: 2640-3498.

\bibitem[Rainforth et~al., 2024]{rainforth_modern_2024}
Rainforth, T., Foster, A., Ivanova, D.~R., and Smith, F.~B. (2024).
\newblock Modern {Bayesian} {Experimental} {Design}.
\newblock {\em Statistical Science}, 39(1):100--114.
\newblock Publisher: Institute of Mathematical Statistics.

\bibitem[Rezende and Mohamed, 2015]{rezende_variational_2015}
Rezende, D.~J. and Mohamed, S. (2015).
\newblock Variational inference with normalizing flows.
\newblock In {\em Proceedings of the 32nd {International} {Conference} on {International} {Conference} on {Machine} {Learning} - {Volume} 37}, {ICML}'15, pages 1530--1538, Lille, France. JMLR.org.

\bibitem[Roberts et~al., 1997]{roberts_weak_1997}
Roberts, G.~O., Gelman, A., and Gilks, W.~R. (1997).
\newblock Weak {Convergence} and {Optimal} {Scaling} of {Random} {Walk} {Metropolis} {Algorithms}.
\newblock {\em The Annals of Applied Probability}, 7(1):110--120.

\bibitem[Roberts and Rosenthal, 2001]{roberts_optimal_2001}
Roberts, G.~O. and Rosenthal, J.~S. (2001).
\newblock Optimal scaling for various {Metropolis}-{Hastings} algorithms.
\newblock {\em Statistical science}, 16(4):351--367.
\newblock Publisher: Institute of Mathematical Statistics.

\bibitem[Rue et~al., 2009]{rue_approximate_2009}
Rue, H., Martino, S., and Chopin, N. (2009).
\newblock Approximate {Bayesian} {Inference} for {Latent} {Gaussian} models by using {Integrated} {Nested} {Laplace} {Approximations}.
\newblock {\em Journal of the Royal Statistical Society Series B: Statistical Methodology}, 71(2):319--392.

\bibitem[Rue et~al., 2017]{rue_bayesian_2017}
Rue, H., Riebler, A., Sørbye, S.~H., Illian, J.~B., Simpson, D.~P., and Lindgren, F.~K. (2017).
\newblock Bayesian {Computing} with {INLA}: {A} {Review}.
\newblock {\em Annual Review of Statistics and Its Application}, 4(1):395--421.

\bibitem[Tierney, 1994]{tierney_markov_1994}
Tierney, L. (1994).
\newblock Markov chains for exploring posterior distributions.
\newblock {\em the Annals of Statistics}, pages 1701--1728.
\newblock Publisher: JSTOR.

\bibitem[van~de Schoot et~al., 2021]{van_de_schoot_bayesian_2021}
van~de Schoot, R., Depaoli, S., King, R., Kramer, B., Märtens, K., Tadesse, M.~G., Vannucci, M., Gelman, A., Veen, D., Willemsen, J., and Yau, C. (2021).
\newblock Bayesian statistics and modelling.
\newblock {\em Nature Reviews Methods Primers}, 1(1):1--26.
\newblock Publisher: Nature Publishing Group.

\bibitem[van Erp, 2020]{van_erp_tutorial_2020}
van Erp, S. (2020).
\newblock A tutorial on {Bayesian} penalized regression with shrinkage priors for small sample sizes.
\newblock {\em Small sample size solutions}, pages 71--84.
\newblock Publisher: Routledge.

\bibitem[van Erp et~al., 2019]{van_erp_shrinkage_2019}
van Erp, S., Oberski, D.~L., and Mulder, J. (2019).
\newblock Shrinkage priors for {Bayesian} penalized regression.
\newblock {\em Journal of Mathematical Psychology}, 89:31--50.

\bibitem[Vats and Knudson, 2021]{vats_revisiting_2021}
Vats, D. and Knudson, C. (2021).
\newblock Revisiting the {Gelman}?{Rubin} {Diagnostic}.
\newblock {\em Statistical Science}, 36(4):518--529.
\newblock Publisher: Institute of Mathematical Statistics.

\bibitem[Vaughan and Dancho, 2022]{davis_vaughan_and_matt_dancho_furrr_2022}
Vaughan, D. and Dancho, M. (2022).
\newblock {\em furrr: Apply Mapping Functions in Parallel using Futures}.
\newblock R package version 0.3.1.

\bibitem[Vehtari et~al., 2024a]{Vehtari_loo_2024}
Vehtari, A., Gabry, J., Magnusson, M., Yao, Y., Bürkner, P.-C., Paananen, T., and Gelman, A. (2024a).
\newblock loo: Efficient leave-one-out cross-validation and waic for bayesian models.
\newblock R package version 2.8.0.9000.

\bibitem[Vehtari et~al., 2021]{vehtari_rank-normalization_2021}
Vehtari, A., Gelman, A., Simpson, D., Carpenter, B., and Bürkner, P.-C. (2021).
\newblock Rank-normalization, folding, and localization: {An} improved {R}{\textasciicircum} for assessing convergence of {MCMC} (with discussion).
\newblock {\em Bayesian analysis}, 16(2):667--718.
\newblock Publisher: International Society for Bayesian Analysis.

\bibitem[Vehtari et~al., 2024b]{vehtari_pareto_2024}
Vehtari, A., Simpson, D., Gelman, A., Yao, Y., and Gabry, J. (2024b).
\newblock Pareto smoothed importance sampling.
\newblock {\em Journal of Machine Learning Research}, 25(72):1--58.

\bibitem[Vowels, 2024]{vowels_typical_2024}
Vowels, M.~J. (2024).
\newblock Typical {Yet} {Unlikely} and {Normally} {Abnormal}: {The} {Intuition} {Behind} {High}-{Dimensional} {Statistics}.
\newblock {\em Statistics, Politics and Policy}, 15(1):87--113.

\bibitem[Welandawe et~al., 2024]{welandawe_framework_2024}
Welandawe, M., Andersen, M.~R., Vehtari, A., and Huggins, J.~H. (2024).
\newblock A framework for improving the reliability of black-box variational inference.
\newblock {\em Journal of Machine Learning Research}, 25(219):1--71.

\bibitem[Wickham, 2016]{Wickham_ggplot2_2016}
Wickham, H. (2016).
\newblock {\em ggplot2: Elegant Graphics for Data Analysis}.
\newblock Springer-Verlag New York.

\bibitem[Wickham et~al., 2023]{Wickham_dplyr_2023}
Wickham, H., François, R., Henry, L., Müller, K., and Vaughan, D. (2023).
\newblock {\em dplyr: A Grammar of Data Manipulation}.
\newblock R package version 1.1.4.

\bibitem[Yao et~al., 2018]{yao_yes_2018}
Yao, Y., Vehtari, A., Simpson, D., and Gelman, A. (2018).
\newblock Yes, but {Did} {It} {Work}?: {Evaluating} {Variational} {Inference}.
\newblock In {\em Proceedings of the 35th {International} {Conference} on {Machine} {Learning}}, pages 5581--5590. PMLR.
\newblock ISSN: 2640-3498.

\bibitem[Zhang et~al., 2019]{zhang_advances_2019}
Zhang, C., Bütepage, J., Kjellström, H., and Mandt, S. (2019).
\newblock Advances in {Variational} {Inference}.
\newblock {\em IEEE Transactions on Pattern Analysis and Machine Intelligence}, 41(8):2008--2026.
\newblock Conference Name: IEEE Transactions on Pattern Analysis and Machine Intelligence.

\bibitem[Zhang and Stephens, 2009]{zhang_new_2009}
Zhang, J. and Stephens, M.~A. (2009).
\newblock A {New} and {Efficient} {Estimation} {Method} for the {Generalized} {Pareto} {Distribution}.
\newblock {\em Technometrics}, 51(3):316--325.

\bibitem[Zhang et~al., 2022]{zhang_pathfinder_2022}
Zhang, L., Carpenter, B., Gelman, A., and Vehtari, A. (2022).
\newblock Pathfinder: {Parallel} quasi-{Newton} variational inference.
\newblock {\em Journal of Machine Learning Research}, 23(306):1--49.

\bibitem[Zhu et~al., 1997]{zhu_algorithm_1997}
Zhu, C., Byrd, R.~H., Lu, P., and Nocedal, J. (1997).
\newblock Algorithm 778: {L}-{BFGS}-{B}: {Fortran} subroutines for large-scale bound-constrained optimization.
\newblock {\em ACM Transactions on Mathematical Software}, 23(4):550--560.

\end{thebibliography}

\newpage

\section*{Appendix}

\subsection*{A. Leapfrog algorithm}

The leapfrog logarithms approximates the path of the Hamiltonian in finite steps. Table \ref{tab:theta_m_evolution} shows the functions used to calculate the position of the m and $\theta$ over time, where $\epsilon$ indicates the stepsize. At time t, a sample for the moment variable m is taken and the initial value of $\theta$ is used. These values are used to approximate the value of m at time point $t + \epsilon/2$, where the gradient with regards to the posterior is used. The value of $\theta$ is then estimated at time point $t + \epsilon$ using the updated value of the momentum, $m (t + \epsilon/2) $, after which the value of m at $t + \epsilon$ can be estimated. The value of m is thus updated at halve steps, whereas $\theta$ is only updated for hole steps. For a deeper dive into the leapfrog algorithm see \cite{brooks_mcmc_2011}. 

\begin{table}[h!]
\centering
\begin{tabular}{c|c|c|c}
 \textbf{Time:} & $t$ & $t + \frac{\epsilon}{2}$ & $t + \epsilon$ \\
\hline
$\theta$ & $\theta_t$ & — & $\theta_{t + \epsilon} = \theta(t) + \epsilon \cdot m_{t + \frac{\epsilon}{2}}$ \\
$m$ & $m_t \sim \text{MVN}(0, I)$ & $m_{t + \frac{\epsilon}{2}} = m(t) - \frac{\epsilon}{2} \cdot \frac{\partial H}{\partial \theta} (\theta_t)$ & $m_{t + \epsilon} = m_{t + \frac{\epsilon}{2}} - \frac{\epsilon}{2} \cdot \frac{\partial H}{\partial \theta} (\theta_{t + \epsilon})$ \\
\end{tabular}
\caption{Evolution of parameters over time for the leapfrog algorithm.}
\label{tab:theta_m_evolution}
\end{table}

\subsection*{B. Additional results}

\begin{table}[H]
\centering
\begin{tabular}{lrrrrrrrr}
  \hline
method  scenario & 1a & 1b & 2a & 2b & 3a & 3b & 4a & 4b \\ 
  \hline
MC & 499 & 500 & 499 & 498 &  57 & 129 & 184 &  93 \\ 
  Pathfinder -$>$ HMC & 500 & 500 & 500 & 500 &  28 & 194 & 190 & 201 \\ 
  Mean-field VI & 492 & 497 & 500 & 500 & 234 & 250 & 250 & 149 \\ 
  Full-rank VI &  50 &  36 &   0 &   0 &   0 &   0 &   0 &   0 \\ 
  Laplace & 500 & 500 &   0 &   0 &   0 &   0 &   0 &   0 \\ 
   \hline
\end{tabular}
\caption{Number of datasets which complied with the quality criteria}
\label{tab:n_conv}
\end{table}

\newpage

\begin{table}[ht]
\small
\centering
\begin{tabular}{llrllllr}
  \hline
Scenario & Method & Output (\%) & Runtime & Bias $\beta$ & Mse & Cov \\ 
  \hline
1a & HMC & 1.00 & 4.94 & 0.41 & 0.56 & 0.95 \\ 
  1a & Pathfinder -$>$ HMC & 1.00 & 5.58 (1.29) & 0.41 & 0.56 & 0.95 \\ 
  1a & Mean-field VI & 0.98 & 0.72 & 0.62 & 0.59 & 0.97 \\ 
  1a & Full-rank VI & 0.10 & 0.81 & 0.48 & 0.59 & 0.98 \\ 
  1a & Laplace & 1.00 & 0.7 & 0.5 & 0.64 & 0.97 \\ 
    \hline
  1b & HMC & 1.00 & 5.26 & 0.48 & 0.55 & 0.95 \\ 
  1b & Pathfinder -$>$ HMC & 1.00 & 5.66 (1.32) & 0.48 & 0.55 & 0.95 \\ 
  1b & Mean-field VI & 0.99 & 0.75 & 0.62 & 0.57 & 0.97 \\ 
  1b & Full-rank VI & 0.07 & 0.89 & 0.63 & 0.6 & 0.98 \\ 
  1b & Laplace & 1.00 & 0.71 & 0.53 & 0.64 & 0.98 \\ 
    \hline
  2a & HMC & 1.00 & 10.81 & 0.57 & 0.59 & 0.95 \\ 
  2a & Pathfinder -$>$ HMC & 1.00 & 12.18 (2.57) & 0.57 & 0.59 & 0.95 \\ 
  2a & Mean-field VI & 1.00 & 0.99 & 1.05 & 0.84 & 0.95 \\ 
    \hline
  2b & HMC & 1.00 & 12.42 & 0.72 & 0.58 & 0.95 \\ 
  2b & Pathfinder -$>$ HMC & 1.00 & 11.7 (2.47) & 0.72 & 0.58 & 0.95 \\ 
  2b & Mean-field VI & 1.00 & 1.08 & 0.91 & 0.66 & 0.98 \\ 
    \hline
  3a & HMC & 0.23 & 85.61 & 0.9 & 0.55 & 0.95 \\ 
  3a & Pathfinder -$>$ HMC & 0.11 & 78.06 (8.55) & 0.98 & 0.56 & 0.95 \\ 
  3a & Mean-field VI & 0.94 & 2.88 & 1.05 & 0.95 & 0.95 \\ 
    \hline
  3b & HMC & 0.52 & 65.2 & 0.95 & 0.51 & 0.95 \\ 
  3b & Pathfinder -$>$ HMC & 0.78 & 58.55 (8.25) & 0.96 & 0.51 & 0.95 \\ 
  3b & Mean-field VI & 1.00 & 3.28 & 0.92 & 0.54 & 0.98 \\ 
    \hline
  4a & HMC & 0.74 & 99.34 & 0.56 & 0.85 & 0.95 \\ 
  4a & Pathfinder -$>$ HMC & 0.76 & 101.48 (15.75) & 0.55 & 0.86 & 0.95 \\ 
  4a & Mean-field VI & 1.00 & 6.97 & 0.98 & 1.01 & 0.95 \\ 
    \hline
  4b & HMC & 0.37 & 95.18 & 0.97 & 0.82 & 0.95 \\ 
  4b & Pathfinder -$>$ HMC & 0.80 & 90.87 (15.12) & 0.99 & 0.82 & 0.95 \\ 
  4b & Mean-field VI & 0.60 & 5.81 & 0.96 & 0.87 & 0.96 \\ 
   \hline
\end{tabular}
\caption{Full results for all algorithms with filtering out the models with a bad criteria part 1. For the Pathfinder -$>$ HMC the runtime of Pathfinder is added in brackets next to the total. }
\label{tab:full_res1}
\end{table}

\newpage

\begin{table}[ht]
\centering
\small
\begin{tabular}{llrrrrrr}
  \hline
scenario & method & n\_div & $\hat{R}$\_mean & $\hat{R}$\_max & ess\_bulk\_mean & ess\_bulk\_min & Pareto $\hat{k}$ \\ 
  \hline
1a & HMC & 0.10 & 1.00 & 8433.44 &  \\ 
  1a & Pathfinder -$>$ HMC & 0.01 & 1.00 & 9440.49 &  \\ 
  1a & Mean-field VI &  &  &  & 1.20 \\ 
  1a & Full-rank VI &  &  &  & 1.72 \\ 
  1a & Laplace &  &  &  & 2.01 \\ 
    \hline
  1b & HMC & 0.10 & 1.00 & 7577.60 &  \\ 
  1b & Pathfinder -$>$ HMC & 0.01 & 1.00 & 9511.47 &  \\ 
  1b & Mean-field VI &  &  &  & 1.18 \\ 
  1b & Full-rank VI &  &  &  & 1.52 \\ 
  1b & Laplace &  &  &  & 1.96 \\ 
    \hline
  2a & HMC & 0.06 & 1.00 & 7598.02 &  \\ 
  2a & Pathfinder -$>$ HMC & 0.01 & 1.00 & 8879.26 &  \\ 
  2a & Mean-field VI &  &  &  & 1.57 \\ 
    \hline
  2b & HMC & 0.10 & 1.00 & 8369.83 &  \\ 
  2b & Pathfinder -$>$ HMC & 0.00 & 1.00 & 8936.79 &  \\ 
  2b & Mean-field VI &  &  &  & 2.00 \\ 
    \hline
  3a & HMC & 0.61 & 1.00 & 7293.81 &  \\ 
  3a & Pathfinder -$>$ HMC & 0.26 & 1.00 & 7280.83 &  \\ 
  3a & Mean-field VI &  &  &  & 2.97 \\ 
    \hline
  3b & HMC & 3.00 & 1.00 & 5249.27 &  \\ 
  3b & Pathfinder -$>$ HMC & 0.55 & 1.00 & 6370.71 &  \\ 
  3b & Mean-field VI &  &  &  & 2.88 \\ 
    \hline
  4a & HMC & 0.43 & 1.00 & 6627.99 &  \\ 
  4a & Pathfinder -$>$ HMC & 0.30 & 1.00 & 7323.29 &  \\ 
  4a & Mean-field VI &  &  &  & 3.35 \\ 
    \hline
  4b & HMC & 3.29 & 1.00 & 4232.88 &  \\ 
  4b & Pathfinder -$>$ HMC & 0.69 & 1.00 & 6771.56 &  \\ 
  4b & Mean-field VI &  &  &  & 3.58 \\ 
   \hline
\end{tabular}
\caption{Full results for all algorithms with filtering out the models with a bad criteria part 2. }
\label{tab:full_res2}
\end{table}

\newpage

\begin{table}[ht]
\small
\centering
\begin{tabular}{llrllllr}
  \hline
Scenario & Method & Output (\%) & Runtime & Bias $\beta$ & Mse & Cov \\ 
  \hline
1a & HMC & 1.00 & 4.94 & 0.41 & 0.56 & 0.95 \\ 
  1a & Pathfinder -$>$ HMC & 1.00 & 5.58 (1.29) & 0.41 & 0.56 & 0.95 \\ 
  1a & Mean-field VI & 0.98 & 0.72 & 0.62 & 0.59 & 0.97 \\ 
  1a & Full-rank VI & 0.98 & 0.92 & 1.8 & 5.46 & 1.00 \\ 
  1a & Laplace & 1.00 & 0.7 & 0.5 & 0.64 & 0.97 \\ 
    \hline
  1b & HMC & 1.00 & 5.26 & 0.48 & 0.55 & 0.95 \\ 
    1b & Pathfinder -$>$ HMC & 1.00 & 5.66 (1.32) & 0.48 & 0.55 & 0.95 \\ 
  1b & Mean-field VI & 0.99 & 0.75 & 0.62 & 0.57 & 0.97 \\ 
  1b & Full-rank VI & 1.00 & 0.94 & 3.56 & 7.02 & 1.00 \\ 
  1b & Laplace & 1.00 & 0.71 & 0.53 & 0.64 & 0.98 \\ 
    \hline
  2a & HMC & 1.00 & 10.81 & 0.57 & 0.59 & 0.95 \\ 
  2a & Pathfinder -$>$ HMC & 1.00 & 12.18 (2.57) & 0.57 & 0.59 & 0.95 \\ 
  2a & Mean-field VI & 1.00 & 0.99 & 1.05 & 0.84 & 0.95 \\ 
  2a & Full-rank VI & 0.42 & 6.59 & 3.88 & 71.91 & 1.00 \\ 
  2a & Laplace & 0.94 & 1.21 & $>$10\verb|^|4 & $>$10\verb|^|4 & 0.99 \\ 
    \hline
  2b & HMC & 1.00 & 12.42 & 0.72 & 0.58 & 0.95 \\ 
  2b & Pathfinder -$>$ HMC & 1.00 & 11.7 (2.47) & 0.72 & 0.58 & 0.95 \\ 
  2b & Mean-field VI & 1.00 & 1.08 & 0.91 & 0.66 & 0.98 \\ 
  2b & Full-rank VI & 0.42 & 6.49 & 12.24 & 79.5 & 1.00 \\ 
  2b & Laplace & 0.94 & 1.17 & $>$10\verb|^|4 & $>$10\verb|^|4 & 0.99 \\ 
    \hline
  3a & HMC & 1.00 & 90.03 & 1.11 & 0.54 & 0.95 \\ 
  3a & Pathfinder -$>$ HMC & 1.00 & 68.92 (8.63) & 1.1 & 0.54 & 0.95 \\ 
  3a & Mean-field VI & 1.00 & 2.88 & 1.04 & 0.94 & 0.96 \\ 
  3a & Full-rank VI & 0.68 & 589.16 & 7.09 & 31.03 & 1.00 \\ 
  3a & Laplace & 0.99 & 4.16 & $>$10\verb|^|4 & $>$10\verb|^|4 & 1.00 \\ 
    \hline
  3b & HMC & 1.00 & 66.08 & 0.98 & 0.52 & 0.95 \\ 
  3b & Pathfinder -$>$ HMC & 1.00 & 60.46 (8.23) & 0.98 & 0.52 & 0.95 \\ 
  3b & Mean-field VI & 1.00 & 3.28 & 0.92 & 0.54 & 0.98 \\ 
  3b & Full-rank VI & 0.14 & 571.17 & 27.87 & 102.6 & 1.00 \\ 
  3b & Laplace & 1.00 & 4.1 & $>$10\verb|^|4 & $>$10\verb|^|4 & 0.98 \\ 
    \hline
  4a & HMC & 1.00 & 101.7 & 0.58 & 0.86 & 0.95 \\ 
  4a & Pathfinder -$>$ HMC & 1.00 & 103.98 (15.71) & 0.58 & 0.86 & 0.95 \\
  4a & Mean-field VI & 1.00 & 6.97 & 0.98 & 1.01 & 0.95 \\ 
  4a & Full-rank VI & 0.05 & 2717.97 & 110.01 & 1953.55 & 0.99 \\ 
  4a & Laplace & 0.98 & 14.86 & $>$10\verb|^|4 & $>$10\verb|^|4 & 0.99 \\ 
    \hline
  4b & HMC & 1.00 & 96.01 & 1.07 & 0.82 & 0.95 \\ 
  4b & Pathfinder -$>$ HMC & 1.00 & 90.93 (15.05) & 1.03 & 0.82 & 0.95 \\ 
  4a & Full-rank VI & 0 &  &  & &  \\ 
  4b & Mean-field VI & 1.00 & 5.72 & 0.95 & 0.86 & 0.96 \\ 
  4b & Laplace & 0.99 & 15.49 & $>$10\verb|^|4 & $>$10\verb|^|4 & 0.97 \\ 
   \hline
\end{tabular}
\caption{Full results for all algorithms without filtering part 1. For the Pathfinder -$>$ HMC the runtime of Pathfinder is added in brackets next to the total. }
\label{tab:full_res1_all}
\end{table}

\newpage

\begin{table}[ht]
\centering
\small
\begin{tabular}{llrrrrrr}
  \hline
scenario & method & n\_div & $\hat{R}$\_mean & $\hat{R}$\_max & ess\_bulk\_mean & ess\_bulk\_min & Pareto $\hat{k}$ \\ 
  \hline
1a & HMC & 0.11 & 1.00 & 1.00 & 8424.63 & 4473.05 &  \\ 
  1a & Mean-field VI &  &  &  &  &  & 1.20 \\ 
  1a & Full-rank VI &  &  &  &  &  & 12.54 \\ 
  1a & Laplace &  &  &  &  &  & 2.01 \\ 
  1a & Pathfinder -$>$ HMC & 0.01 & 1.00 & 1.00 & 9440.49 & 5297.42 &  \\ 
    \hline
  1b & HMC & 0.10 & 1.00 & 1.00 & 7577.60 & 4253.26 &  \\ 
  1b & Mean-field VI &  &  &  &  &  & 1.18 \\ 
  1b & Full-rank VI &  &  &  &  &  & 15.47 \\ 
  1b & Laplace &  &  &  &  &  & 1.96 \\ 
  1b & Pathfinder -$>$ HMC & 0.01 & 1.00 & 1.00 & 9511.47 & 5430.34 &  \\ 
    \hline
  2a & HMC & 0.07 & 1.00 & 1.00 & 7591.07 & 2667.35 &  \\ 
  2a & Mean-field VI &  &  &  &  &  & 1.57 \\ 
  2a & Full-rank VI &  &  &  &  &  & 24.25 \\ 
  2a & Laplace &  &  &  &  &  & Inf \\ 
  2a & Pathfinder -$>$ HMC & 0.01 & 1.00 & 1.00 & 8879.26 & 2976.89 &  \\ 
    \hline
  2b & HMC & 0.11 & 1.00 & 1.00 & 8355.26 & 3779.31 &  \\ 
  2b & Mean-field VI &  &  &  &  &  & 2.00 \\ 
  2b & Full-rank VI &  &  &  &  &  & 24.53 \\ 
  2b & Laplace &  &  &  &  &  & Inf \\ 
  2b & Pathfinder -$>$ HMC & 0.00 & 1.00 & 1.00 & 8936.79 & 3599.07 &  \\ 
    \hline
  3a & HMC & 0.97 & 1.00 & 1.01 & 7141.84 & 627.39 &  \\ 
  3a & Mean-field VI &  &  &  &  &  & 3.06 \\ 
  3a & Full-rank VI &  &  &  &  &  & 26.81 \\ 
  3a & Laplace &  &  &  &  &  & Inf \\ 
  3a & Pathfinder -$>$ HMC & 0.71 & 1.00 & 1.02 & 6985.60 & 451.09 &  \\ 
    \hline
  3b & HMC & 6.24 & 1.01 & 1.04 & 3444.91 & 1406.26 &  \\ 
  3b & Mean-field VI &  &  &  &  &  & 2.88 \\ 
  3b & Full-rank VI &  &  &  &  &  & 50.79 \\ 
  3b & Laplace &  &  &  &  &  & Inf \\ 
  3b & Pathfinder -$>$ HMC & 1.47 & 1.00 & 1.01 & 5349.28 & 2691.00 &  \\ 
    \hline
  4a & HMC & 1.11 & 1.00 & 1.01 & 5664.95 & 1733.19 &  \\ 
  4a & Mean-field VI &  &  &  &  &  & 3.35 \\ 
  4a & Full-rank VI &  &  &  &  &  & Inf \\ 
  4a & Laplace &  &  &  &  &  & Inf \\ 
  4a & Pathfinder -$>$ HMC & 0.90 & 1.00 & 1.01 & 6507.61 & 1812.44 &  \\ 
    \hline
  4b & HMC & 6.83 & 1.01 & 1.04 & 2375.39 & 913.54 &  \\ 
  4b & Mean-field VI &  &  &  &  &  & 3.90 \\ 
  4b & Full-rank VI &  &  &  &  &  &  \\ 
  4b & Laplace &  &  &  &  &  & Inf \\ 
  4b & Pathfinder -$>$ HMC & 1.60 & 1.00 & 1.02 & 5787.01 & 3297.05 &  \\ 
   \hline
\end{tabular}
\caption{Full results for all algorithms without filtering part 2. }
\label{tab:full_res2_all}
\end{table}

\newpage

\subsection*{C. VI with the regularized horseshoe}

The posterior of the regression coefficient, up to a normalizing constant, can be written as:

\begin{equation}
    p(\beta_0, beta, \sigma^2, \lambda | X, y) \propto \underbrace{p(y|\beta_0, x, \beta, \sigma^2)}_{\text{likelihood}} * \underbrace{p(\beta_0) * p(\beta | \sigma^2, \lambda) * p(\sigma^2) p(\lambda)}_\text{prior}
\end{equation}

The prior for the regression coefficients, disregarding $\beta_0$, depends on the shrinkage prior $\lambda$. In case of the regularized horseshoe prior, the following prior is used as stated in equation \ref{eq:hs_r1}:

\begin{equation*}
        \beta_j | \lambda_j, \tau, c \sim Normal(0, \tau^2  \Tilde{\lambda^2_j}),
\end{equation*}

This can be rewritten as:

\begin{equation*}
        \beta_j \sim Normal(0, 1) * \tau^2  \Tilde{\lambda^2_j},
\end{equation*}

In the VI approximations, the posterior for all parameters will be normal in the unconstrained space. The mean of the posterior for the regression parameters has no bounds, so the contained space is the uncontrained space. The variance terms, however, have a lower bound of 0. These will first be transformed to the log scale (uncontained space) and then normally approximated. These normals are then back transformed to the constrained space, which can results in non-normal posteriors. 

The posterior for a regression coefficient can be written as:

\begin{equation}
      \beta_j \sim Normal(\mu, 1) * \tau^2  \Tilde{\lambda^2_j}.
      \label{eq:post_reg}
\end{equation}

\begin{figure}[H]
    \centering
    \includegraphics[width=0.7\linewidth]{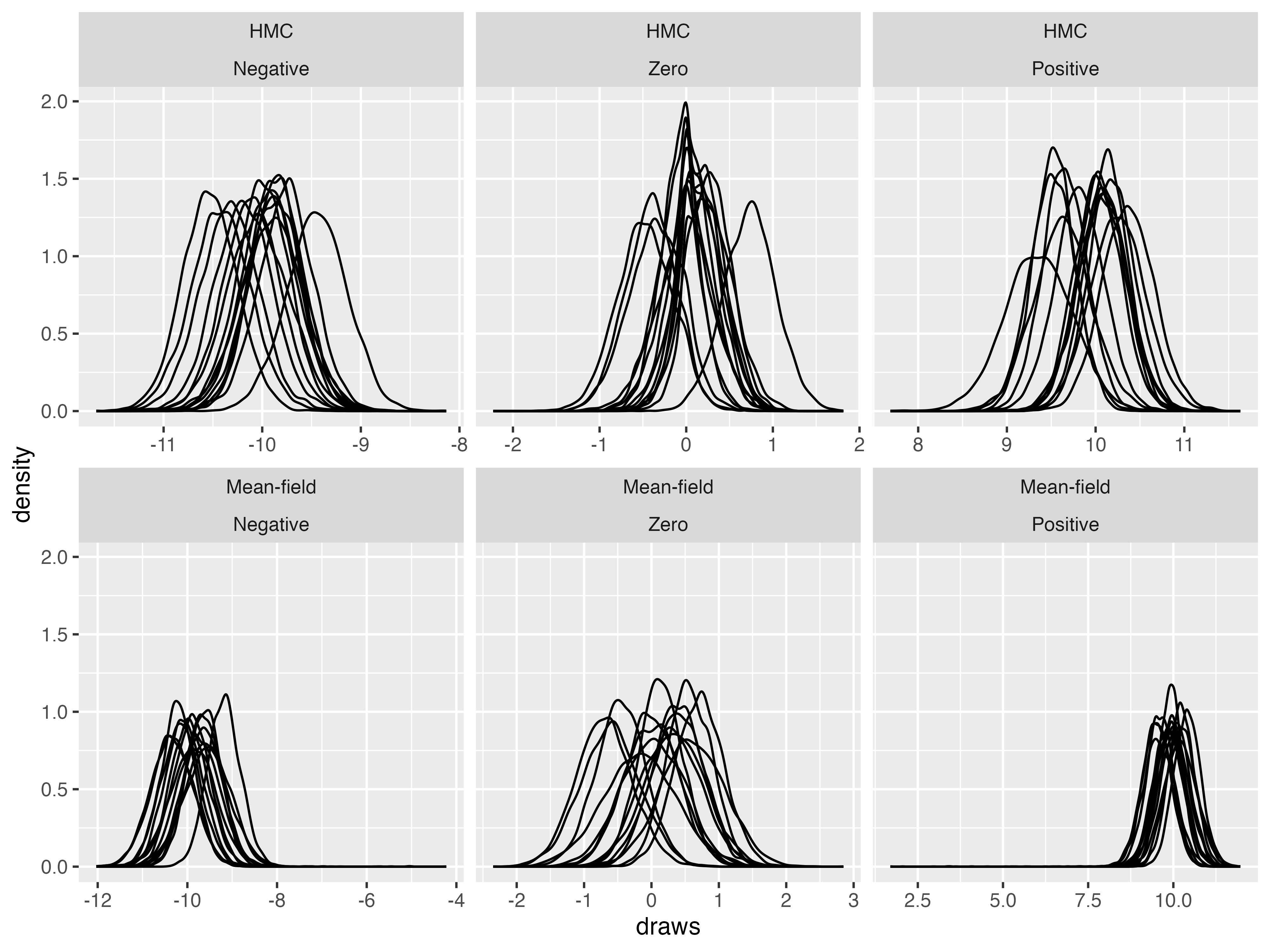}
    \caption{Posteriors obtain using mean-field VI for the regression coefficients (intercept excluded).}
    \label{fig:stan_ex_a}
\end{figure}

The resulting distribution does not have to be normal, take the following example. The regularized horseshoe prior is used in a regression model with 15 negative parameters (set to -10), 15 zero parameters and 10 positive parameters (set to 10). In Figure \ref{fig:stan_ex_a}, the posterior densities are shown. The posteriors using mean-field VI clearly show non-normal posteriors. 

\begin{figure}[H]
    \centering
    \includegraphics[width=0.7\linewidth]{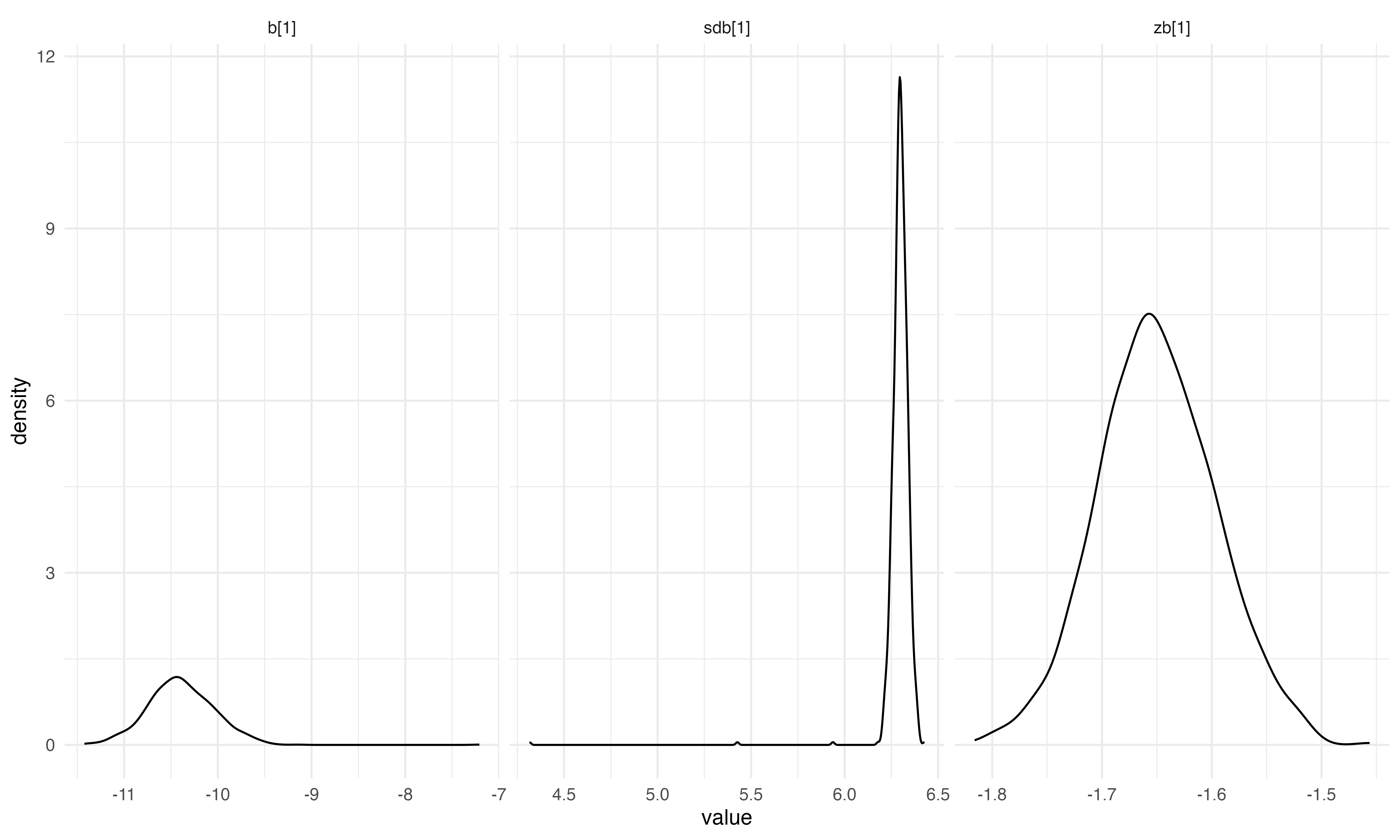}
    \caption{The posterior draws for a single negative parameters (set to 0) split by components.}
    \label{fig:stan_ex_b}
\end{figure}

By zooming in on a negative parameter, the two parts of Equation \ref{eq:post_reg} can be visualized in Figure \ref{fig:stan_ex_b}. There sbs is $\tau^2  \Tilde{\lambda^2_j}$ and zb is $Normal(\mu, \sigma) $. The resulting product b is clearly non normal, due to the heavy tails towards zero. 

\subsection*{D. Prior sensitivity analysis}

In a sensitive analysis four different prior were used:

\begin{itemize}
    \item regularized Horseshoe, df = 1
    \item regularized Horseshoe, df = 3
    \item Normal(0,0.01)
    \item Student t (3, 0, 0.1)
\end{itemize}

There were 14 runs from the mean-field omitted from the figure, because the results had a MSE of \> 1000. Of the 14, 9 used the student-t prior. 

The coverage for the mean-field is overestimated in every scenario with the regularized horseshoe; that is not the case for HMC. The normal prior for mean-field seems to have the least overestimation. 

\begin{figure}[H]
    \centering
    \includegraphics[width=0.7\linewidth]{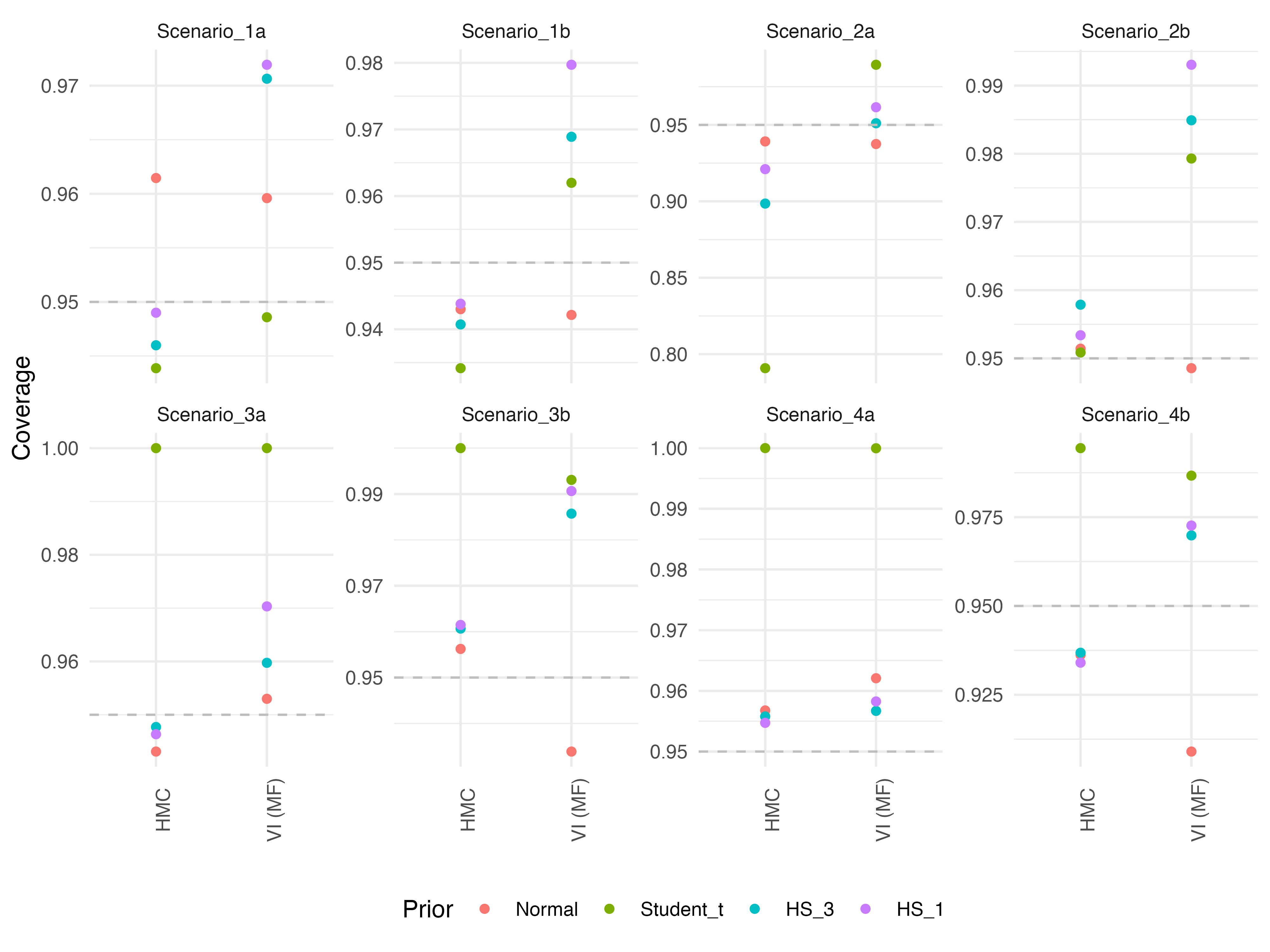}
    \caption{Coverage of the prediction interval for HMC and mean-field VI with different priors. }
    \label{fig:sens_cov}
\end{figure}

With regards to the predictive performance, it can be seen that the horseshoe priors work the best across all scenarios. The student-t and normal prior results in much higher mse for most scenarios.

\begin{figure}[H]
    \centering
    \includegraphics[width=0.7\linewidth]{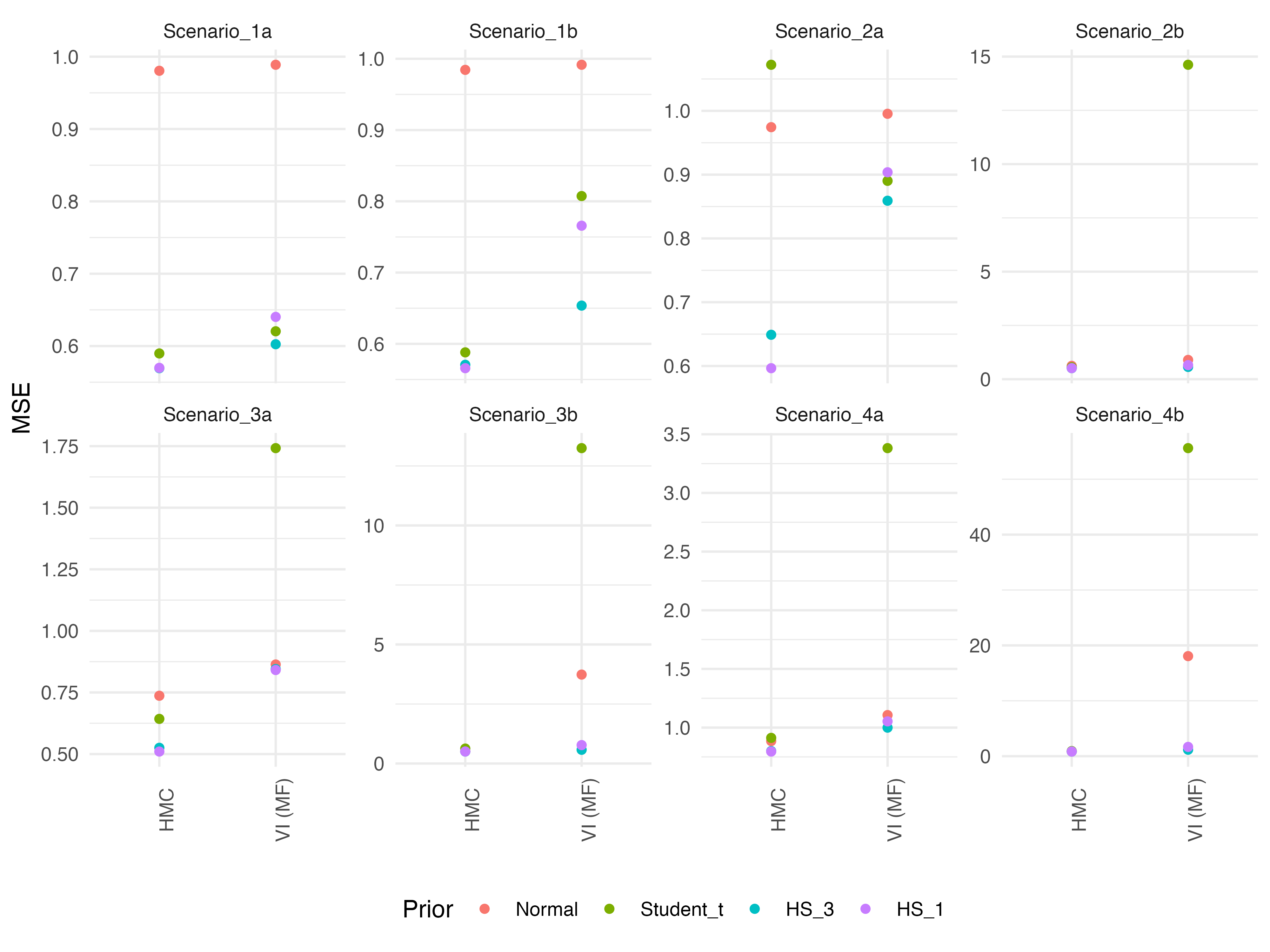}
    \caption{MSE of the prediction interval for HMC and mean-field VI with different priors. }
    \label{fig:sens_mse}
\end{figure}

There is thus a tradeoff. Using the horseshoe prior with the mean-field method results in low estimation error, but an overestimation of the uncertainty. The overestimation can be, partly, corrected at the cost of the mse. 

\newpage

\subsection*{E. Pareto $\hat{k}$ and model predictive performance}

\begin{figure}[H]
    \centering
    \includegraphics[width=\linewidth]{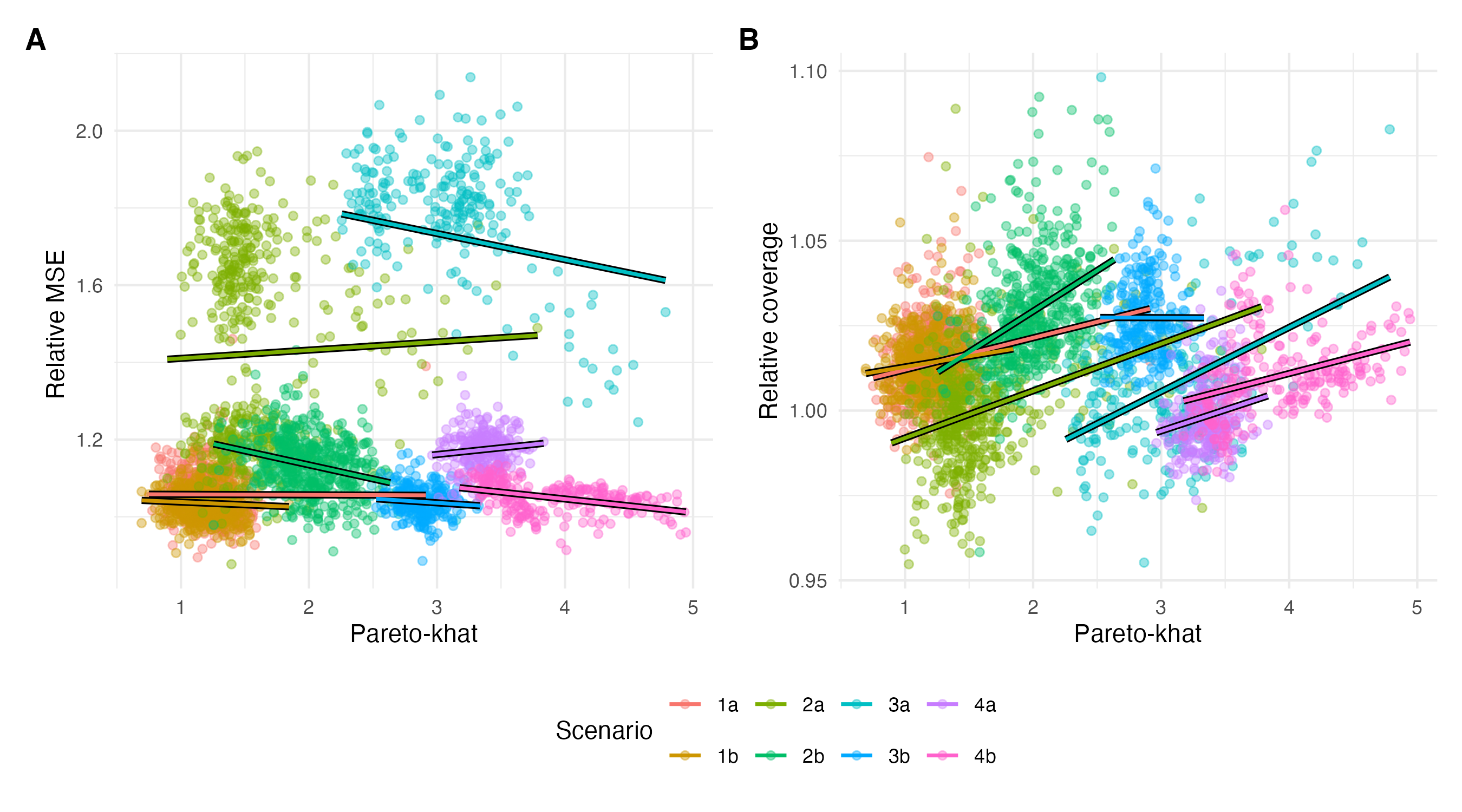}
    \caption{The MSE and coverage on the test set for mean-field VI relative to HMC, split by the value for the Pareto $\hat{k}$. A higher value for the relative value indicates that the estimate was higher for mean-field VI than for HMC. Plot \textbf{A} indicates that worse predictive performance does not seem to be connected to higher Pareto $\hat{k}$ values. In plot \textbf{B} the width of the prediction interval seems to be positively related to the Pareto $\hat{k}$ value.}
    \label{fig:mse_pareto}
\end{figure}

For the non-MCMC methods, the convergence of the model can be assessed with the Pareto $\hat{k}$ value. Higher Pareto $\hat{k}$ values indicate a larger divergence between the variational approximation and the unnormalized posterior. To evaluate if the Pareto $\hat{k}$ can also be used to assess specific attributes of the posterior predictive distribution, outcome metrics between HMC and mean-field VI are compared for different Pareto $\hat{k}$ values. The Pareto $\hat{k}$ value and relative MSE distinctly separate the simulation scenarios, and there does not seem to be a clear visible relationship between the Pareto $\hat{k}$ value and the relative performance in terms of MSE (Figure \ref{fig:mse_pareto}). For the coverage it can be seen that higher Pareto $\hat{k}$ values indicate overcoverage, which in turn is caused by a variational approximation of the latent variables that is too wide. 

One potential reason for the relationship between the Pareto $\hat{k}$ values and the coverage can come from the long tail behavior of the mean-field approximation (Appendix C). The Pareto $\hat{k}$ is calculated based on area's of the variational approximation where the posterior has little density. If the variational approximation has very long tails towards zero, then there will be less density at the opposite tail, where we do expect there to be density. The difference in density at the "short" tail can cause higher importance weights, and thus a higher Pareto $\hat{k}$ value. 



\newpage

\subsection*{F. Additional empirical results}

Additional results continuous outcome:

\begin{table}[H]
\centering
\begin{tabular}{llrr}
  \hline
dataset & method & n\_div & Pareto $\hat{k}$ \\ 
  \hline
math & HMC & 0.10 &  \\ 
  math & Pathfinder -$>$ HMC & 0.07 &  \\ 
  math & VI (mean-field) &  & 1.58 \\ 
  math & VI (full-rank) &  & Inf \\ 
  math & Laplace &  & 2.66 \\ 
    \hline
  loan & Pathfinder -$>$ HMC &  &  \\ 
  loan & VI (mean-field) &  & 48.70 \\ 
  loan & VI (full-rank) &  &  \\ 
  loan & Laplace &  &  \\ 
    \hline
  topo & HMC & 0.85 &  \\ 
  topo & Pathfinder -$>$ HMC & 0.00 &  \\ 
  topo & VI (mean-field) &  & 8.60 \\ 
  topo & VI (full-rank) &  &  \\ 
  topo & Laplace &  & Inf \\ 
    \hline
  qsar & HMC & 0.64 &  \\ 
  qsar & Pathfinder -$>$ HMC & 0.00 &  \\ 
  qsar & VI (mean-field) &  & 7.80 \\ 
  qsar & VI (full-rank) &  &  \\ 
  qsar & Laplace &  & 100.52 \\ 
    \hline
  cristalli & HMC & 5.81 &  \\ 
  cristalli & Pathfinder -$>$ HMC & 24.03 &  \\ 
  cristalli & VI (mean-field) &  & 3.81 \\ 
  cristalli & VI (full-rank) &  &  \\ 
  cristalli & Laplace &  & Inf \\ 
   \hline
\end{tabular}
\caption{Results from empirical datasets with an continuous outcome.}
\label{tab:emp_res_cont3}
\end{table}

Additional results binary outcome:

\begin{table}[H]
\centering
\begin{tabular}{llrr}
  \hline
dataset & method & n\_div & Pareto $\hat{k}$ \\ 
  \hline
ALLAML & HMC & 0.133 &  \\ 
  ALLAML & Pathfinder -$>$ HMC &  &  \\ 
  ALLAML & VI (mean-field) &  & 9.602 \\ 
  ALLAML & VI (full-rank) &  &  \\ 
  ALLAML & Laplace &  & 58.716 \\ 
    \hline
  parkinson & HMC & 5.013 &  \\ 
  parkinson & Pathfinder -$>$ HMC &  &  \\ 
  parkinson & VI (mean-field) &  & 61.936 \\ 
  parkinson & VI (full-rank) &  &  \\ 
  parkinson & Laplace &  &  \\ 
    \hline
  prostate & HMC & 0.017 &  \\ 
  prostate & Pathfinder -$>$ HMC &  &  \\ 
  prostate & VI (mean-field) &  & 7.754 \\ 
  prostate & VI (full-rank) &  &  \\ 
  prostate & Laplace &  & 19.314 \\ 
    \hline
  annomolies & HMC & 7.415 &  \\ 
  annomolies & Pathfinder -$>$ HMC &  &  \\ 
  annomolies & VI (mean-field) &  & 5.243 \\ 
  annomolies & VI (full-rank) &  & 42.194 \\ 
  annomolies & Laplace &  &  Inf \\ 
   \hline
\end{tabular}
\caption{Additional results from empirical datasets with a binary outcome.}
\label{tab:emp_res_bin2}
\end{table}

\end{document}